\documentclass[useAMS,usenatbib]{mn2e}
\usepackage{amssymb, graphicx}
\usepackage{hyperref}

\def\aap{A{\&}A}

\def\aj{AJ}
\def\apj{ApJ}
\def\apjl{ApJL}

\def\icarus{Icarus}
\def\mnras{MNRAS}

\title{Possible scenarios for eccentricity evolution in the extrasolar 
planetary system HD 181433}
\author[G. Campanella, R.P. Nelson, C.B. Agnor]{Giammarco Campanella$^{1}$\thanks{E-mail:
g.campanella@qmul.ac.uk}, Richard P. Nelson$^1$ and Craig B. Agnor$^1$\\
$^{1}$Astronomy Unit, School of Physics and Astronomy, Queen Mary University of London, 
Mile End Road, London E1 4NS, U.K.}

\begin{document}

\date{Accepted 2013 May 31. Received 2013 May 30; in original form 2012 December 22}

\pagerange{\pageref{firstpage}--\pageref{lastpage}} \pubyear{2013}

\maketitle

\label{firstpage}

\begin{abstract}
We analyse the dynamics of the multiple planet system HD 181433. This consists
of two gas giant planets (bodies $c$ and $d$) with $m \sin{i}=0.65$ M$_{\rm Jup}$ and
0.53 M$_{\rm Jup}$ orbiting with periods 975 and 2468 days, respectively. 
The two planets appear to be in a 5:2 mean motion resonance, as this is required
for the system to be dynamically stable. A third planet with mass 
$m_b \sin{i} = 0.023$ M$_{\rm Jup}$ orbits close to the star with 
orbital period 9.37 days. Each planet orbit is significantly eccentric, with current 
values estimated to be $e_b=0.39$, $e_c=0.27$ and $e_d=0.47$.
In this paper we assess different scenarios that may explain the origin of these eccentric
orbits, with particular focus on the innermost body, noting that
the large eccentricity of planet $b$ cannot be explained through
secular interaction with the outer pair.
We consider a scenario in which the system previously contained 
an additional giant planet that was ejected during a period of dynamical instability 
among the planets. N-body simulations are presented that demonstrate
that during scattering and ejection among the outer planets a
close encounter between a giant and the inner body can raise $e_b$ to its observed 
value. Such an outcome occurs with a frequency of a few percent. We also demonstrate, however, 
that obtaining the required value of $e_b$ and having the two surviving outer planets land in 
5:2 resonance is a rare outcome, leading us to consider alternative scenarios involving
secular resonances.  We consider the possibility that an
  undetected planet in the system increases the secular forcing of 
planet {\it b} by the exterior giant planets, but we find that the
resulting eccentricity is not large enough to agree with the
observed one.
We also consider a scenario in which the spin-down of the central star causes the system 
to pass through secular resonance. 
In its simplest form this latter scenario fails to produce the system observed today, 
with the mode of failure depending sensitively on the rate of stellar spin-down. 
For spin-down rates above a critical value, planet $b$ passes through the resonance too quickly, 
and the forced eccentricity only reaches maximum values $e_b \simeq 0.25$. 
Spin-down rates below the critical value lead to long-term capture of planet $b$ in secular 
resonance, driving the eccentricity toward unity. If additional short-period low mass planets 
are present in the system, however, we find that mutual scattering can release planet $b$ 
from the secular resonance, leading to a system with orbital parameters similar to those observed 
today.

\end{abstract}


\section{Introduction}
\label{Sec:Intro}

Discovery of the first extrasolar planet around a sun-like star, 51 Peg b \citep{1995Natur.378..355M}, 
and subsequent detection of more than 850 exoplanets raises many questions about the formation 
and dynamics of planetary systems.  Now that we know of more than 120 multi-planet 
systems\footnote{``The Extrasolar Planets Encyclopaedia'' \url{http://www.exoplanet.eu}}, 
a broad range of planetary system architectures have been sampled. 
This variegate ensemble offers a wide range of features that are not displayed by our Solar System: 
giant planets in short period orbits (``hot Jupiters''), the wide eccentricity distribution of 
exoplanets, the relatively common detection of mean motion resonances (MMRs) in multiple planet 
systems offer interesting examples against which theories of planetary formation and evolution 
need to be tested \citep[e.g.][]{b1}.

Determining the physical processes that lead to the observed diversity of planetary systems remains an
area of active research and debate. The discovery of dynamically quiescent, short-period multiplanet 
systems by the Kepler mission, such as Kepler 11 \citep{b39}, provides clear evidence 
for disc-driven type I migration \citep{1997Icar..126..261W} of bodies that probably formed beyond 
the ice line in their protoplanetary discs. Similarly, giant planet systems in 2:1 mean-motion 
resonance such as GJ 876 \citep{2001ApJ...556..296M} provide evidence for disc-driven type II migration 
of gap-forming planets because of the slow convergent migration rates required for resonant capture
\citep{2001A&A...374.1092S, 2002ApJ...567..596L, 2008A&A...483..325C}, as do giant planets 
on short to intermediate period orbits that are too remote from their central stars to have
undergone significant tidal evolution.

The relatively large eccentricities observed for the extrasolar planet population as a whole, however, 
indicate that disc-planet interactions do not tell the whole story. A plausible explanation for
many of these systems is that they formed on near-circular orbits in the disc, but after disc 
dispersal dynamical instability led to planet-planet gravitational scattering \citep{b7}. Evidence
is provided by the fact that numerical planet scattering experiments can reproduce the observed
eccentricity distribution successfully \citep{b5,2008ApJ...686..603J}, and the observation that 
a number of short-period
transiting planets have orbit planes inclined significantly with respect to the stellar equatorial plane \citep{2010ApJ...718L.145W}.
Further ideas for explaining eccentric orbits include eccentricity driving through disc-driven 
migration of a resonant pair of planets \citep{b29}, and sweeping secular 
resonances caused by various
combinations of protoplanetary disc dispersal, spin-down of an initially rapidly spinning oblate 
central star, and general relativistic precession \citep{b31,b8}.

The three-planet extrasolar system HD 181433 has an architecture that makes it
an interesting test-bed for exploring dynamical processes during and after planetary system 
formation. \cite{b11} report that the system consists of two outer gas giant planets 
(`$c$' and `$d$') on eccentric orbits with periods $962 \pm 15$ and $2172 \pm 158$ days, 
and an inner super-Earth on a $9.3743 \pm 0.0019$ day orbit with eccentricity $e_b = 0.396 \pm 0.062$. 
Dynamical analysis of this system indicates that the parameters reported by \cite{b11} lead to 
global instability on very short time scales \citep{b3}. Long-term stability requires the two 
outer planets to be in 5:2 mean motion resonance. Parameters that lead to a minimum $\chi^2_{red}$ value of 4.96
for the orbital fits to the data are given in table~1, and these are the parameters of the system that 
we adopt in this paper.
It is noteworthy that planet $b$ is dynamically isolated from the outer planets
such that gravitational interaction over secular time scales is unable to excite the
observed eccentricity of this body.

The best fit parameters presented in \citet{b3} and reported in table~1
lead to the 5:2 resonant angles librating with large amplitude. Disc driven
migration scenarios for resonant capture of giant planets normally result in 
capture in low-order resonances with relatively small libration amplitudes 
\citep{2002MNRAS.333L..26N}. Planet-planet scattering, however, may also result
in resonant capture into higher order resonances with large libration amplitudes
\citep{b2}, and this would appear to be a more probable explanation for the currently
inferred orbital parameters for the two outer planets $c$ and $d$.

In this paper, we present a case study of the evolution of the planetary system around 
HD 181433, and explore plausible scenarios of how the observed large eccentricities may 
have been generated after depletion of the protoplanetary disk. In particular we 
focus on scenarios that can lead to the current observed orbital configuration of planet \textit{b}.
In Section \ref{Properties}, we discuss the properties of the HD 181433 system, and study the 
locations of secular resonances and regions of stability. In Section \ref{tides}, we consider 
how tides may have influenced the evolution of the semi-major axis and eccentricity of 
planet \textit{b} over the life-time of the system. In Section \ref{Scattering}, we explore 
the planet-planet scattering mechanism as a means of generating large eccentricities. We simulate
the evolution in the presence of an additional giant planet whose
role is to destabilise the system. We estimate
the probability for planets \textit{c} and \textit{d} to be captured into the 5:2 MMR
after scattering, and
thus consider the joint probability for both resonant capture into 5:2 and excitation of
the eccentricity of planet $b$. In Section \ref{spin}, we examine an alternative scenario in 
which sweeping of secular resonances induced by stellar spin-down force the eccentricity
of HD 181433\textit{b}. We study a range of initial stellar rotation periods and spin-down rates,
and examine the effect of including additional terrestrial planets -- whose presence
we find to be necessary for the model to be successful. In Section \ref{concl} we briefly 
discuss and summarize our findings.

\section[]{Properties of the HD 181433 system}
\label{Properties}

\begin{table*}
 \centering
 \begin{minipage}{140mm}
  \caption{Orbital and physical parameters of HD 181433 planets.}
\centering
  \begin{tabular}{@{}lccc@{}}
  \hline
   \textbf{Parameter} & \textbf{HD 181433 b} & \textbf{HD 181433 c} & \textbf{HD 181433 d} \\
 \hline
 P (days) & 9.37 & 975 & 2468 \\
$T_{peri}$ (BJD-2450000) & 2788.92 & 2255.6  & 1844 \\
e & 0.39 & 0.27 & 0.47 \\
$\varpi$ ($^{\circ}$) & 202.04 & 22 & 319 \\
$m\sin i$ ($M_{\rm Jup}$) & 0.023 & 0.65 & 0.53 \\
$m\sin i$ ($M_\oplus$) & 7.4 & 208 & 167 \\
a (AU) & 0.080 & 1.77 & 3.29 \\
\hline
\label{param}
\end{tabular}
\end{minipage}
\end{table*}

The mass of the host star is reported to be $M_{*} = (0.781 \pm 0.10) M_\odot $ \citep{b21}, 
its inferred radius is $R_{*} = (1.01 \pm 0.07) R_\odot $ \citep{b22}, while an age of 
$(6.7 \pm 1.8)$ Gyr has been reported \citep{b23}. HD 181433 is believed to be of spectral type 
K3IV, and has a spin period of 54 days \citep{b11}.
The orbital elements of the HD 181433 planets adopted in this paper are shown in Table \ref{param}. 
These values are quoted from the best-fit orbital solution that is dynamically stable
obtained by \citet{b3}.

The orbital period of HD 181433\textit{b} is 9.4 days with an eccentricity of 0.39. The longitude 
of HD 181433\textit{b}'s periastron advances mostly due to 
the relativistic correction of the Newtonian potential. Being a close-in planet in eccentric orbit, 
HD 181433\textit{b} is undergoing tidal circularization. Given the age of the
system and the measured $m \sin i$ value, this suggests that planet \textit{b}
has a tidal $Q$ factor substantially larger than inferred for terrestrial planets
in the Solar System, for which $Q \sim 100$ (e.g. Murray \& Dermott 1999). This point is discussed further in Section~\ref{tides}. 

The orbital parameter values quoted in Table \ref{param} are specific to a particular moment in time.
In fact, resonant and secular interactions cause $e_c$ and $e_d$ to reach values as large as 0.52 and 0.50,
respectively \citep{b3}. In this paper, when discussing the eccentricities acquired by planets 
during dynamical evolution we will report the maximum values
acquired during numerical computations once final stable systems have been established.

\subsection{Formation and evolution scenarios}
\label{formation} 
Before embarking on a detailed examination of the HD 181433 system, it is useful
to present a qualitative description of the possible formation and evolution history.
We envisage that the planets formed within a protoplanetary disc exterior to the snow-line. 
Planet \textit{b} appears to have characteristics of a Neptune-like planet based on analysis 
of its tidal evolution (see later in the paper).
It seems likely that this body formed largely from material beyond the snow-line and
migrated in by type I migration, arriving at its observed location on a near-circular orbit. 
It is possible that it accreted during migration, possibly through giant impacts 
with other large bodies. If these occurred after migration was essentially complete then 
the body would be close to the star and the impactors would need to be at least as massive as 
the inferred minimum mass of \textit{b} to generate significant eccentricity through scattering
during the giant impacts phase. Interactions and giant impacts with bodies in the Earth-mass 
regime at semi-major axes $\sim 0.1$ AU would lead to $e_b < 0.1$. Therefore, we anticipate
that planet \textit{b} was formed with a small free eccentricity.

We anticipate that the outer giant planets \textit{c} and \textit{d}
formed somewhat later than planet \textit{b}, but were able to grow larger cores 
that enabled rapid gas accretion and formation of gas giant planets. 
Although disc driven migration may have brought the planets into
close proximity, it is unlikely that it was responsible for establishing the 5:2 resonance.
Disc-driven resonant locking tends to cause the resonant angles to librate with
small amplitudes because the planets are pushed deep into the resonance, but
planets \textit{c} and \textit{d} appear to be weakly embedded in the resonance
\citep{b3}.
Furthermore, \cite{2002MNRAS.333L..26N} undertook a study of disc-driven resonant capture
and only obtained a 5:2 resonance when one of the two planets already had a high
eccentricity equal to 0.3. A more plausible explanation for the resonance is
capture as a result of planet-planet scattering, possibly involving an additional
planet that was ejected during the dynamical instability \citep{b2}. Given the stabilisation
provided by the eccentricity damping influence of the gas disc, this instability most
likely occurred during the final stages of the disc life time or after it was
dispersed altogether.

The rest of this paper largely concerns the question of how planet \textit{b}'s
eccentricity was excited, and when addressing this we consider the following
possibilities: scattering by a giant planet during the previously mentioned
dynamical instability that created the 5:2 resonance;  by invoking the presence of an additional undetected planet 
in the system that enhances the forced eccentricity of planet \textit{b}; passing planet \textit{b} through a secular resonance through
the process of stellar spin-down from an initially rapidly rotating state.

\subsection{Secular evolution}
For the purpose of undertaking a simplified analysis of the dynamical evolution
of the HD 18433 planetary system, we have implemented a Laplace-Lagrange secular model 
based on the discussion in chapter 7 of \citet{b24}, assuming that all 
planets are coplanar. Given the standard nature of
this secular theory, we refer the interested reader to \citet{b24} for description
of the mathematical details. The disturbing function includes secular terms for the 
mutual gravitational interaction between all three planets that are 
second order in the eccentricities and first order in the masses. Precession 
due to general relativity (GR) is also included \citep[e.g. see][]{b8}, but our model
does not take account of the 5:2 MMR occupied by the two outer planets.
Although techniques have been developed to include first-order MMRs in secular treatments 
of planetary system evolution \citep{1989A&A...221..348M,
  ChristouMurray1999, 2012ApJ...745..143A}, 
this becomes significantly more involved technically for higher order MMRs because of the 
requirement to include terms that are higher order in the eccentricities. Our purpose here 
is to use the secular theory as a rough guide to the dynamics of the system, and not to 
accurately predict its long term evolution, so this omission is not crucial here. 
This point is illustrated later in the paper where we demonstrate, by means of direct
numerical simulations, that the eccentricity evolution of the two outer planets
maintains highly regular and periodic behaviour, reminiscent of secular evolution
(see Figure~\ref{ecc} for example).
It appears that the effect of the 5:2 MMR is rather weak, with the main influence 
being an alteration of the precession frequencies.
The secular model, therefore,  allows us to identify the existence and approximate 
locations of secular resonances, in addition to estimating the magnitudes of forced 
eccentricities experienced by additional low mass planets that may be present in the 
system today, or which were present in the past.
All detailed analyses of the HD 181433 system dynamics undertaken in this
paper use direct numerical simulations for accurate modelling of the mutual interactions,
and all firm conclusions drawn about the past and future evolution are based on
those simulations.

\begin{figure*}
\includegraphics[width=0.33\textwidth]{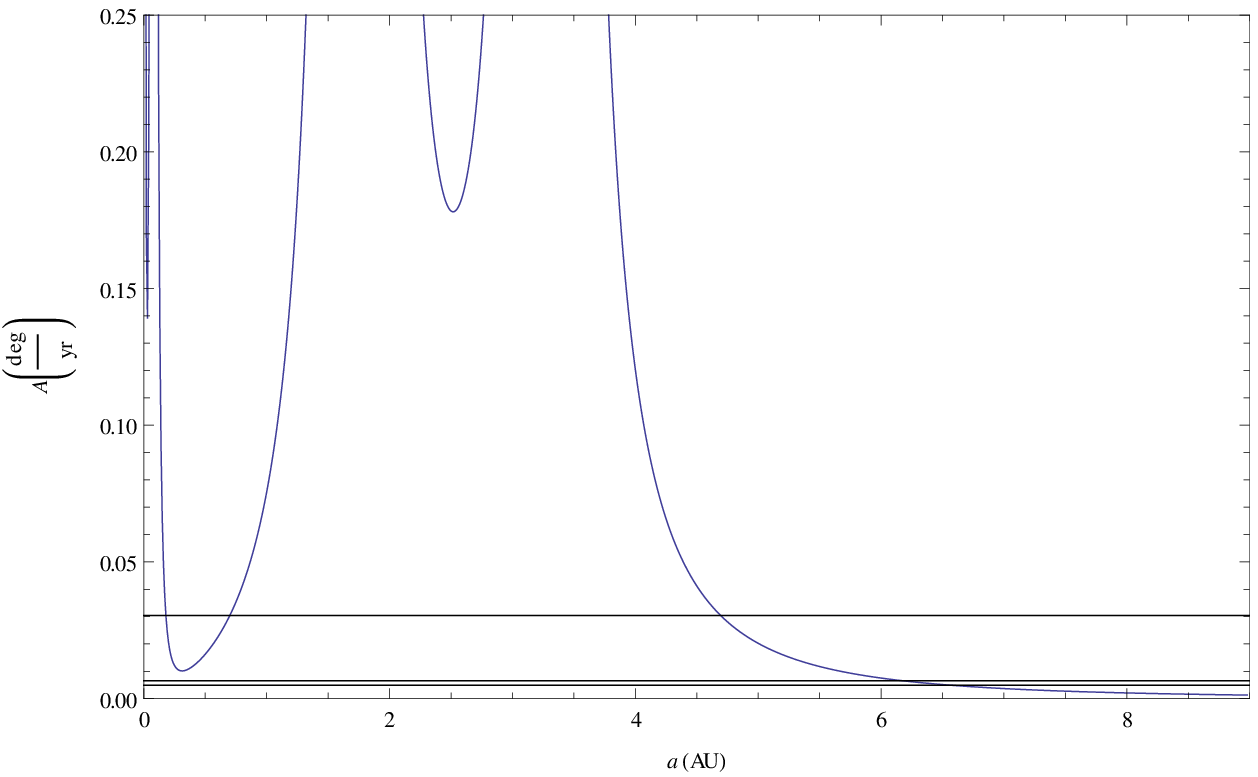}\includegraphics[width=0.33\textwidth]{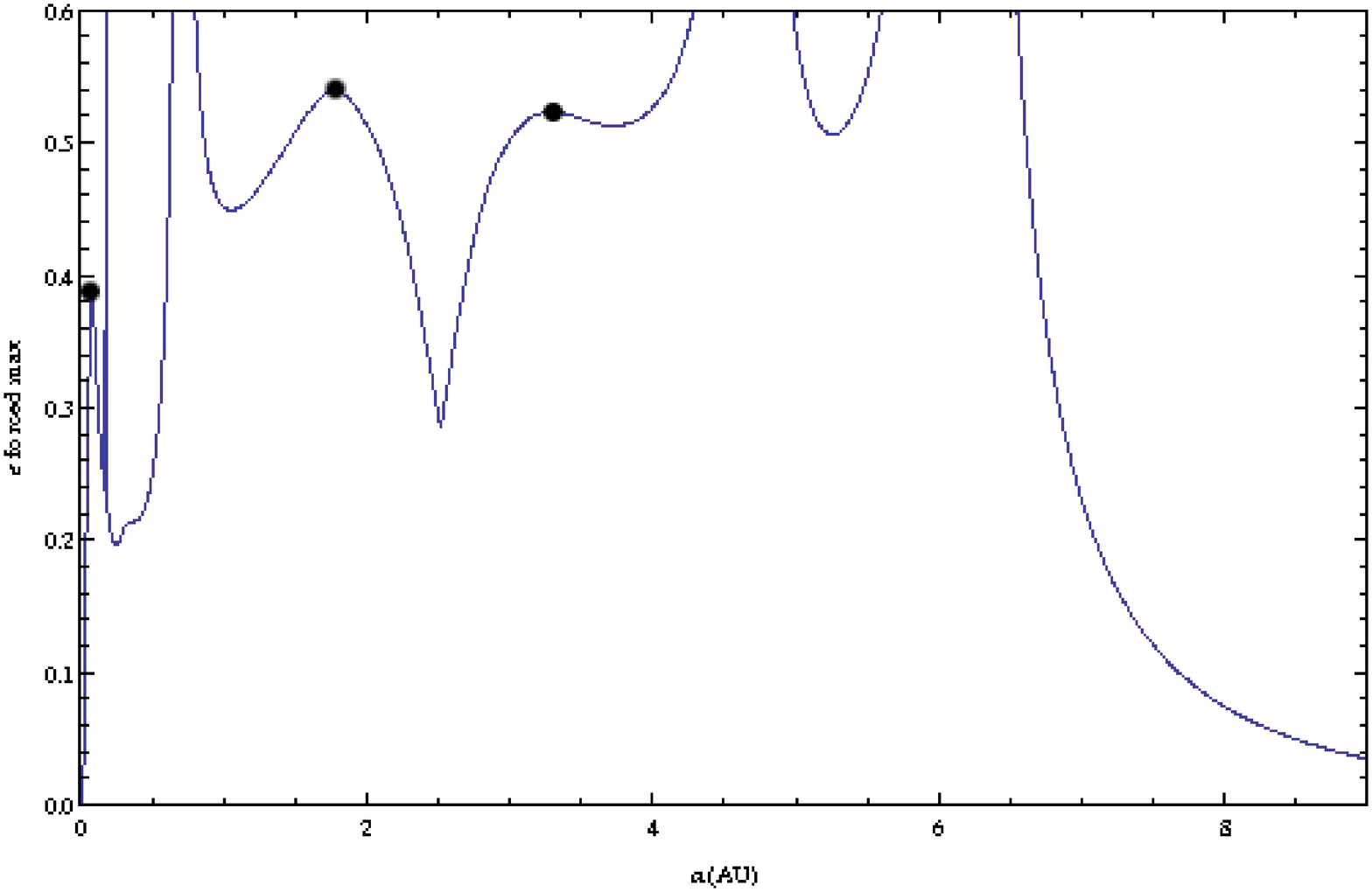}\includegraphics[width=0.33\textwidth]{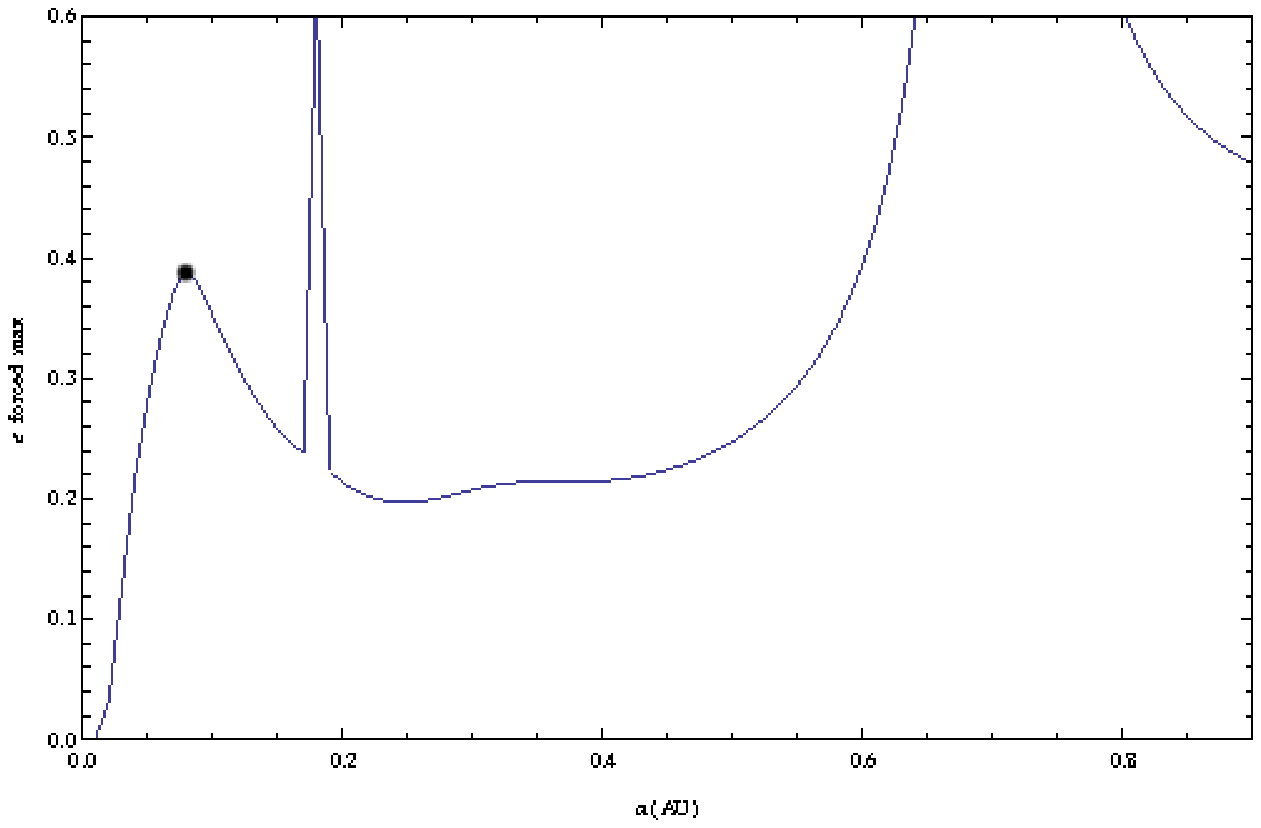}
\caption{Secular model for the planetary system HD 181433. Left: Precession frequency $A$ of a test particle as a function of semi-major axis, derived from perturbations by the planets and considering GR. The horizontal solid lines denote the values of the three eccentricity-pericentre eigenfrequencies. Singularities in the plot correspond to the orbital semi-major axis of the three planets. Centre: The variation in the maximum forced eccentricity as a function of semi-major axis. The dots indicate the eccentricity for the planets at their semi-major axes. The four singularities indicate regions of secular resonance where $A$ is equal to one of the $g_i$. Right: The maximum forced eccentricity within 0.9 AU.}
\label{analytic}
\end{figure*}

The secular system we examine is based on the osculating orbital elements
listed in Table~\ref{param}. The left-hand panel of 
Figure \ref{analytic} shows the variation of the free precession frequency $A$ experienced 
by a test particle located between 0 to 9 AU. 
On the same diagram the three eccentricity-pericentre 
eigenfrequencies of the system are denoted by solid horizontal lines. They are 
(in descending order) $g_2 = 0.0304 ^{\circ}$yr$^{-1}$, 
$g_3 = 0.0065 ^{\circ}$yr$^{-1}$ and $g_1 = 0.0049 ^{\circ}$yr$^{-1}$. 
Note that in the absence of relativistic precession $g_1 = 0.0009 ^{\circ}$yr$^{-1}$, 
with the other eigenfrequencies that are largely determined by the outer planets
being essentially unaffected. The intersections of the lines with the curve 
show where the eigenfrequencies $g_i$ equal $A$, and identify the semi-major axes 
where large forced eccentricities can be expected. 
In the central panel of Figure \ref{analytic}, the values of the maximum forced eccentricity 
are shown as a function of semi-major axis. The singularities close to 0.18, 0.7, 4.7 and 
6.3 AU represent locations where the value of $A$ is equal to one of the $g_i$ 
eigenfrequencies of the secular system, as anticipated above.
This plot indicates that the system is more-or-less dynamically packed out to distances of 
$\sim 7$ AU, in the sense that additional planets located in the system will experience 
large forced eccentricities that may be destabilising. Even in the region between
0.2--0.6 AU (focus of the right-hand panel of Figure \ref{analytic}), where a clear minimum exists in the forced eccentricity, a value of
$e_{\rm forced} \ge 0.2$ is predicted. It is noteworthy
that the habitable zone for this K3IV star is centred at $\sim 0.55$ AU \citep{b3}.

\subsection{Full integration of the system}
\label{sec:fullintegr}
We use the symplectic N-body code Mercury-6 \citep{b25}, augmented to include the effects
of GR \citep{b26}, to perform a direct integration of the system using initial
conditions from Table~\ref{param}.
Unless stated otherwise, all simulations in this paper adopted the
hybrid integrator option which utilises the second-order mixed-variable symplectic (MVS) algorithm 
for systems that are not strongly interacting, but switches to the Bulirsch-Stoer (BS) integrator 
when the minimum separation between objects $\le 4$ Hill radii. A time step of
1/20 of the orbital period of the innermost planets is adopted for the MVS integrator,
and the Bulirsch-Stoer integrator is employed with an accuracy parameter of $10^{-11}$.

\begin{figure}
\centering
\includegraphics[width=\columnwidth]{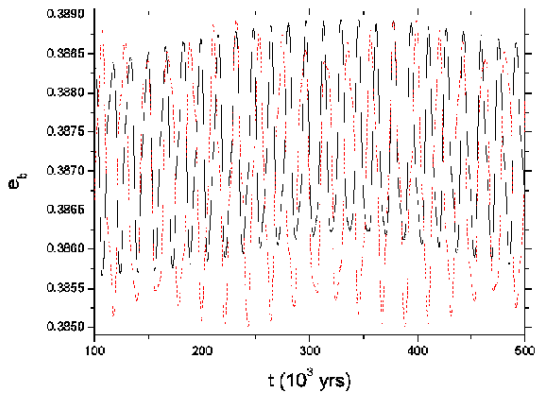}
\caption{Eccentricity evolution of HD 181433 \textit{b}. Solid line (black) not considering GR, dashed line (red) including GR.}
\label{GR}
\end{figure}

Figure \ref{GR} shows two integrations of the system, one including the relativistic correction
and one without it, and underlines how the relativistic correction for planet \textit{b}
is not relevant with regard to the amplitude of $e_b$, even if it is important with respect to the
precession period.
This differs from the $\upsilon Andromed{\ae}$ system, where the forced eccentricity of the inner planet depends sensitively on whether or not the post-Newtonian force is accounted for in secular evolution \citep{b10}
because of the proximity of a secular resonance.

\begin{figure*}
\includegraphics[width=0.5\textwidth]{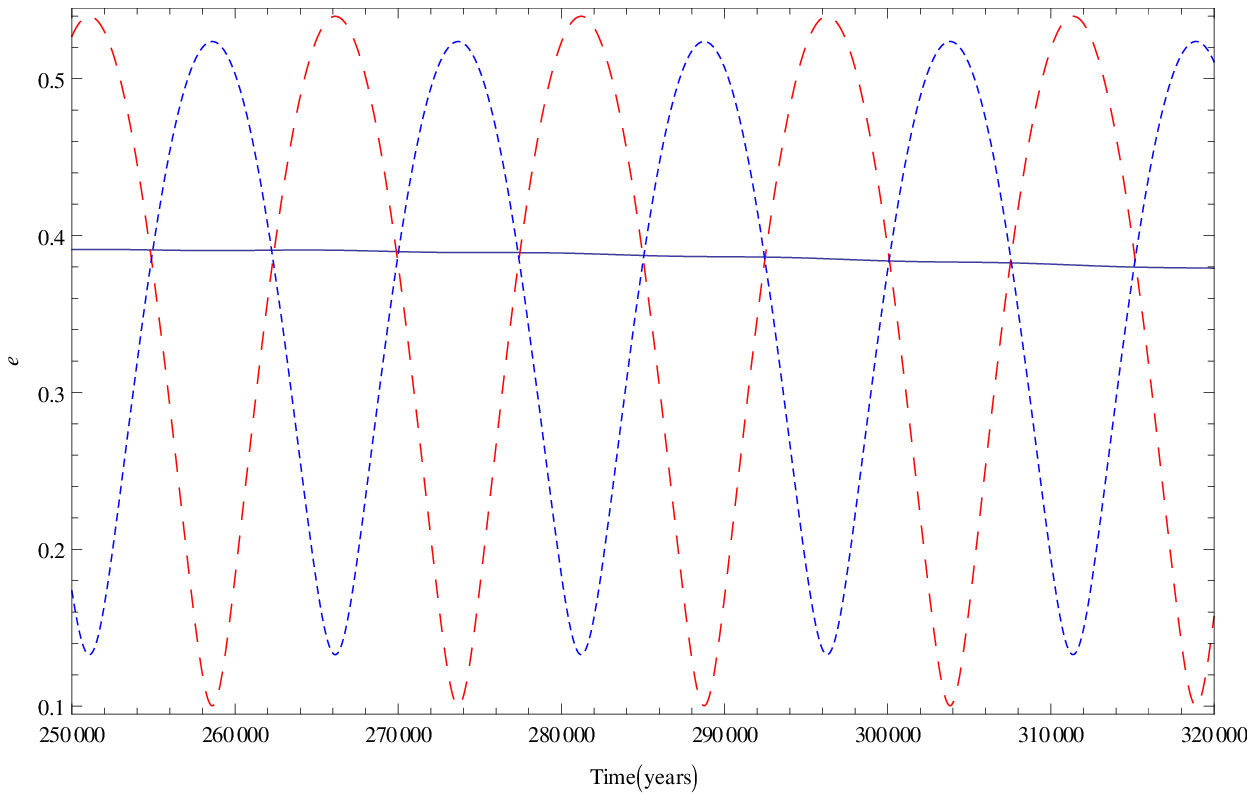}\includegraphics[width=0.5\textwidth]{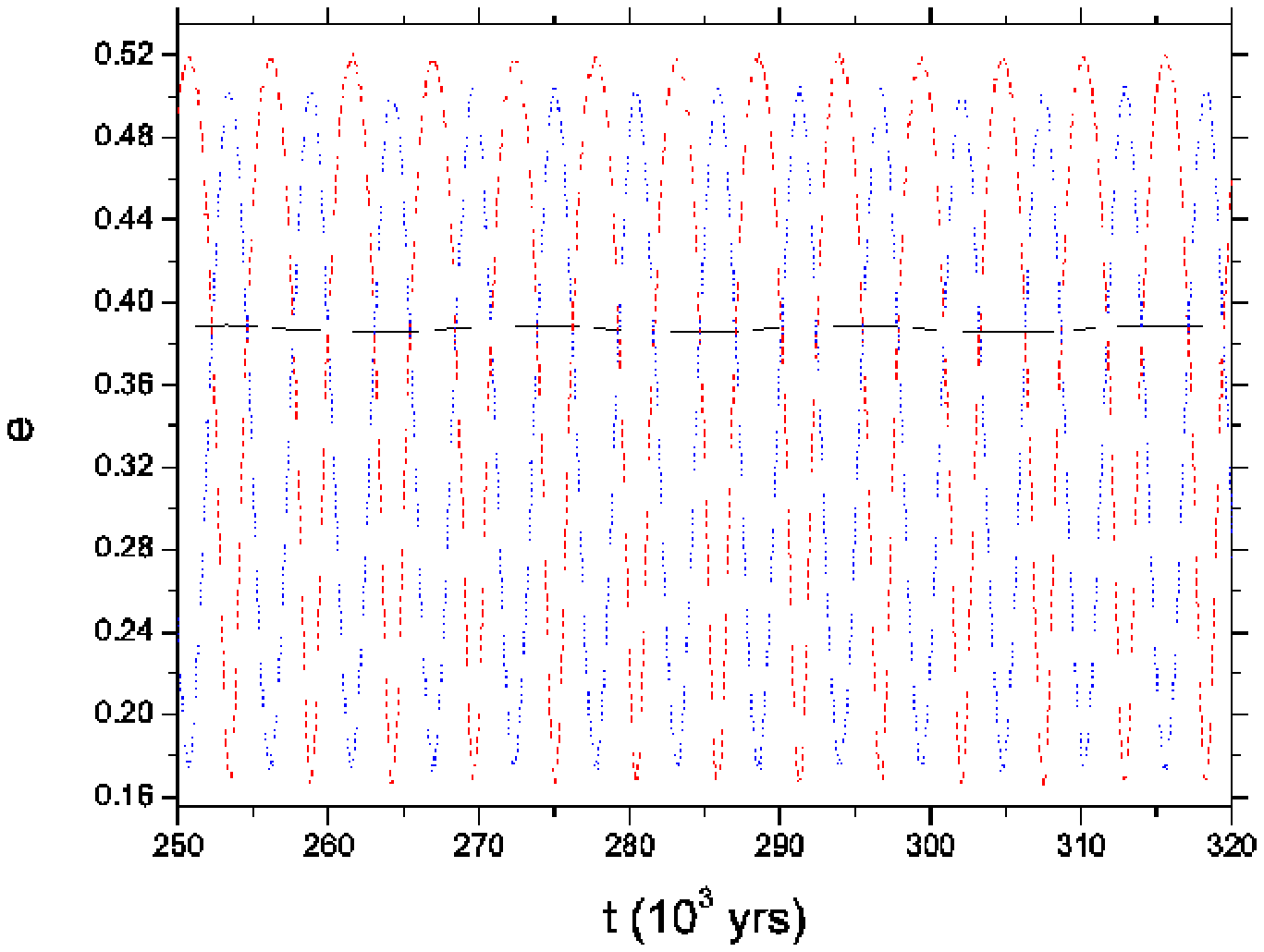}
\caption{Eccentricity evolution of three planets around HD 181433 including the post-Newtonian effect. The solid, dashed and dotted lines show the evolution of planet b, c and d, respectively. Left: Secular model. Right: Numerical simulation.}
\label{ecc}
\end{figure*}

While the loose residence of
planets c and d in the 5:2 mean motion resonances introduces some
short-term oscillations, the eccentricity evolution of the system over
longer periods appears regular and  well-described by a simple
three frequency secular model.  Fourier analysis of the system's
apsidal behaviour indicates that the main effect of the 5:2 resonance
is to simply modify the values of the principal precession frequencies that describe the giant planets (i.e., $g_2 = 0.0304^{\circ}\mbox{
  /yr}\rightarrow 0.02305^{\circ}\mbox{
  /yr}$ and $g_3 = 0.00652 ^{\circ}\mbox{
  /yr}\rightarrow -0.04354^{\circ}\mbox{
  /yr}$ making it inaccessible for resonant interaction with $g_1$).
%Table \ref{period} presents the precession periods of the pericentres of the three planets of HD 181433
%obtained from the numerics and secular calculations. In the full numerical integrations the precession
%periods for \textit{c} and \textit{d} are not constant, as indicated by the range of values
%quoted in Table \ref{period}, and moreover it is found to be considerably shorter for
%\textit{d} because of the 5:2 MMR.
Figure \ref{ecc} shows the evolution of the eccentricities of the HD 181433 planets retrieved from the
Laplace-Lagrange secular model (left panel) and the numerical integration (right panel).
The amplitudes of the eccentricity variation are very similar, while the periodicity of the eccentricity of
\textit{c} and \textit{d} is approximately reduced of two thirds to $\sim 5000$ years as result of the 5:2 MMR.
This illustrates the accuracy with which the secular model estimates the locations of the secular
apsidal resonance, in addition to the strong regularity displayed by the full integration
even though the outer two planets are in the 5:2 MMR. It is clear that the effects of the
5:2 MMR in this system are rather weak and make only a modest change to the secular behaviour,
namely through shifting the precession frequencies of the outermost planet \textit{d}.

\subsection{Stability of an additional low mass planet}
\label{stabadd}
To examine the consequences for an additional low mass planet in the system arising from
the large forced eccentricities predicted in Figure \ref{analytic}, we have performed
N-body simulations that include an additional planet.
We place an Earth-mass object (which would be undetectable in current radial velocity
surveys) at different distances between planets \textit{b} and \textit{c} 
to check for the stability of that region. The sampling used was of 0.05 AU for semi-major axes in the range $0.15-0.6$ AU and 0.1 AU for semi-major axes in the range $0.7-1.7$ AU.

We find that in full N-body simulations, the secular resonance at 0.18 AU moves inward leaving the 
region between 0.1-- 0.35 AU stable for a single body, but unstable outside of this region. 
In particular, for semi-major axes $a > 0.26$ AU the forced eccentricity attains values 
$e_{\rm forced} \gtrsim 0.2$. Later in the paper we consider the influence of additional low mass
planets, and we find that inserting two or three extra planets with at least one of them
between 0.26 and 0.35 AU makes the system unstable (see Section \ref{addpl}).

\section{Tidal evolution}
\label{tides}
The close proximity of planet \textit{b} to its host star suggest it should
be undergoing tidal evolution. Tidal effects circularize planetary orbits on a time scale given 
by \citep{b36}
\begin{equation}
\tau_{e}=\frac{4}{63}Q\left(\frac{a^3}{GM_*}\right)^{1/2}\frac{m_p}{M_*}\left(\frac{a}{R_p}\right)^5.
\end{equation}
The value of $Q$ calculated for rocky objects in the Solar System is of the order of $10^2$ \citep{b24}. 
Such a value would imply a circularization time scale of the order of $10^7$ years for planet \textit{b},
much shorter than any reasonable estimate for the age of the system. If we reject the hypothesis that a 
recent event has generated the observed eccentricity, then the tidal factor $Q$ must be considerably greater 
than expected for a terrestrial body of mass $\lesssim 10$ M$_{\oplus}$. If we assume a $Q$ value and density 
typical of an ice giant such as Neptune ($Q=10^4$, $\rho_b = \rho_{Nep} = 1.638$ g cm$^{-3}$), then the 
circularization time scale becomes $\tau_{e} = 8.2$ Gyr, comfortably longer than the estimated
system age. This would make HD 181433\textit{b} a member of the class of low-mass, low-density, close-in planets 
exemplified by Kepler 11c and Kepler 11e \citep{b39, 2013arXiv1303.0227L}.
To uncover the history of planet \textit{b} and make an estimate for the orbital parameters
that existed prior to significant tidal evolution, it is necessary to examine 
the influence of tides on past orbital evolution and estimate their influence of future evolution.

To study the orbital evolution of planet \textit{b}
we integrate the coupled tidal evolution equations for changes in $a$ and $e$ (see \citet{b17} 
and references therein) backward and forward in time: 
\begin{eqnarray}
  \frac{1}{a} \frac{da}{dt} &=& -\left[ \frac{63 \sqrt{G M_{*}^3} R_p^5}{2 Q_{\mathrm{p}} m_p} e^2 \right.\nonumber \\
    &&+ \left.\frac{9 \sqrt{G/M_{*}} R_{*}^5 m_p}{2  Q_{*}} \left(1 + \frac{57}{4} e^2\right)\right] a^{-13/2} \\
  \frac{1}{e} \frac{de}{dt} &=& -\left[ \frac{63 \sqrt{G M_{*}^3} R_{\mathrm{p}}^5}{4 Q_{\mathrm{p}} m_p} + 
    \frac{225 \sqrt{G/M_{*}} R_{*}^5 m_p}{16 Q_{*}} \right] a^{-13/2}.\textrm{~~~~}
\label{eq:tides}
\end{eqnarray}
The effects of the tide raised on the star $Q_*$ as well as on the planet $Q_{\rm p}$ are both included. 
The stellar $Q$ value is typically estimated through the observed circularization of binary star orbits. 
The rate of tidal effects may be a very strong function of orbital frequency \citep{b38}. If this is the case, 
the planet may spend a lot of time at certain states where tidal effects are slow and rapidly pass through 
states where tidal effects are faster. This model assumes that the star is rotating slowly relative to the 
orbit of the planet (so when the star has already evolved toward the Main Sequence) and is second order in 
eccentricity. It describes tidal evolution associated with orbit circularization, as this operates most rapidly,
and ignores other sources of tidal interaction which operate on longer time scales. 
Our intention is to use this simple tidal theory to develop an approximate picture for the likely orbital 
history for HD 181433\textit{b}, and its possible future evolution.

\begin{figure}
\centering
\includegraphics[width=\columnwidth]{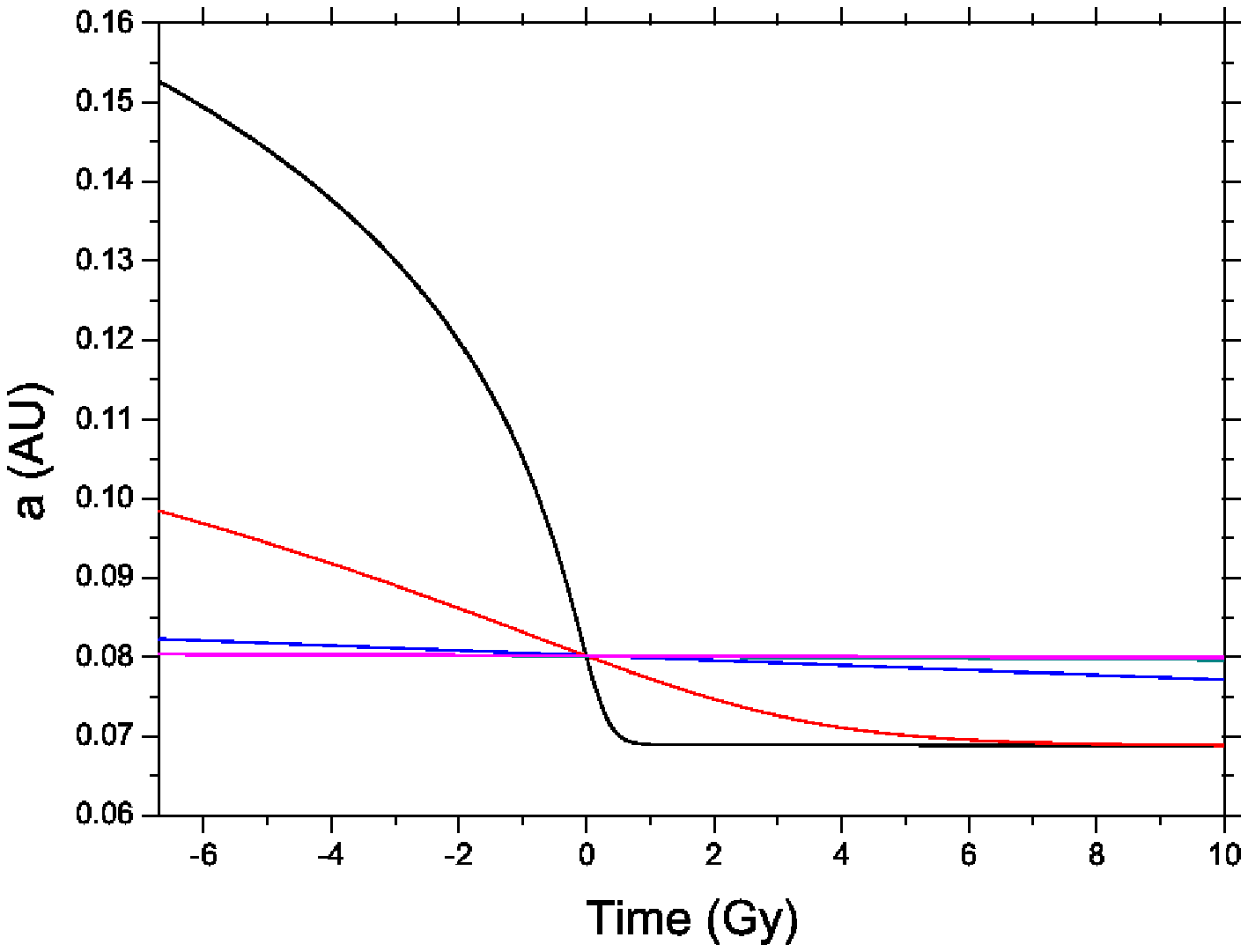} \includegraphics[width=\columnwidth]{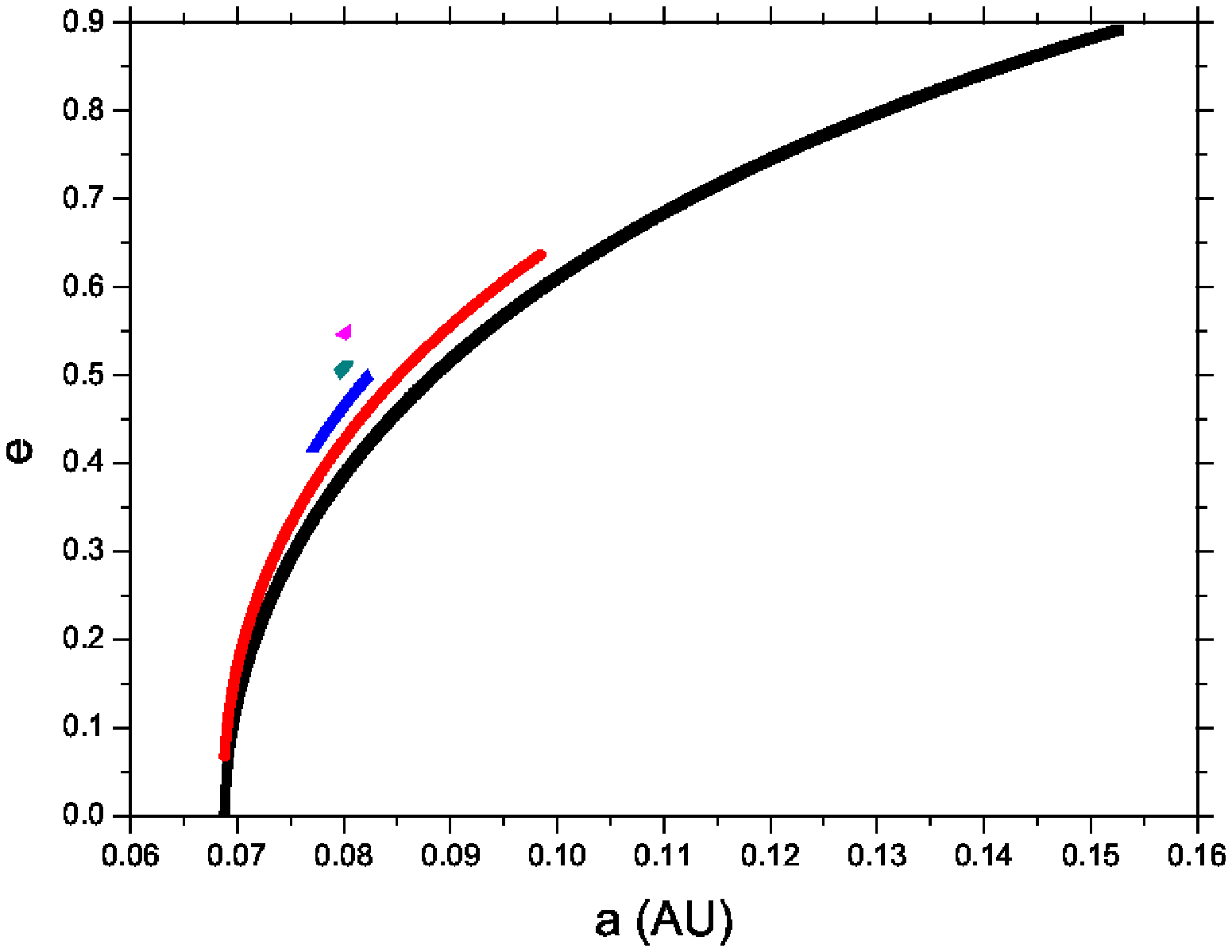}
\caption{Tidal evolution for HD 181433\textit{b} for the case $Q_*=10^{5.5}$. Top panel: Orbital migration for a range of 
$Q_{\rm p}$. Initially, a greater semi-major axis is expected in the case $Q_{\rm p}=10^{3}$, followed by the values 
$10^{4}$, $10^{5}$, $10^{6}$ and $10^{6.5}$. The point of intersection between all the lines represents the 
present state. Bottom: Evolution in $a-e$ space for the cases $Q_{\rm p}$ equal to $10^{3}$, $10^{4}$, 
$10^{5}$, $10^{6}$ and $10^{6.5}$. A smaller value of $Q_{\rm p}$ implies greater fraction of the $a-e$ space has
been spanned. The evolution moves from right to left. For clarity, evolutionary trajectories for $Q_{\rm p}$ equal to $10^{4}$, 
$10^{5}$, $10^{6}$ and $10^{6.5}$ have been off-set by $\Delta e = + 0.04$ from each previous case.}
\label{migration}
\end{figure}

We computed the orbital evolution for various pairs of values of $Q_{\rm p}$ 
(ranging between $10^3$ and $10^{6.5}$) and $Q_*$ (between $10^4$ and $10^{7}$). 
The chosen $Q$ values span the range that are plausible. The rate of 
tidal damping may depend on the interior structure of the planet and is likely
to be different for different planets. We integrate the equations for the age of the star backward in time (6.7 Gyr) 
and for 10 Gyr into the future. Results show that only when $Q_*$ falls below $10^4$ does the behaviour change
in a noticeable manner. We conduct our analysis using $Q_*=10^{5.5}$, which corresponds to the value for which 
\citet{b13} have found the distribution of initial \textit{e} values of close-in planets to match that of the 
general population. The top panel of Figure \ref{migration} shows the migration of HD 181433\textit{b} 
for different values of $Q_p$, while in the bottom panel the evolution in $a$-$e$ space is presented
(obtained by assuming orbital angular momentum conservation). 
Since both $de/dt$ and $da/dt$ scale $\propto a^{-13/2}$, the rate of evolution is slower for larger initial
values of $a$, but speeds up dramatically as $a$ decreases.

A smaller $Q_{\rm p}$ means a larger range of values is spanned during the evolution for a fixed age of the system.
For example, $Q_{\rm p}=10^3$ implies that during the early main-sequence stage planet \textit{b} 
would have had a semi-major axis of 0.15 AU and an eccentricity almost equal to $0.9$. However, assuming a current age of
6.7 Gyr, complete circularization of the orbit will take much less than one more gigayear to be achieved 
(with the migration terminating at about 0.07 AU). Without further detailed information about the system
allowing us to constrain its physical nature, this evolutionary track and associated $Q_p$ value
is equally probable as any others. Our purpose here is to define a plausible initial state of the system prior
to significant tidal evolution, and in doing so we make the assumption that we are observing the system close
to the midpoint of its tidal evolution. A value of $Q_{\rm p}=10^4$ means that planet \textit{b} would have had initial
values $a_b \simeq 0.1$ AU and $e_b \simeq 0.6$, and circularization would be completed within the next 10 Gyr.
As noted earlier, $Q_{\rm p}=10^4$ is similar to the values inferred for Uranus and Neptune \citep{b24}.
We find that a value $Q_p=10^5$ leads to tidal effects that are almost negligible such that the
observed values of $a_b$ and $e_b$ would be very similar to those 6.7 Gyr ago. When considering
mechanisms for exciting the eccentricity of planet \textit{b}, we aim to achieve values between
$0.4 \le e_b \le 0.6$, implying modest or essentially no tidal evolution has taken place since the
excitation occurred.

\begin{figure}
\centering
\includegraphics[width=\columnwidth]{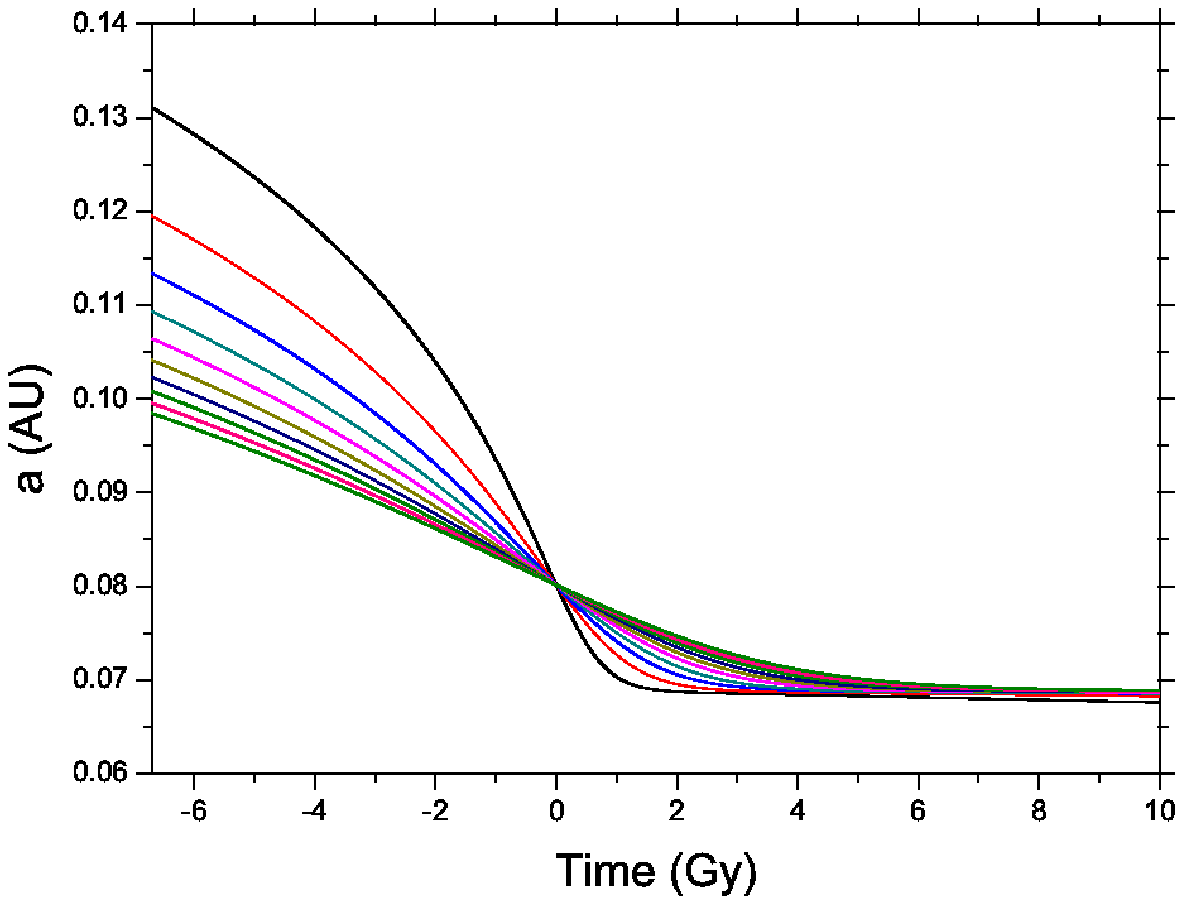} \includegraphics[width=\columnwidth]{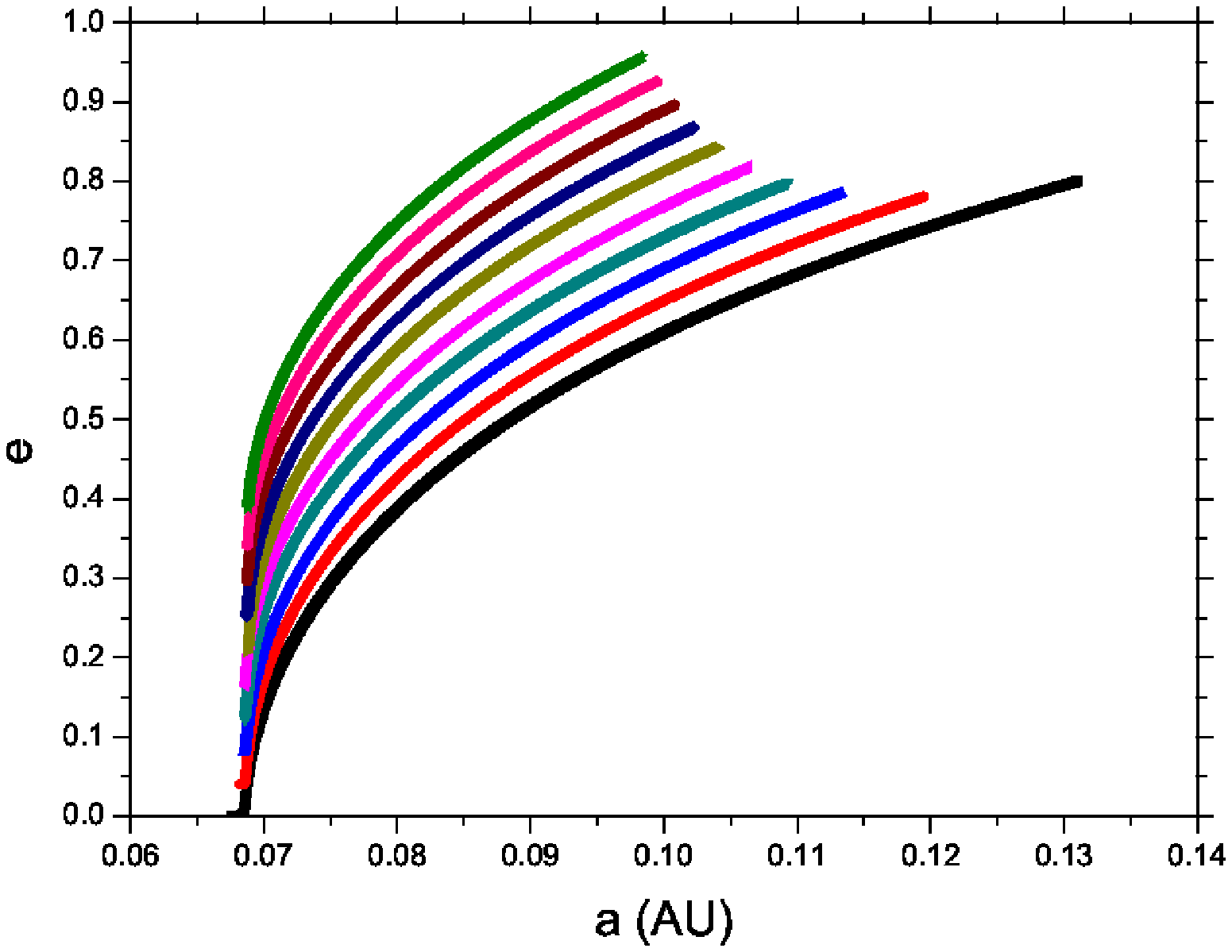}
\caption{Tidal evolution of HD 181433\textit{b} for the range of $\sin i$ between 0.1 and 1 when $Q_*=10^{5.5}$ and 
$Q_{\rm p}=10^4$. The case $\sin i = 1$ can also be found in Figure \ref{migration}. Top panel: Orbital migration. A
smaller $\sin i$ corresponds to a greater initial semi-major axis and a quicker circularization. The point of 
intersection between all the lines represents the present state. Bottom: Evolution in $a-e$ space. A smaller 
value of $\sin i$ corresponds a greater fraction of spanned space. The evolution goes from right to left. For clarity, evolutionary trajectories for values of $\sin i $ from 0.2 to 1 have been off-set by $\Delta e = + 0.04$ from each previous case.}
\label{sini}
\end{figure}

For completeness, we have also investigated how the tidal evolution may depend on the actual mass of \textit{b}. 
We span a range for $\sin i$ between 0.1 and 1 and we assume a density $\rho_{Nep}$ in all cases. 
A smaller $\sin i$ implies a more massive planet: when $\sin i = 0.5$ planet {\it b} is about the mass of 
Uranus/Neptune. 
This is purely a qualitative analysis as a small $\sin i$ would imply a larger mass for planets 
\textit{c} and \textit{d} which could destabilize the system. 
As expected from equation ~\ref{eq:tides}, more massive planets experience significantly more rapid migration than 
less massive ones. Figure \ref{sini} displays the case for the instance $Q_{\rm p}=10^4$. In the extreme case where 
$\sin i = 0.1$, planet \textit{b} should have been found initially at around $0.13$ AU with an eccentricity of 
about 0.8; the circularization process speeds up significantly with time and would be fully accomplished in the next 
billion years for this particular example.

We end our discussion of tidal evolution by noting that substantial tidal heating of a
gaseous planet can in principle cause it to undergo Roche lobe overflow
and orbital expansion \citep{b15,b16,b17}. As yet calculations have not been performed 
with parameter sets that would allow us to determine whether or not such a scenario
is compatible with the current orbital configuration of HD 181433 \textit{b}.

\section[]{Planet-planet scattering}
\label{Scattering}
We now consider the origin of the eccentricities of the planets in the HD 181433 system, with our
primary focus being the eccentricity of the inner-most body. Recent work has demonstrated that
planet-planet scattering can explain the eccentricity distribution of the extrasolar planet population as a
whole \citep[e.g.][]{b5,2008ApJ...686..603J}. Here we explore a scenario in which an additional giant planet
was present in the system originally, orbiting close to the two existing outer giant planets,
but shortly after dispersal of the protoplanetary disc planet-planet scattering caused this extra planet
to be ejected from the system, leaving behind a three-planet system with eccentricities and semi-major
axes similar to those observed today. Given the non-linear dynamics involved, the likelihood of
producing a system with parameters very close to those of the observed HD 181433 system is exceedingly small.
We therefore define two requirements that must be met for a simulation outcome to be deemed a success:
the inner-most planet has a close encounter with one of the giant planets during the period of dynamical
instability leading to significant growth of its eccentricity; two giant planets remain at the end of
the simulation in 5:2 resonance. We are able to estimate the probability of each of these separate
outcomes from our simulations, and hence the joint probability of both requirements being satisfied.

We use the Mercury-6 N-body code to study this problem, and employ the hybrid integrator 
with the characteristics described already in Section \ref{sec:fullintegr}. The physical size for the bodies is determined by their mean 
densities. For low mass planets we adopt a value equal to 3 g cm$^{-3}$, and for giant planets we adopt 
the Jovian value (1.326 g cm$^{-3}$).

\subsection{Scattering between three giant planets}
\label{sec:3p-scattering}

In order for one of the giant planets to undergo a close encounter with 
planet \textit{b} and scatter it onto an eccentric orbit with $0.4 \le e_b \le 0.6$, as required,
the perturbing body needs to have a mass large enough to generate the required velocity
perturbation. The perturbing body could be the additional planet `\textit{x}' or either of
the planets \textit{c} or \textit{d}. We approximate the eccentricity as $e \simeq v_r/v_{\rm orb}$,
where the $v_r$ is the perturbed radial velocity and $v_{\rm orb}=\sqrt{GM_*/a_b}=88$ km/s 
is the Keplerian orbital velocity of \textit{b}. Assuming that $v_r$ due to an encounter is of the 
same order as the escape velocity from the perturbing planet gives 
$v_r \simeq \sqrt{2Gm_{\rm p}/R_{\rm p}}$. The mass required to generate eccentricity $e$ can then
be written as 
$$m_{\rm p} \ge e^3 v_{\rm orb}^3 \left(\frac{3}{32 \pi G^3 \rho_{\rm p}} \right)^{\frac{1}{2}}$$
where $\rho_{\rm p}$ is the mean density of the perturber. Assuming a Jovian mean density
gives a required mass for planet \textit{x} in the range $0.19 \le m_x \le 0.64$ M$_{\rm Jup}$
if it plays the role of planet \textit{b}'s perturber.
We have performed simulations with $m_x=0.3$ M$_{\rm Jup}$ which is large enough to produce
values of $e_b > 0.4$.  If planets \textit{c} or \textit{d} act as the perturbers then
larger values of $e_b$ are possible given their larger masses.

In addition to requiring that the perturbing planet can excite a large enough value of $e_b$,
we also require that the initial conditions of our simulations at least in principle allow
the observed configuration of HD 181433 to be attained once the dynamical stability 
has caused ejection of the additional planet. The combined specific energy of the two outer 
planets is given by $E_{\rm tot} =-G M_*/(2 a_c) - G M_*/(2 a_d)$, and we note that ejection of 
planet \textit{x} requires a loss of specific energy from the system equal to $E_x=G M_*/(2 a_x)$.
We therefore ensure that our initial conditions are such that if energy $E_x$ is lost from
the system of outer giant planets the remaining energy equals $E_{\rm tot}$.
We consider a number of basic initial configurations for our simulations, and each
simulation set corresponds to a particular stellocentric ordering of the outer giant planets:
\textit{cdx}, \textit{cxd}, \textit{xcd}, \textit{dcx} and \textit{xdc}. The first and last letters 
in the labels correspond to the closest and further planets from the star. For each set, 
planet \textit{x} is initially placed randomly between 2 and 5 mutual Hill radii ($R_{H,m} = 0.5(a_{b} + a_{x})[(m_{b} + m_{x})/3M_{*}]^{1/3}$) from its neighbour (for the set \textit{cxd} planet \textit{x} is initially closest to planet \textit{d}),
and the initial mean longitudes are also set randomly. The planets are initially on circular
orbits with mutual inclinations $i\le 1$ degree. We note that with the ice line defined by
$a_{\rm ice} = 2.7 \sqrt{L_*/L_\odot}$ AU gives $a_{\rm ice} = 1.50$ AU for HD 181433.
Our initial set-up therefore concurs with the general expectation that giant planets
emerge from the disk at locations beyond the ice line. 

Throughout the integrations, close encounters and collisions between any two bodies were logged, 
as well as ejections and collisions with the parent star. We ran the simulations for 250 Myr but instabilities usually arise over much shorter time-scales. If dynamical instability resulted in ejection of one or more planet we calculated the 
orbital elements of the remaining planets. In particular, we are interested in the possibility that
the two outer planets are in 5:2 resonance after the scattering, and the value
attained for the eccentricity of planet \textit{b} when the system has stabilised. 
In all simulations energy was conserved
to a level better than 1 part in $10^4$, which is adequate for testing stability 
(e.g. Barnes \& Quinn 2004). 300 simulations were performed overall, 50 for each of the
sets described above except for set \textit{cdx} where we ran 100 simulations.

\begin{figure}
\centering
\includegraphics[width=\columnwidth]{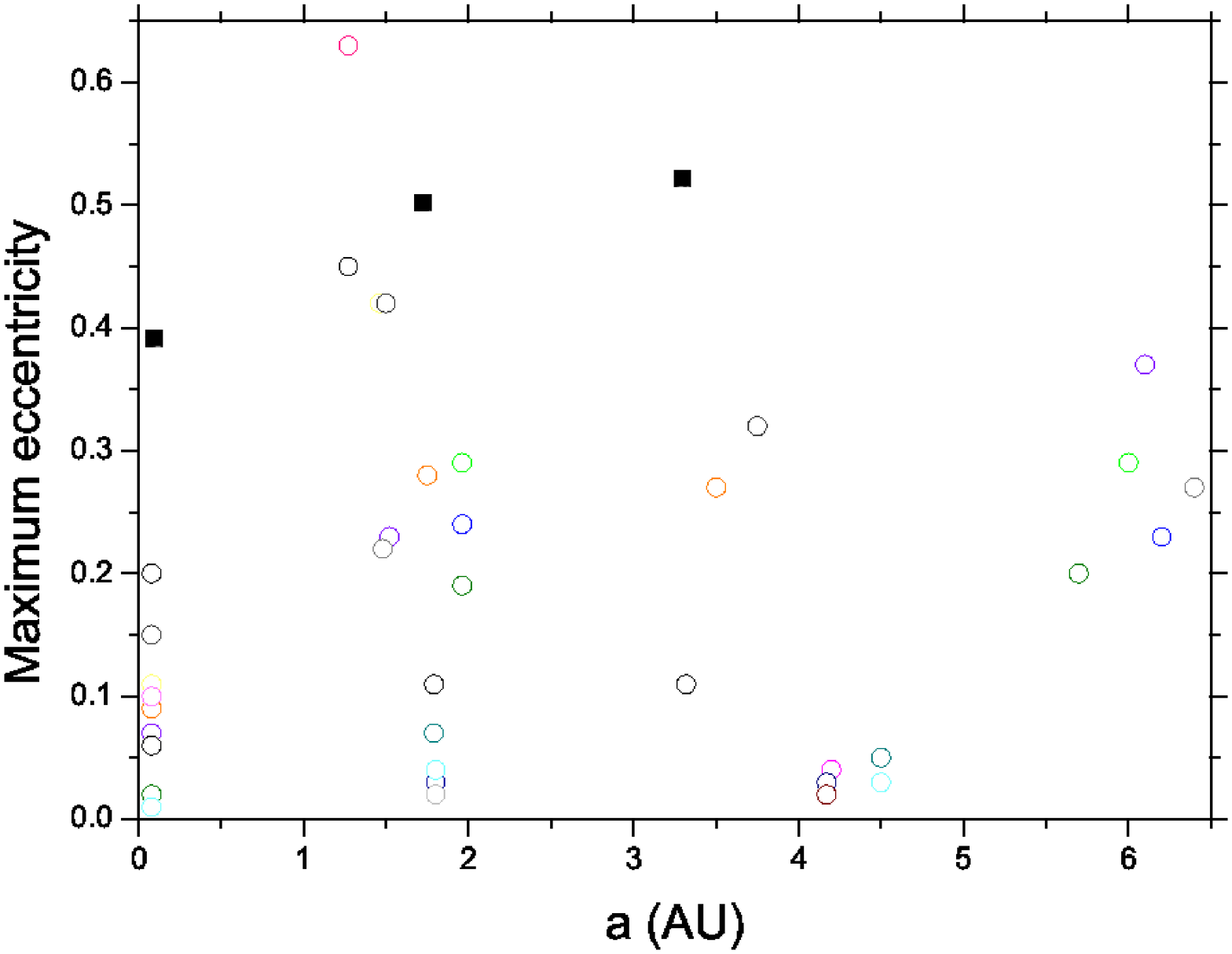} \includegraphics[width=\columnwidth]{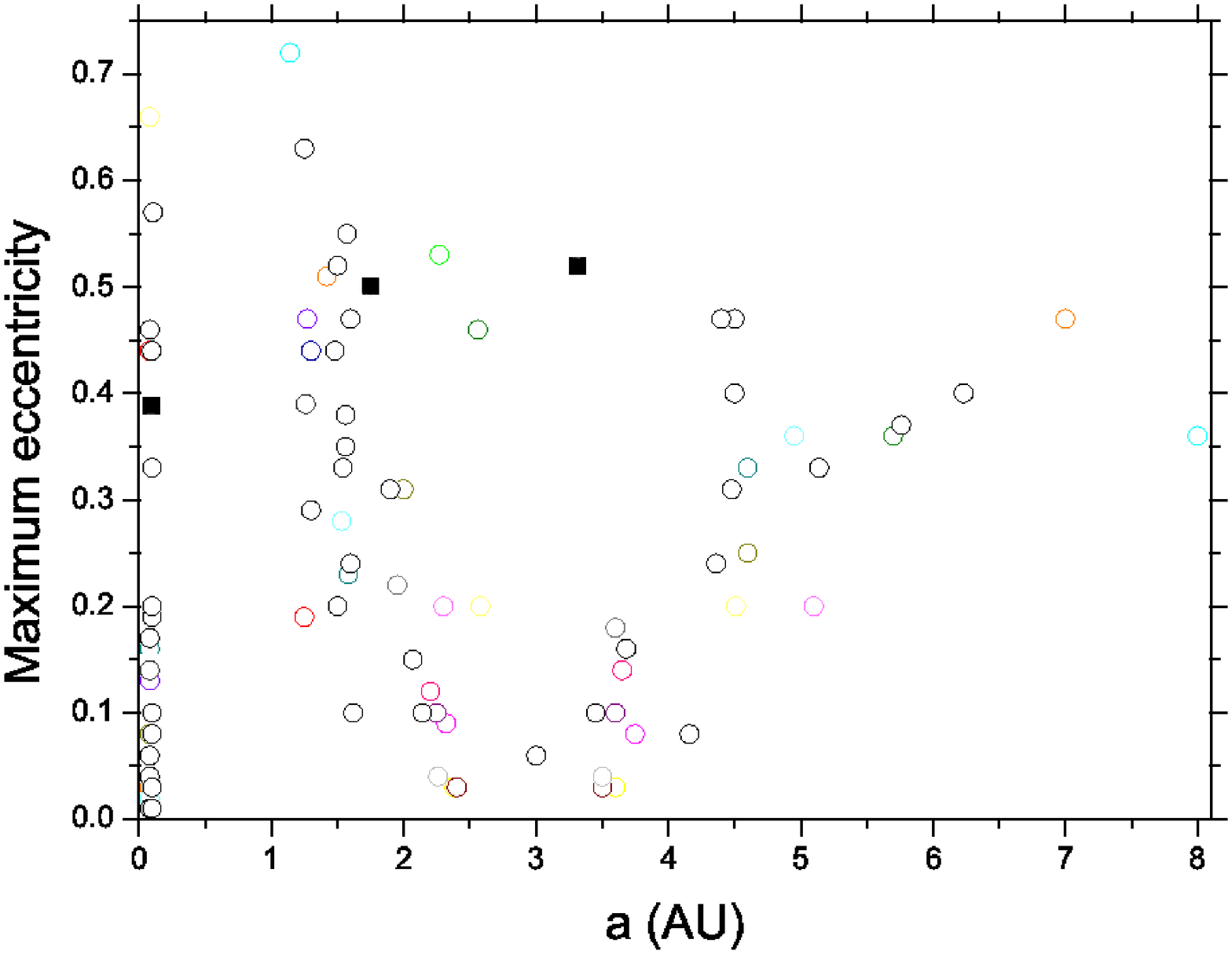}
\caption{Comparison of our synthetic final planetary systems with the HD 181433 system showing
maximum eccentricity versus semi-major axis. The values for the real planets are presented 
with filled black squares. Top panel: Cases from set \textit{cdx}. Bottom: Cases from set 
\textit{xcd}.}
\label{scatt}
\end{figure}

The simulation results show that planet \textit{x} is ejected in almost 50\% of cases, 
but in none of our simulations do we find that planets
\textit{c} and \textit{d} are in 5:2 mean motion resonance. The dynamical instability 
often leaves planets \textit{c} and {d} more highly separated than in the observed 
configuration, with planet \textit{d} in particular orbiting with significantly
larger semi-major axis. Figure \ref{scatt} shows the outcome from sets (\textit{cdx}, \textit{xcd})
that have undergone strong scattering, and we see that in each case the currently observed
values of $e_b$, $e_c$ and $e_d$ are at, or close to, the upper limits of the eccentricities
generated in the simulations. More importantly, these simulations also demonstrate the
eccentricity of the short-period planet \textit{b} can also be excited to the required value.
Inclinations typically remain small in accordance to what was found by \citet{b7}.

\begin{figure*}
\includegraphics[width=0.5\textwidth]{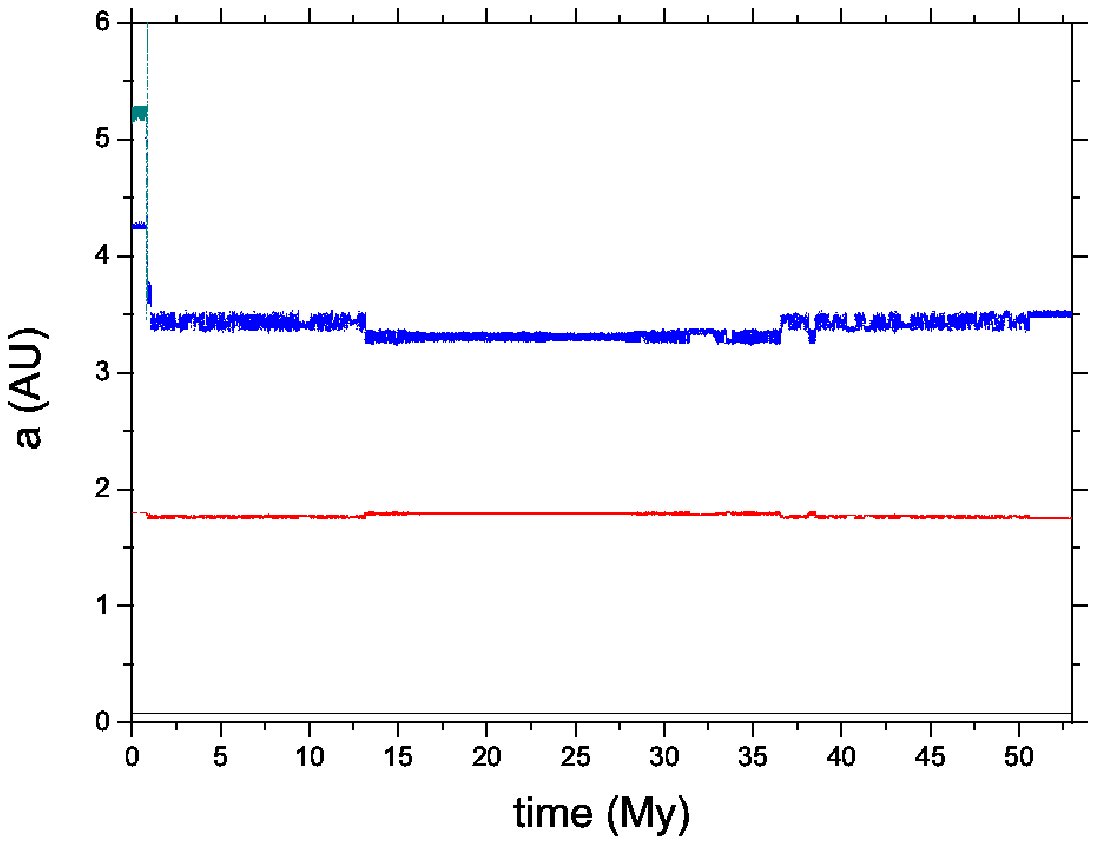}\includegraphics[width=0.5\textwidth]{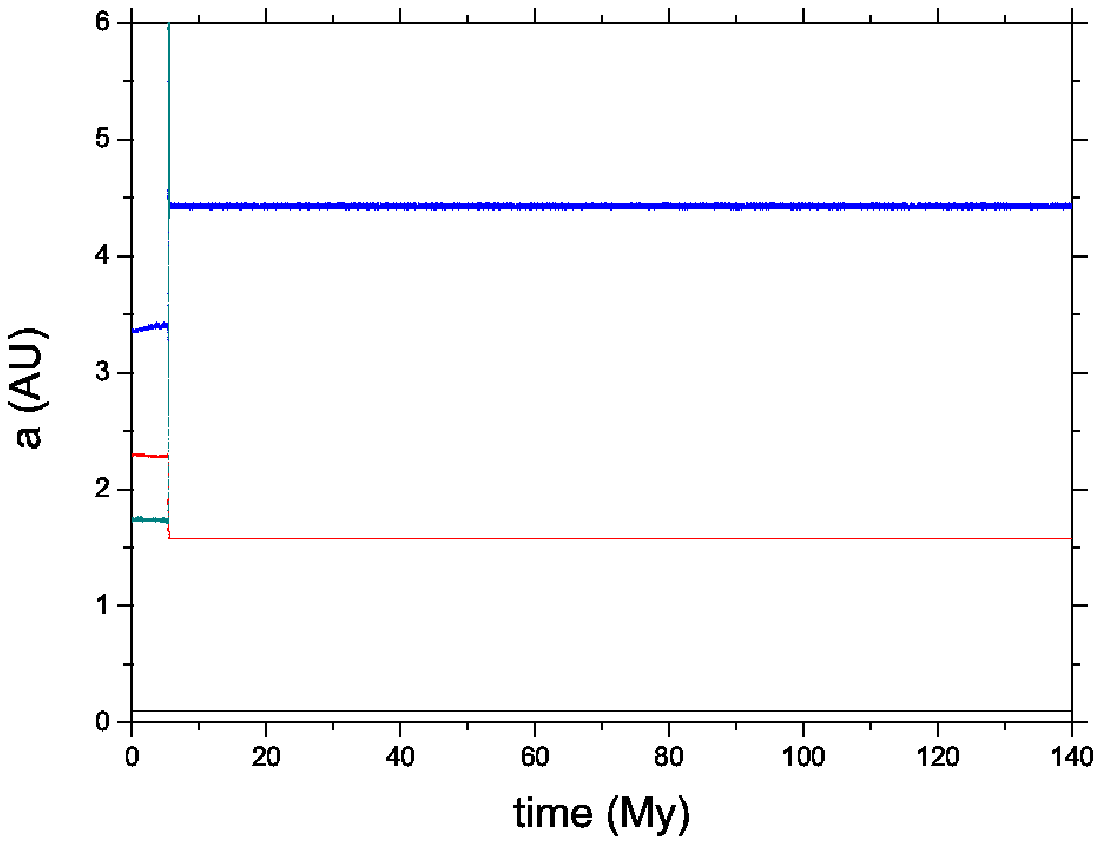}
\includegraphics[width=0.5\textwidth]{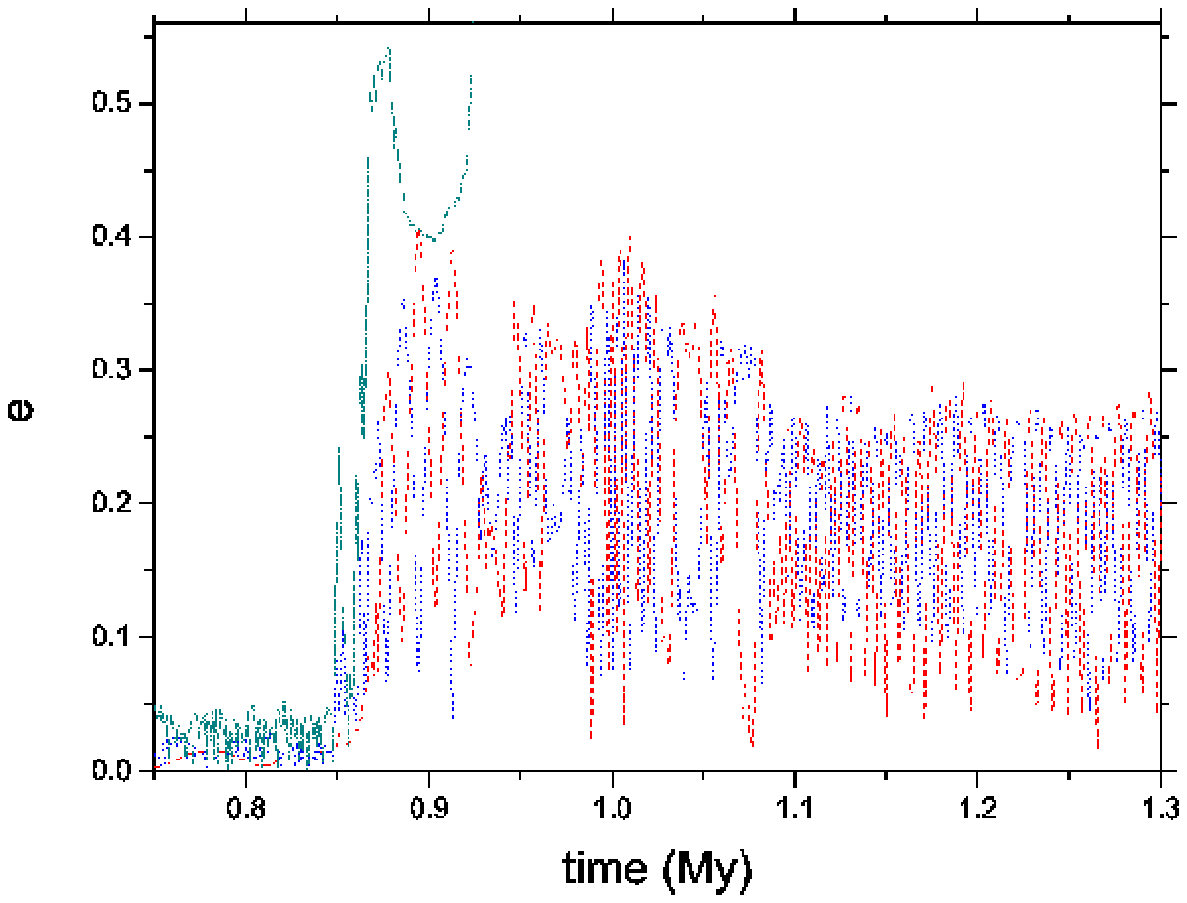}\includegraphics[width=0.5\textwidth]{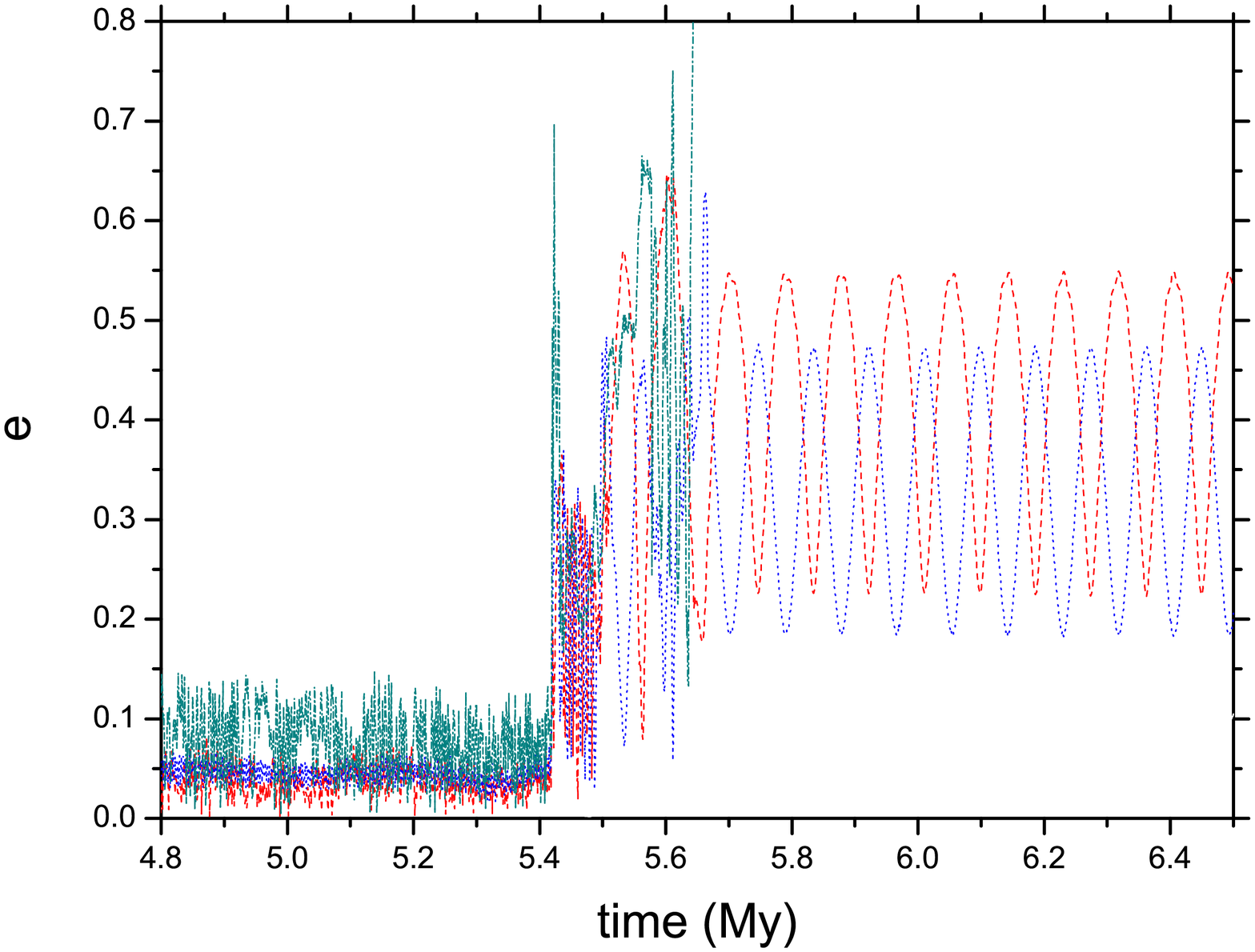}
\caption{Evolution of $a$ and $e$ for two cases of scattering. Top panels: The time evolution of the semi-major 
axes. Initially, in order of increasing distance, we have planets \textit{b}, \textit{c}, \textit{d} and \textit{x} 
(left panel) and planets \textit{b}, \textit{x}, \textit{c} and \textit{d} (right panel). After ejection, 
the planets are in long-term stable orbits. Bottom panels: The time evolution of the eccentricities. 
The dashed line (red) denotes planet \textit{c}, the dotted line (blue) represents planet \textit{d} and the 
dash-dotted line (dark cyan) indicates planet \textit{x}. After the ejection, the eccentricities oscillate 
on a secular time-scale. In the first case, $e_c$ moves in the interval 0.03--0.29 and $e_d$ in the range
0.05--0.28. In the second instance, $e_c$ moves in the interval 0.22--0.55 and $e_d$ in the range 0.18--0.47.}
\label{highecc}
\end{figure*}

The left-hand panels of Figure \ref{highecc} show a case from set \textit{cdx} where planet 
\textit{x} is ejected after close encounters, and planets \textit{c} and \textit{d} land close to
their present locations (but are not in resonance). Planets \textit{b}, \textit{c} and \textit{d} 
achieve maximum eccentricities of 0.09, 0.29 and 0.28, respectively.
A case from set \textit{xcd} is presented in the right panels of Figure \ref{highecc}. Following 
close encounters, planet \textit{x} is ejected with planets \textit{c} and \textit{d} 
landing at 1.6 and 4.4 AU, respectively. Planets \textit{b}, \textit{c} and \textit{d} 
acquire maximum eccentricities of 0.19, 0.55 and 0.47, and maximum inclinations of 12$^{\circ}$, 11$^{\circ}$ and 9$^{\circ}$, respectively. Therefore, in this case planets \textit{c} and \textit{d} 
reach values for the eccentricities that are similar to the observed one, but 
planet \textit{d} orbits at a greater distance.

\begin{figure}
\centering
\includegraphics[width=\columnwidth]{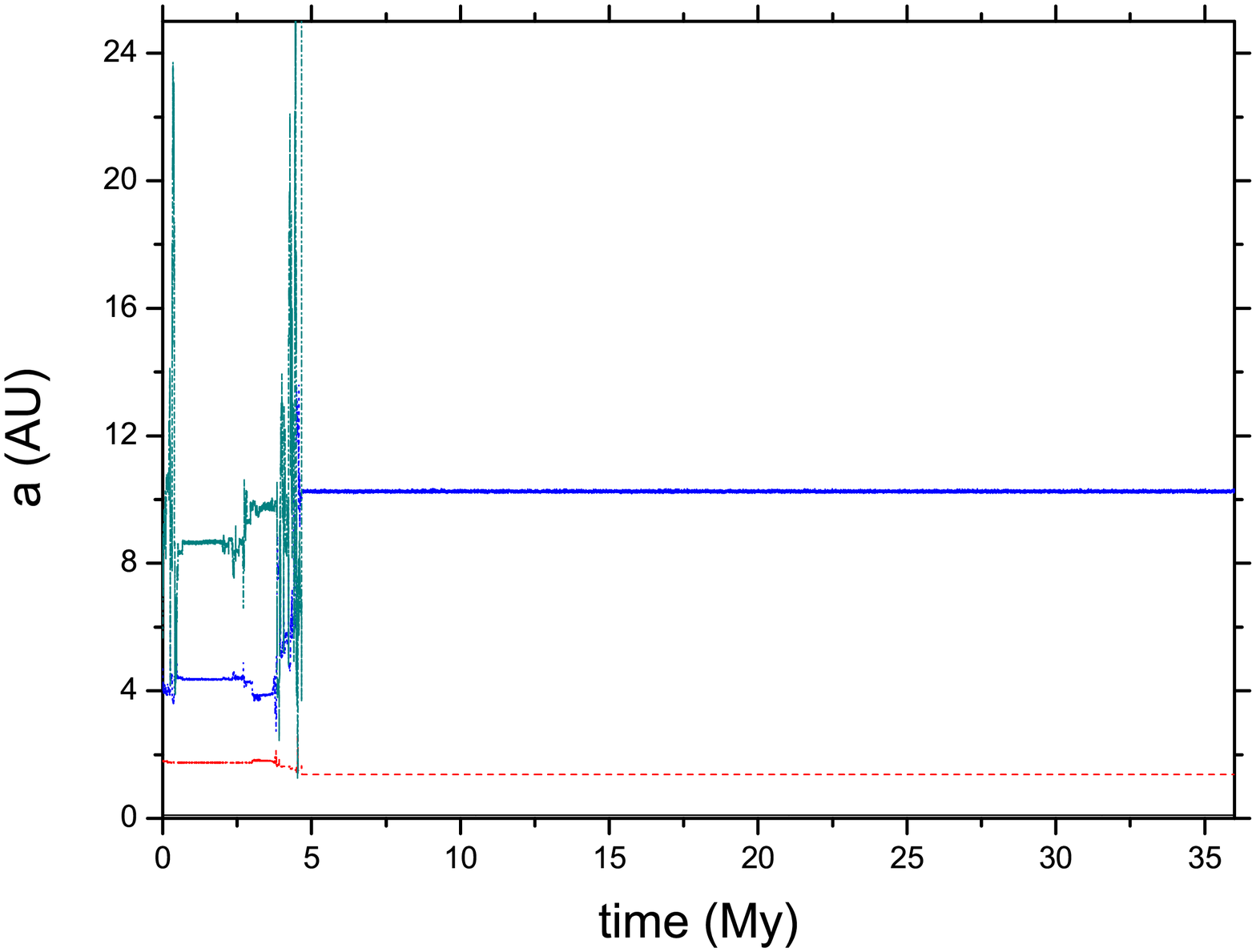} \includegraphics[width=\columnwidth]{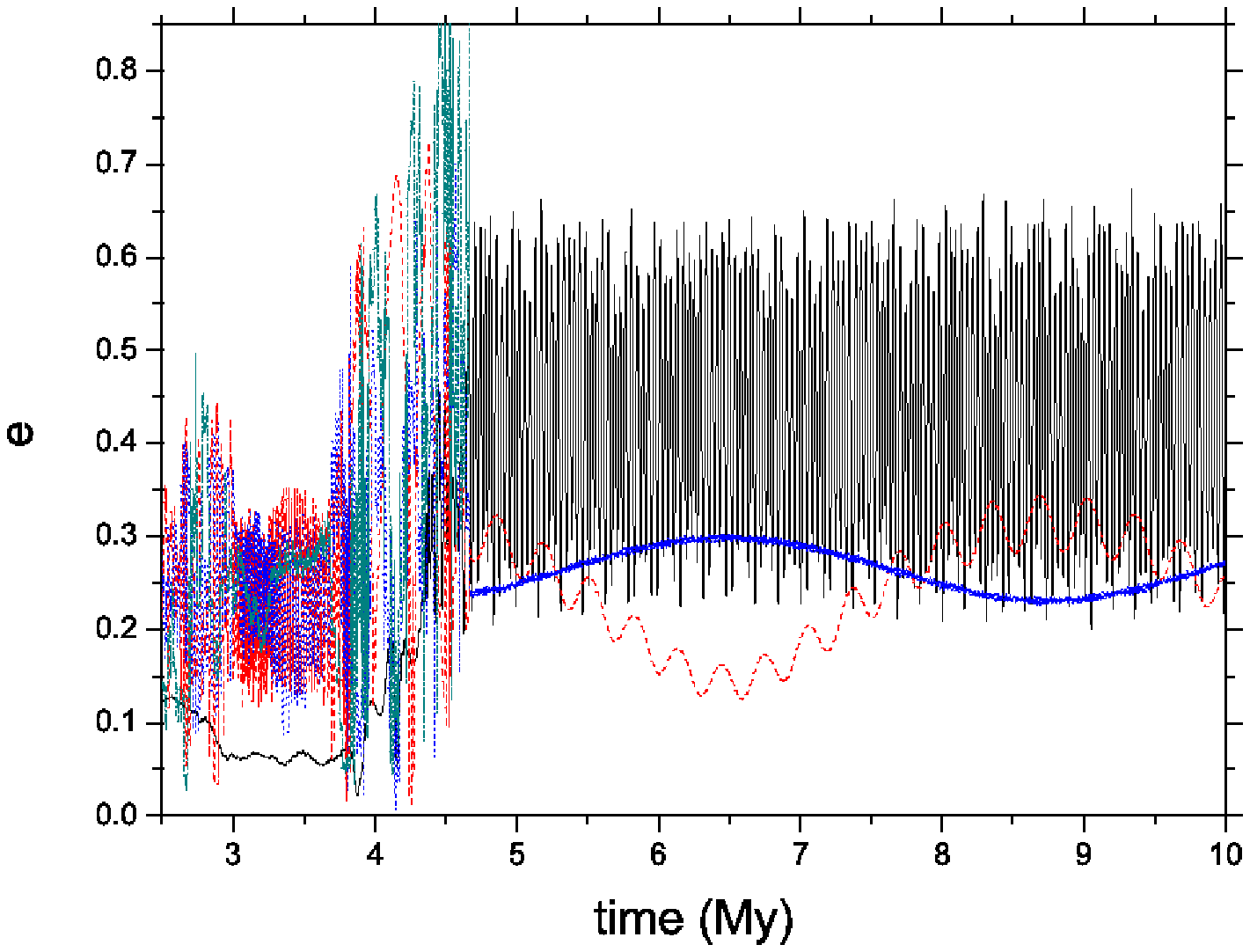}
\caption{Evolution of some orbital elements for a configuration from set cdx. Top panel: Time evolutions of the semi-major axes. Initially, in order of increasing distance planets \textit{b}, \textit{c}, \textit{d} and \textit{x} are located. Bottom: Time evolutions of the eccentricities. The solid (black) line indicates planet \textit{b}, the dashed line (red) represents planet \textit{c}, the dotted line (blue) denotes planet \textit{d} and the dashed-dotted line (dark cyan) is planet \textit{x}. When \textit{x} is ejected around 4.5 My, the eccentricities oscillate stably on a secular time-scale. $e_b$ moves in the interval 0.20-0.65, $e_c$ moves in the interval 0.12-0.34, while $e_d$ in the range 0.23-0.30.}
\label{higheb}
\end{figure}

A case in which $e_b$ grows to the required value is displayed in Figure \ref{higheb} 
where after many close encounters planet \textit{x} is ejected and planet \textit{b} 
gains a maximum eccentricity of $e_b=0.65$, which is a value that becomes relevant when 
considering tidal effects (see Section \ref{tides}). The final semi-major axis for planet \textit{d},
however, is equal to 10.3 AU so the overall final architecture of the system differs
considerably from the observed system. This model is also characterized by orbital inclination 
growth: $i_b$ moves in the interval 25$^{\circ}$-83$^{\circ}$, $i_c$ in the range 
12$^{\circ}$-24$^{\circ}$, while $i_d$ stays in the range 3$^{\circ}$-13$^{\circ}$. In this 
and similar cases we have checked that the large eccentricity arises because of
scattering rather than the Kozai effect which can potentially become active when 
mutual inclinations exceed a value $\simeq 40$ degrees \citep{1962AJ.....67..591K}.
Summing over all simulations, we find that 11\%, 6\% and 2\% of the runs generate eccentricities 
for planet \textit{b} of at least 0.2, 0.4 and 0.6, respectively. These simulations therefore
demonstrate the feasibility of dynamical instability of the outer planet system causing
all planets in the system to develop large eccentricities.

\subsubsection{Probability of resonant capture}
Defining a successful outcome for the planet-planet scattering experiments is not
straightforward. The nonlinear nature of the process clearly means we cannot reasonably
expect that a relatively small number of N-body simulations will result in systems that
are close analogues to the currently-observed HD 181433 system. Instead we use a more
restricted definition of success in which planet \textit{b} experiences an increase in its
eccentricity and planets \textit{c} and {d} end up in 5:2 resonance. As discussed above,
our simulations have demonstrated that the eccentricity of planet \textit{b} can be
raised to the required value, and we also have outcomes in which planets \textit{c} and 
\textit{d} have period ratios that are quite close to 5:2. None of the simulations produce a
system in 5:2 resonance, however, so we now examine the probability of capture in resonance
by considering the width of the 5:2 resonance and the relative mean longitudes and
longitudes of pericentre required for the planets to orbit stably in resonance.

\begin{figure}
\centering
\includegraphics[width=\columnwidth]{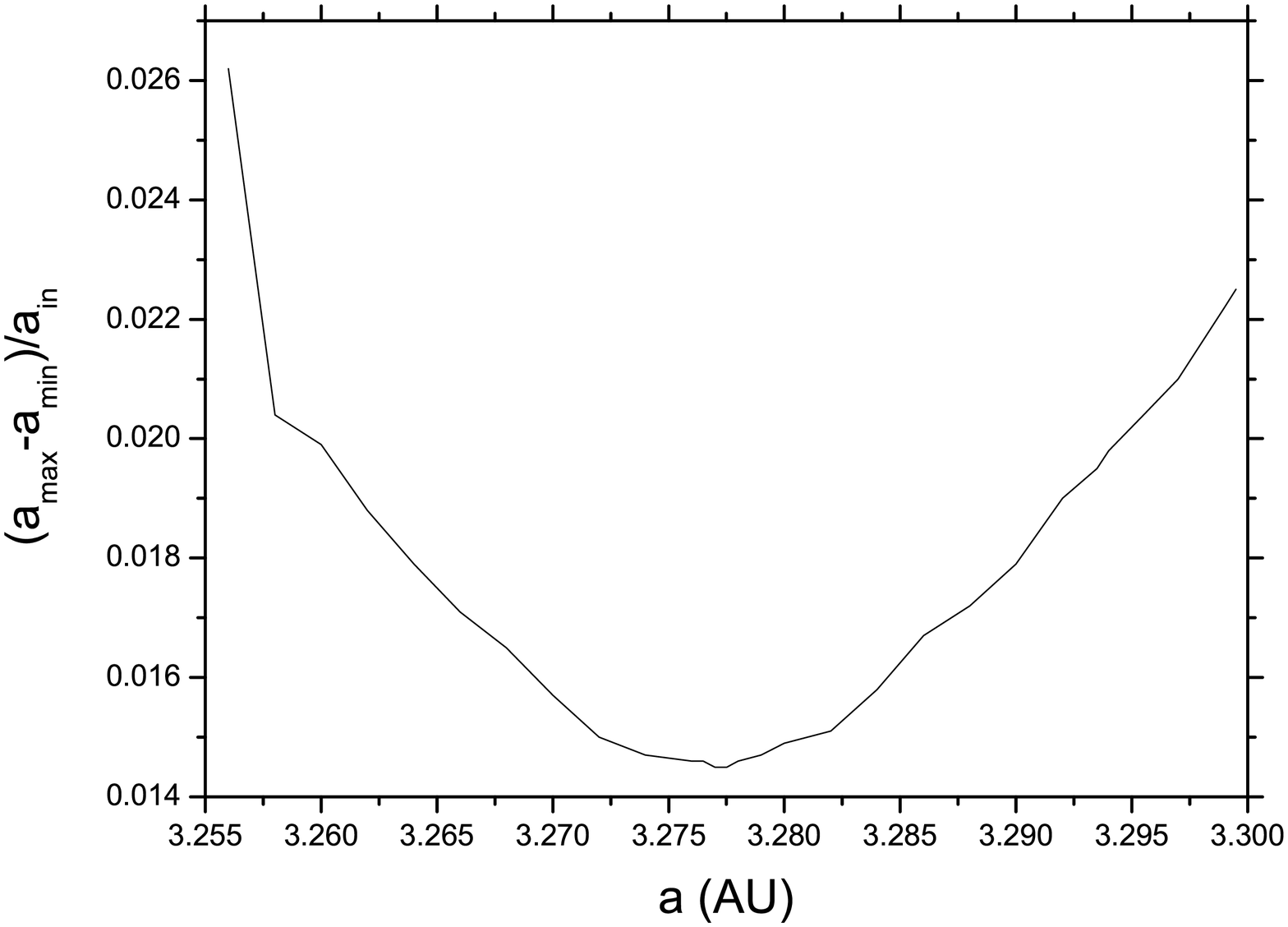} \includegraphics[width=\columnwidth]{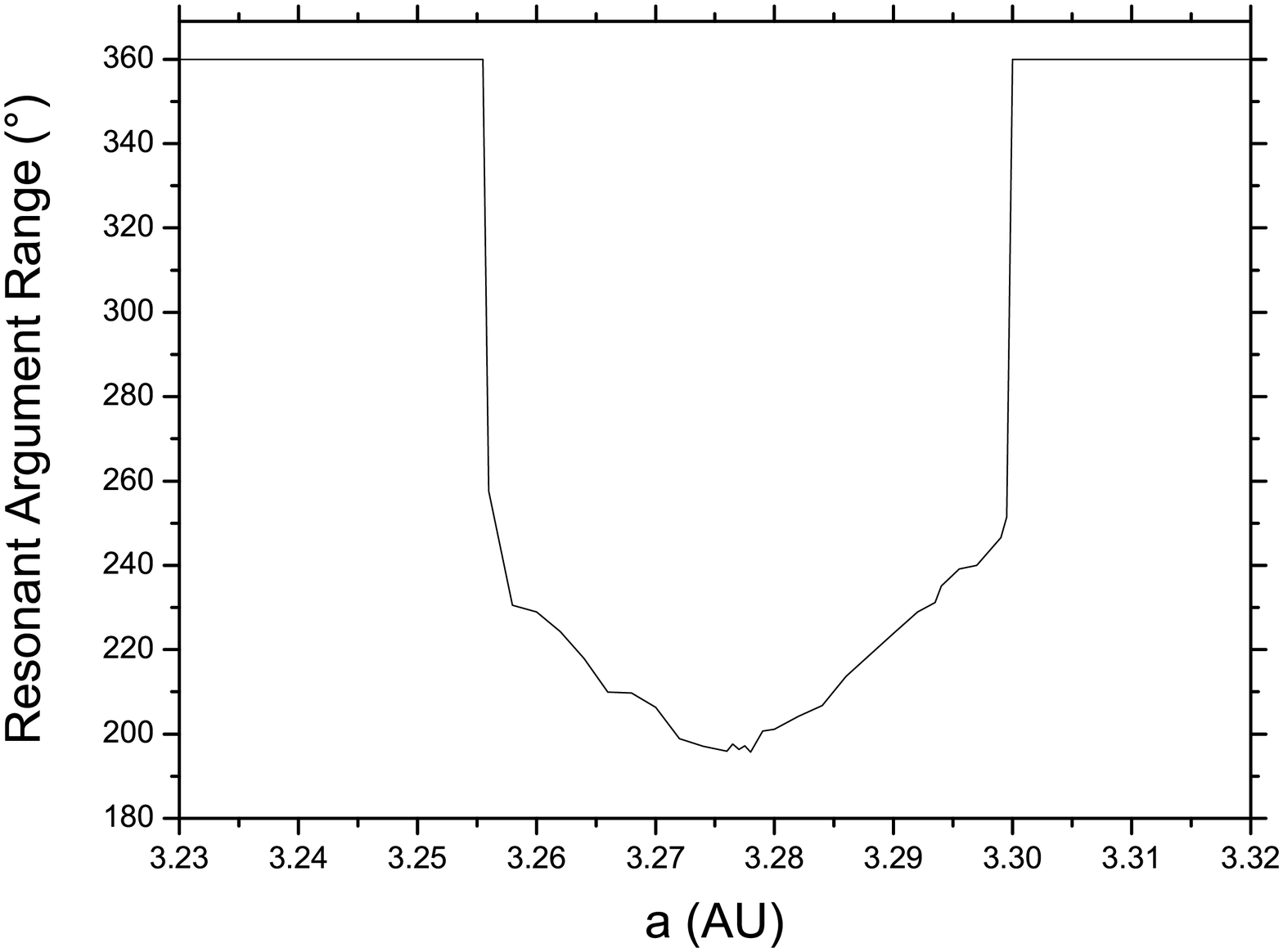}
\caption{The width of the 5:2 MMR. Top panel: The variation in size of resonant semi-major axis oscillations for different locations of planet \textit{d}. Bottom: The variation in the resonant argument. The system is unstable outside the resonance.}
\label{reswidth}
\end{figure}

\citet{b2} show that planet-planet scattering may result in pairs of planets landing in
high order MMRs. The resulting systems tend to have quite high eccentricities and resonant angles
that librate with large amplitudes, characteristics that are displayed by the HD 181433 planets as
described below and in \citet{b3}. The simulations presented by \citet{b2} produced a few 
5:2 MMRs for every  one thousand simulations. Here we examine the probability of two planets being scattered into 
the 5:2 MMR by determining the width of the resonance using N-body simulations that explore the dynamics 
of two bodies in resonance. Our procedure follows that adopted by Soja et al. (2011) in their study of
asteroids in resonance. We consider the presently inferred orbital elements of planets \textit{c} and 
\textit{d} and we study the width of the region inside of which libration occurs by varying the 
semi-major axis of \textit{d} in steps of 0.0005 AU. For each case, we compute the maximum amplitude of the 
oscillations in semi-major axis by taking the difference between the maximum and minimum values over 
50000 years of integration, normalized by the initial semi-major axis. The top panel of Figure \ref{reswidth} 
shows how the amplitude of oscillations in semi-major axis varies in the resonant region.
 Planet \textit{d} survives only in the range $3.2560 \le a \le 3.2995$ AU,
and the system is disrupted when \textit{d} is placed just outside this zone. The resonant argument for the 
5:2 MMR of the planetary system HD 181433 is \citep{b3} 
\begin{equation}\label{resarg}
\psi = 5\lambda_d-2\lambda_c-3\varpi_d
\end{equation}
where $\lambda$ is the mean longitude and $\varpi$ is the longitude of pericentre.
The bottom panel of Figure \ref{reswidth} shows the libration amplitude of this resonant angle as a 
function of the semi-major axis of planet \textit{d}. The libration amplitude at the location of \textit{d}
that corresponds to the best-fit stable orbital solution presented by \citet{b3} agrees well with the value
quoted in that paper.  We find the width of the resonance to be $\Delta_a = 0.0435$ AU centred at $a=3.2775$ AU.

A simple estimate for the probability of the planets landing within the resonance after scattering is
\begin{equation}\label{Pdeltaa}
P(\Delta_a)=\frac{\Delta_a}{\overline{a_d-a_c}}
\end{equation}
where ${\overline{a_d - a_c}}$ is the median value of $a_d - a_c$ at the end of all simulations for which strong plant-planet
interactions occurred (we include only those runs for which final eccentricity of at least one of the planets $e > 0.1$). 
We obtain $P(\Delta_a)=0.0075$. Having planets land within the required $\Delta_a$ after scattering, however, does not 
guarantee that they will be in resonance. It also depends on the angles that define the mutual orientation of their 
eccentric orbits (the difference between their longitudes of pericentre $\varpi_c-\varpi_d$), and also the values of their 
mean anomalies at the beginning of their interaction once they land within the resonance width. To quantify this 
aspect of the problem we ran a set of simulations where we take the semi-major axis values that lie at the centre of 
the resonance, and the eccentricity values for the stable best fit solution. We vary $\varpi_c-\varpi_d$ in steps of 
$90^\circ$ and the mean anomalies $M_c$ and $M_d$ in steps of $45^\circ$, for a total of 256 simulations.
We ran the integrations for 30 Myr. 34 pairs of planets survive, all in resonance and in anti-aligned mode. 
The resonant argument (equation \ref{resarg}) librates with amplitudes that vary from a few degrees up to about 
$240^\circ$, as outlined by the lower panel of Figure \ref{reswidth}.
Finally, we are able to estimate the probability of resonance capture to be $P_{5:2} \sim 0.0075 \times 34/256 \simeq 10^{-3}$,
in decent agreement with the larger sample of numerical simulations presented by \citet{b2}.

\subsubsection{Probability of scattering generating HD 181433 systems}
\label{sec:probscat}
We have determined that planetary systems with global structure similar to
HD 181433, but which originally had an additional gas giant planet orbiting
close to the two outer giant planets, can lead to the excitation of the
eccentricity of the inner super-Earth up to values $e_b \simeq 0.4$ 
during approximately 6 \% of the time when the system experiences a 
global dynamical instability. This eccentricity excitation occurs 
because one or more of the outer planets has a close encounter 
(or a series of close encounters) with the inner planet during the chaotic
phase of evolution. Treating the perturbation of the interior planet
\textit{b} onto an eccentric orbit, and the landing of the two outer planets
in the 5:2 mean motion resonance, as being independent processes, the joint
probability of eccentricity excitation and resonant capture becomes $P \simeq 6 \times 10^{-5}$.
Taken at face value, this result suggests that systems with characteristics similar to HD 181433
occur through planet-planet scattering rather rarely.

\section{Sweeping secular resonances due to stellar spin-down}
\label{spin}
Having determined that the planet-planet scattering hypothesis is a plausible scenario
for excitation of the eccentricity of all planets in the HD 181433 system, but that the
excitation of $e_b$ combined with resonant capture of planets \textit{c} and \textit{d}
is likely to be a rare event, we now consider alternative scenarios for exciting planet 
\textit{b}'s eccentricity. We consider the hypothesis that the eccentricities and resonant 
structure for the orbits of the outer planets were established after a period of dynamical 
instability once gas disc dispersal had occurred, and the eccentricity of planet \textit{b} 
was established through the sweeping of a secular resonance with the outer planets caused by the 
spin-down of the central star from an initial state of very rapid rotation. 

The idea that a short-period planet may experience excitation of its eccentricity and
inclination because of stellar spin-down originates with \citet{b30}. They showed that 
if the primordial gravitational field of the Sun had a larger second-degree harmonic 
[i.e. $J_2 \gtrsim {\cal O}(10^{-3})$], equivalent to a rotation period $\lesssim 5.7$ hours,
then subsequent solar spin-down would drive Mercury's orbit through secular resonances
capable of generating its large mean eccentricity and inclination.
Secular resonances are generated when the orbits of two bodies precess synchronously. A 
small body in secular resonance with a large planet will have its eccentricity and inclination 
modified over relatively short time periods. As the mass distribution of a planetary system evolves, 
for example as a result of gas disc dispersal, orbital migration or stellar spin-down, the locations 
of secular resonances move. Planets located in regions through which the resonance sweeps are
perturbed as the resonance drives the eccentricity and inclination. 
Given that we assume the current orbits of the outer planets were established shortly after gas disc 
dispersal, stellar spin-down provides the resonant sweeping in our model, and
the rate of eccentricity/inclination forcing scales with the square root of the stellar spin-down 
time \citep{b31}. In principle it should therefore be possible to tune the spin-down time scale
to obtain the desired eccentricity for planet \textit{b}.
There are other planetary systems with architectures similar
  to HD 181433 (i.e., a factor of $>10$ in orbit period between
  the inner and outer planets, and an eccentric inner planet) such as
  HD125612 and $\mu$ Arae.  The same arguments used to constrain the
  orbital history of the HD 181433 system may apply to these systems
  too.  We are investigating the more general implications of these
  evolutionary tuning processes and their application elsewhere \citep{AgnorLin2012b}.

Stars are generally believed to lose their primordial angular momentum through the magnetic braking 
action of the stellar wind with a mass-loss rate orders of magnitude greater than that on the main 
sequence (e.g. Skumanich 1972). Pre-main sequence stars with masses over $0.25 M_\odot$ exhibit a bimodal period distribution with observed values clustered around 6--8 days and 2 days. The transition between the two peaks is fairly abrupt. T Tauri stars can have spin periods of the order of hours during their evolution, with a significant fraction of them showing this characteristic \citep{2002A&A...396..513H, b35}. 
In addition to providing a method of exciting
the eccentricity of an interior planet through resonant sweeping \citep{b30}, stellar spin-down
from an initial rapid state of rotation has also been invoked to explain the low eccentricity
of an interior planet. Using early system parameters for the then three planet $\upsilon Andromed{\ae}$ system, \citet{b10}
showed that the rotation period of the parent star had to be shorter than 2 
days during dispersal of the gas disc so that the passage of the sweeping secular resonance near 
the orbit of the short-period planet \textit{b} could have been avoided leaving its eccentricity at 
a low level. Our model differs in that we assume the eccentricities of the outer planets to
have been established after gas disc removal rather than before or during its occurrence
given its role in damping large planetary eccentricities (e.g. Papaloizou et al. 2001).

\subsection{Secular model including stellar spin-down}
We begin our analysis by modifying the Laplace-Lagrange secular model described in Section~\ref{Properties}
to account for stellar spin-down through inclusion of the $J_2$ contribution to the eigenfrequencies
of the system. As previously remarked, the secular model provides only an approximate estimate of 
the locations of secular resonance and can therefore be used to quickly evaluate the 
hypothesis that a secular resonance may have swept the present-day semi-major axis 
of planet \textit{b} at 0.080 AU during stellar spin-down. 
We adopt the method described in \citet{b24} for including the $J_2$ terms, and we use the
relation between the stellar spin period $P_*$ and $J_2$ provided by \citet{b30}.
%taking a tidal Love number \textbf{[DO YOU MEAN THE APSIDAL CONSTANT $k$ HERE?]} for a star with polytrope of index 3 equal to $0.02$.

\begin{figure*}
\includegraphics[width=0.33\textwidth]{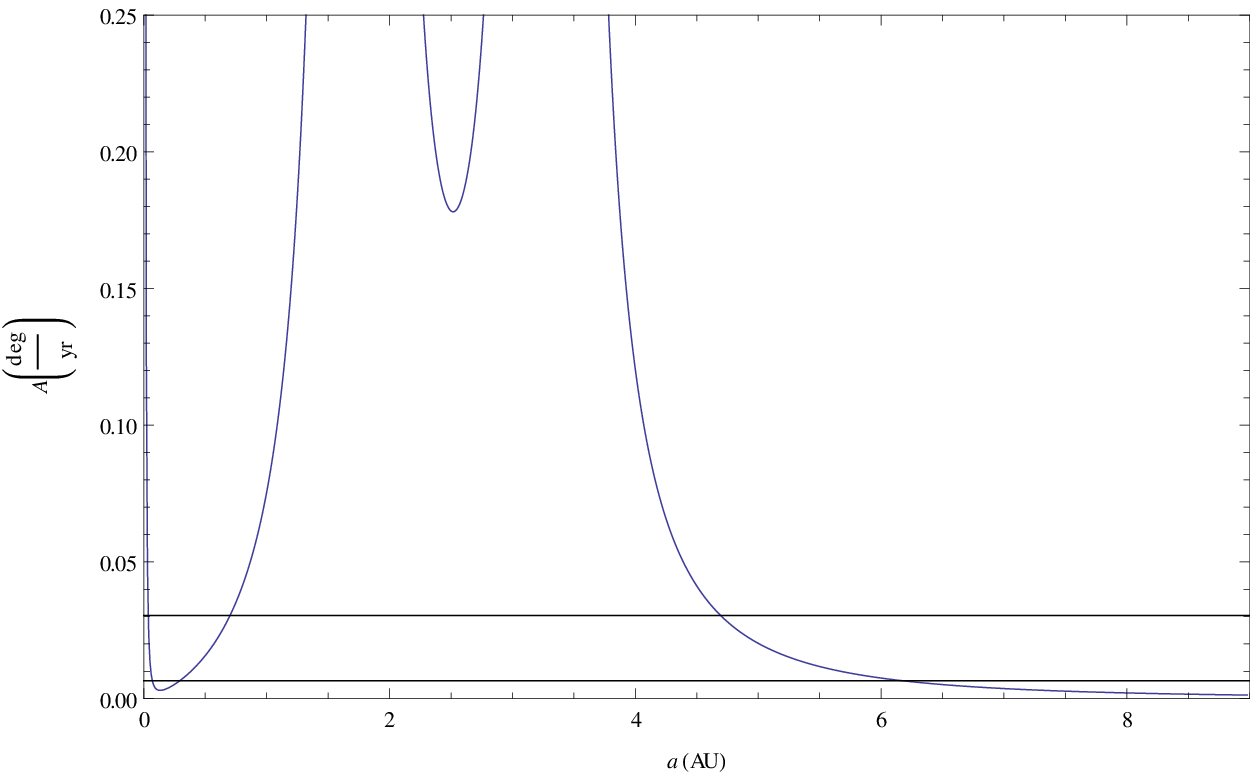}\includegraphics[width=0.33\textwidth]{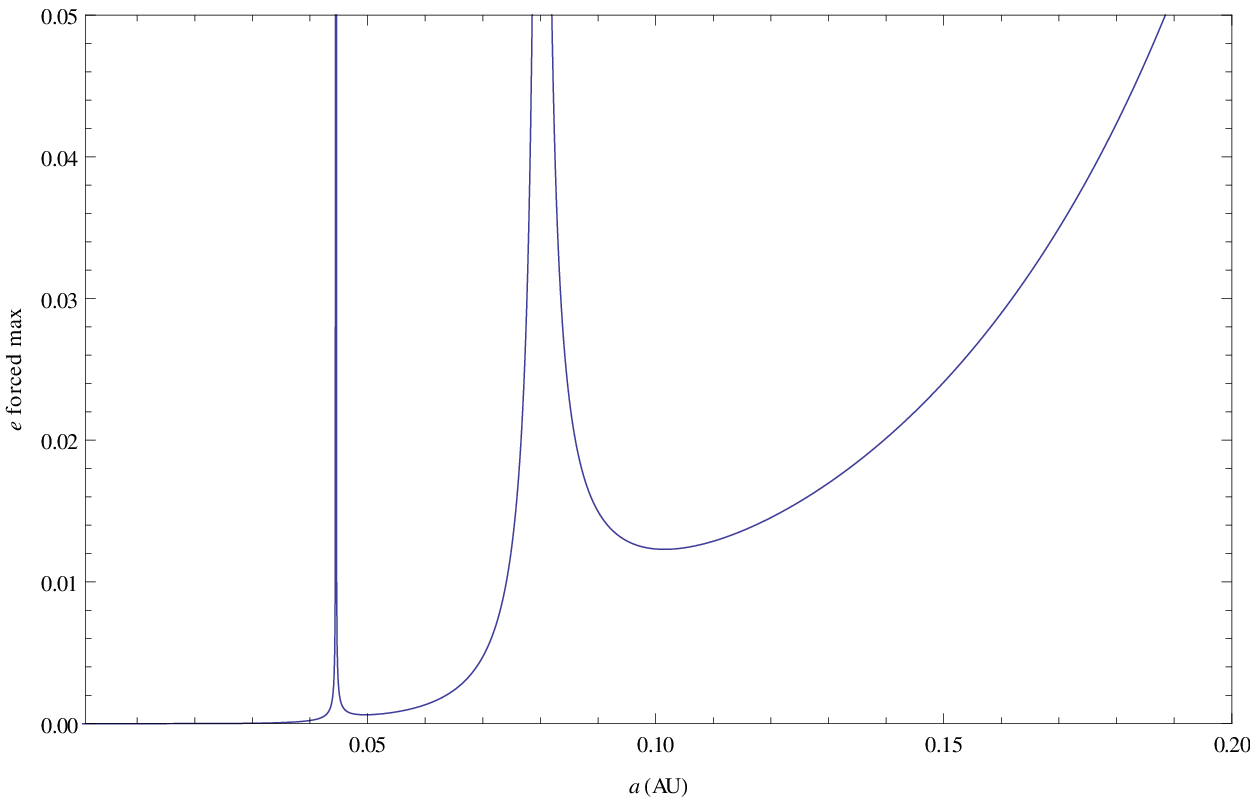}\includegraphics[width=0.33\textwidth]{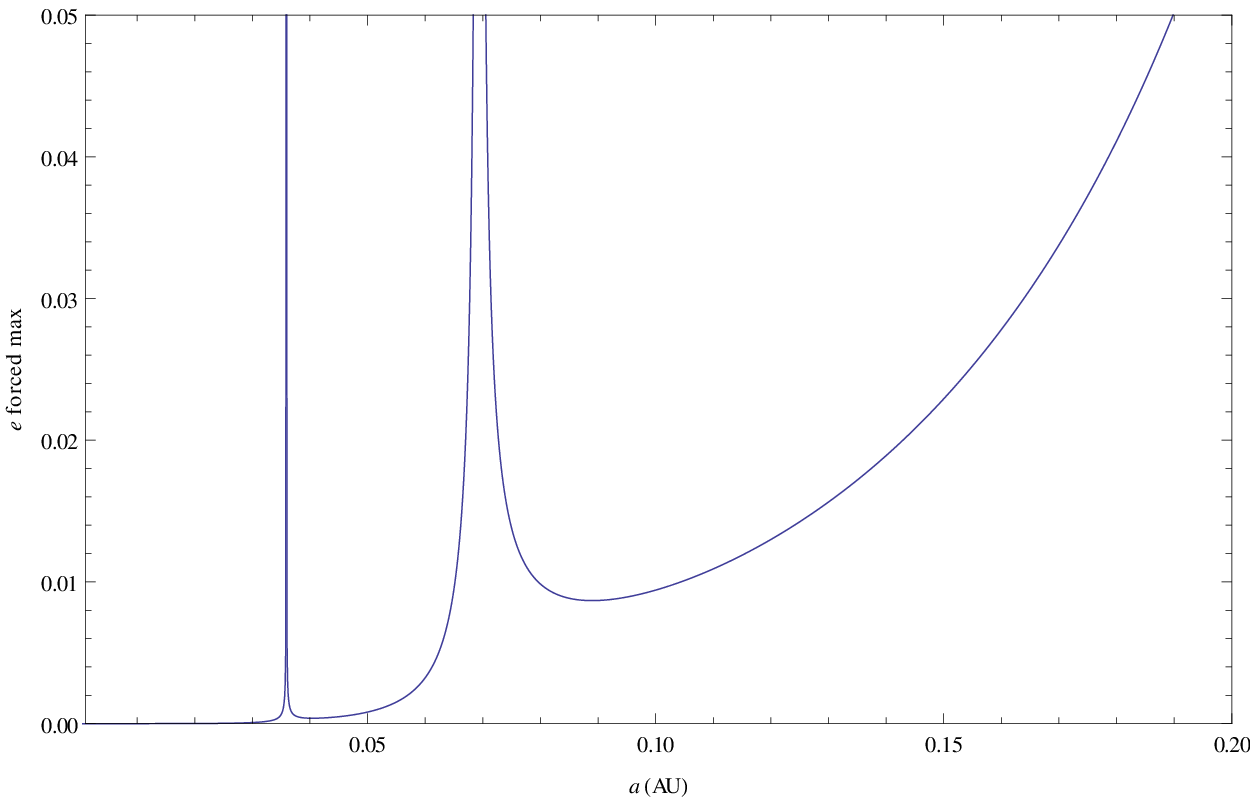}
\caption{The process of sweeping secular resonances for HD 181433. Left: Precession frequency $A$ of a test particle as a function of semi-major axis, derived from perturbations by the two giant planets, considering stellar spin and GR. The horizontal solid lines denote the values of the two eccentricity-pericentre eigenfrequencies. Centre: The maximum forced eccentricity in the very close region of the parent star which is assumed to rotate in 2.1 days. The secular resonances are located at 0.045 and 0.08 AU. Right: The maximum forced eccentricity within 0.2 AU from HD 181433 when it is slowly rotating. The secular resonances have now moved to 0.035 and 0.07 AU, respectively.}
\label{sweep}
\end{figure*}

The mass of planet \textit{b} is much smaller than the outer two giants so we treat it as a
test particle in the secular model. The left panel of Figure \ref{sweep} presents the 
free precession period of a test particle induced by planets \textit{c} and \textit{d}. It is important 
to point out the role of GR-induced precession in promoting secular resonances close to the star:
neglecting GR, the precession rate would fade to zero very near to the 
star making it difficult to match any eigenfrequencies of the outer planet secular system
(see Section \ref{Properties} for how precession rates change due to GR). 
The central and right plots in Figure \ref{sweep} show the sweeping of two secular resonances 
as the parent star spins down from a rotation period of 2 days to 30 days (for which the 
$J_2$ effects become insignificant). According to this simplified model, when the rotation period of 
HD 181433 was $P_* \approx 2.1$ days (equivalent to $J_2 \approx 2.2 \times 10^{-5}$), the 
free precession rate at the present location of planet \textit{b} matched the one of the
eigenfrequencies of the system. 
Later, the secular resonances move inward toward their present-day locations. 
Using the argument that the eccentricity excitation due to the passage of a secular resonance is 
inversely proportional to the square root of the changing rate of the resonant frequency \citep{b31}, 
we can estimate the spin-down time $\tau= \Omega/\dot{\Omega}$, where
$\Omega=2\pi/P_*$ is the rotational angular velocity of the star required to excite the present eccentricity. We have:
\begin{equation}\label{rateprec}
e  \simeq \left(\frac{2\pi}{|\dot{A}|}\right)^{1/2} \mu_d 
\end{equation} 
where, from \citet{b24},
\begin{equation}\label{mud}
\mu_d = \sum_{j=1}^2 A_j e_{jd} 
\end{equation} 
with $\mu_d$ modelling the resonance of interest, $e_{jd}$ being
the components of the scaled eigenvector for planet \textit{d},
$A_j$ defined by equation 7.142 in \citet{b24}, and $\dot{A}$ is the
rate of change of the resonant precession frequency $A=\dot{\varpi}$ at resonance.
Stellar rotation enhances the star's oblateness. The effect of
rotation on stellar oblateness can be encapsulated in the resulting
$J_2$ coefficient which allows for simple inclusion into secular
models and N-body simulations. From \citet{b30}, we have:
\begin{equation}\label{dotj2}
   J_2 = \frac{2}{3}k\frac{\Omega^2 R_*^3}{GM_*}
 \end{equation}
where for the apsidal constant $k$ we take the value $8.16 \cdot 10^{-3}$ calculated for
the Sun \citep{b30}
%[CHECK THIS WITH THE LOVE NUMBER ABOVE -- they should be the same - and put in the value here].
Combining Eq.\ref{dotj2} with secular theory for an oblate primary
\citep{b24} gives the following relation between the spin down rate of
the star $\dot{\Omega}$ and the rate of change in the precession frequency $\dot{A}$
\begin{equation}\label{precMurray}
 \dot{A} \simeq 2\left(\frac{R_*}{a} \right)^2
  \left( \frac{\dot{\Omega}}{\Omega} \right) k\left(\frac{\Omega^2 R_*^3}{GM_*}\right)n.
\end{equation} 
Finally, for HD 181433 b this model suggests that a spin down
timescale of $\tau \approx 1.9 \times 10^{7}$ years may be capable of
accounting for its large observed free eccentricity of 0.39.
This time scale is consistent with estimated mass loss rates from stars in the T-Tauri stage \citep{b30},
suggesting that excitation of planet \textit{b}'s eccentricity through sweeping
secular resonances is a realistic hypothesis worthy of further more detailed exploration.

\subsection{N-body simulations of secular resonance sweeping}

As discussed in Section \ref{Properties}, the precession period for planet \textit{d} is shorter than 
suggested by the secular model because of MMR effects. Therefore, we implement a stellar spin-down
model in the BS algorithm of the integrator MERCURY-6 to account for the time dependent rotational flattening of
the host star. Our model now includes effects due to GR and stellar oblateness, where we adopt
the expression for the acceleration due to the oblate star given by \citet{b10}, in additional
to the gravitational interaction between the three planets. We have
run checks to ensure that the precession periods  that arise from the simulations agree with
analytical expression from secular theory. The stellar spin-down is modelled as a magnetic braking 
torque based on the empirical Skumanich law (Skumanich 1972): 
\begin{equation}\label{down}
\frac{d{\bf \Omega}}{dt}=-\alpha \Omega^2 {\bf \Omega}
\end{equation}
where ${\bf \Omega}$ is the stellar spin vector and $\alpha = 1.5
\times 10^{-14}$ years for a G or K dwarf (e.g.,~Barker \& Ogilvie 2009).
This parameter measures the speed of removal of angular momentum from a rotating star. 
Given the importance of angular momentum in determining the eccentricity evolution
of the system we have confirmed that it is conserved in the simulations
at a level better than 1 part in $10^6$.

\subsubsection{Evolution during resonant sweeping}
\label{reson}
The secular theory  suggests that a stellar spin period of between 11.1--16.7 hours would force planet \textit{b} to precess 
with a frequency similar to one of the system's eigenfrequencies, such
that a resonance may be established.  For comparison, the minimum
rotation period for the star is $P_{cr}=2\pi\sqrt{R_{*}^3/GM_*} \approx 3.2$ hours.

We initiate N-body simulations with planets \textit{c} and \textit{d} in their inferred present
day configuration, and with planet \textit{b} on a circular orbit close to the star
(later on we consider scenarios with differing implications for the long-term tidal
evolution of the system discussed in Section \ref{tides}, and so place planet \textit{b}
at different semi-major axes). Here we consider evolution that implies very little tidal evolution 
of the system has occurred over its life-time, consistent with an adopted
value of $Q_{\rm p} \gtrsim 10^5$ as discussed in Section \ref{tides}. We therefore place planet
\textit{b} with its currently observed semi-major axis $a_b=0.08$ AU. We initiate simulations
with a stellar spin period of 14 hours, and vary the value of the spin-down parameter
$\alpha$, beginning with its nominal value given above.

\begin{figure*}
\includegraphics[width=0.5\textwidth]{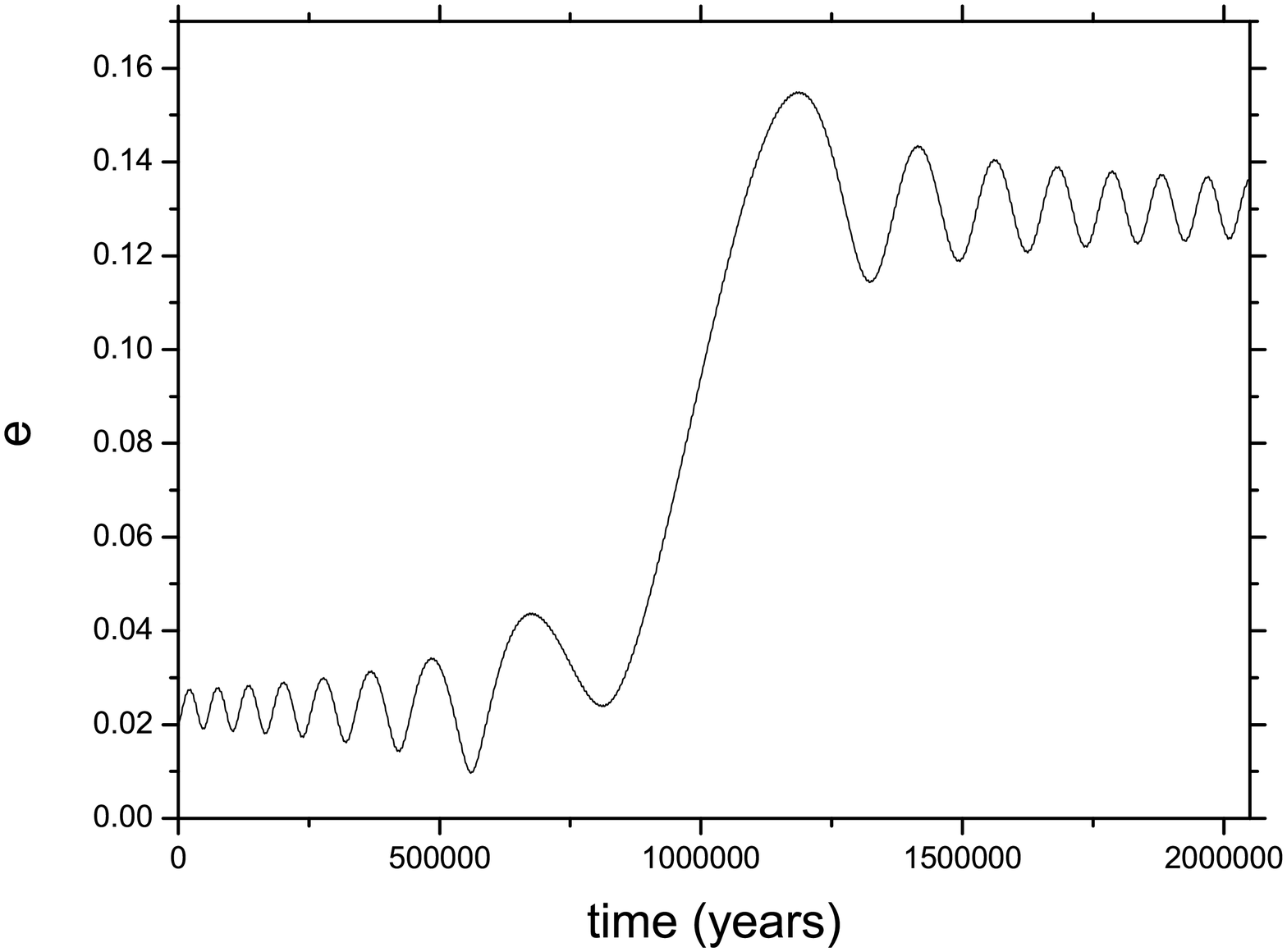}\includegraphics[width=0.5\textwidth]{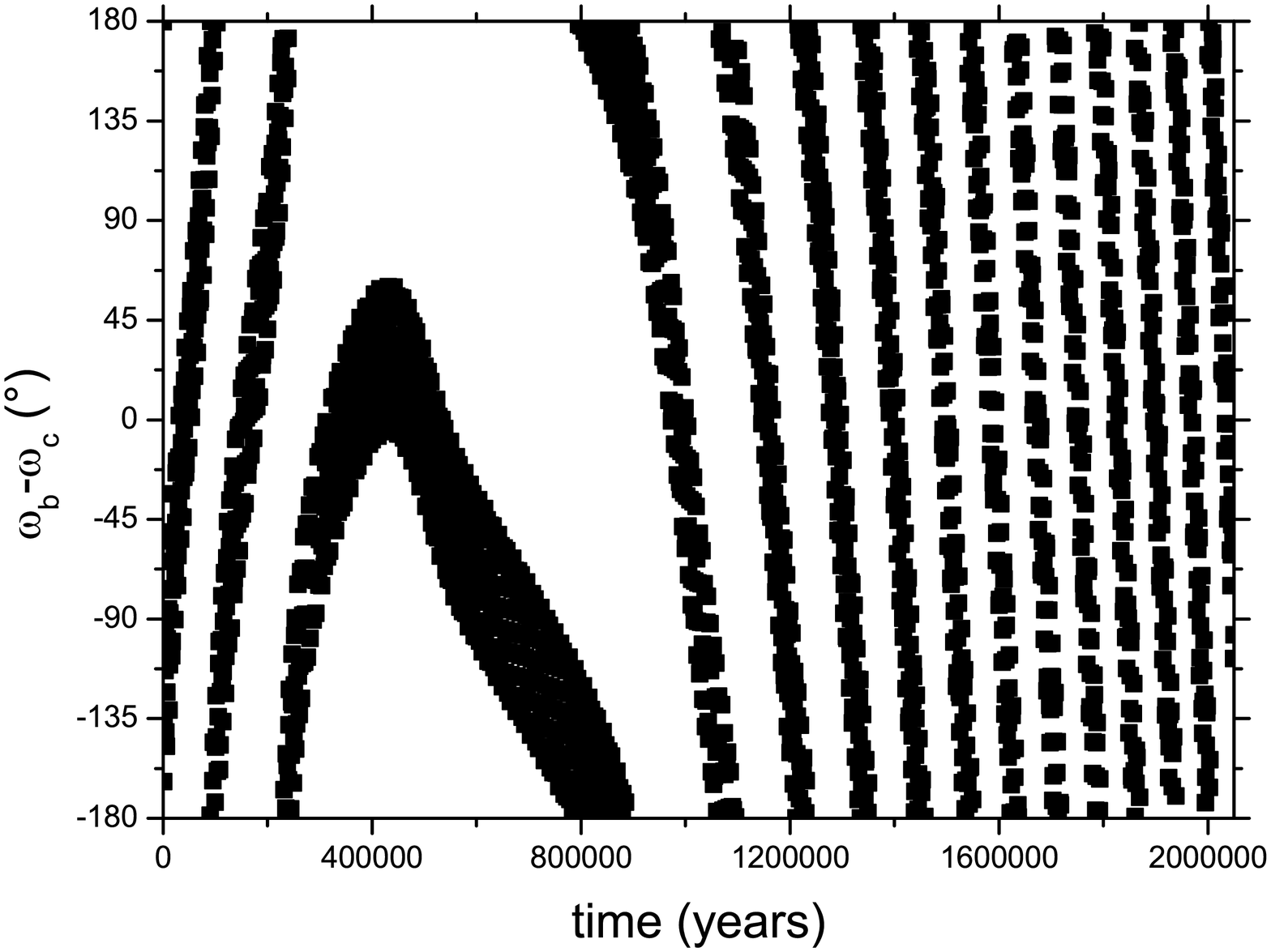}
\includegraphics[width=0.5\textwidth]{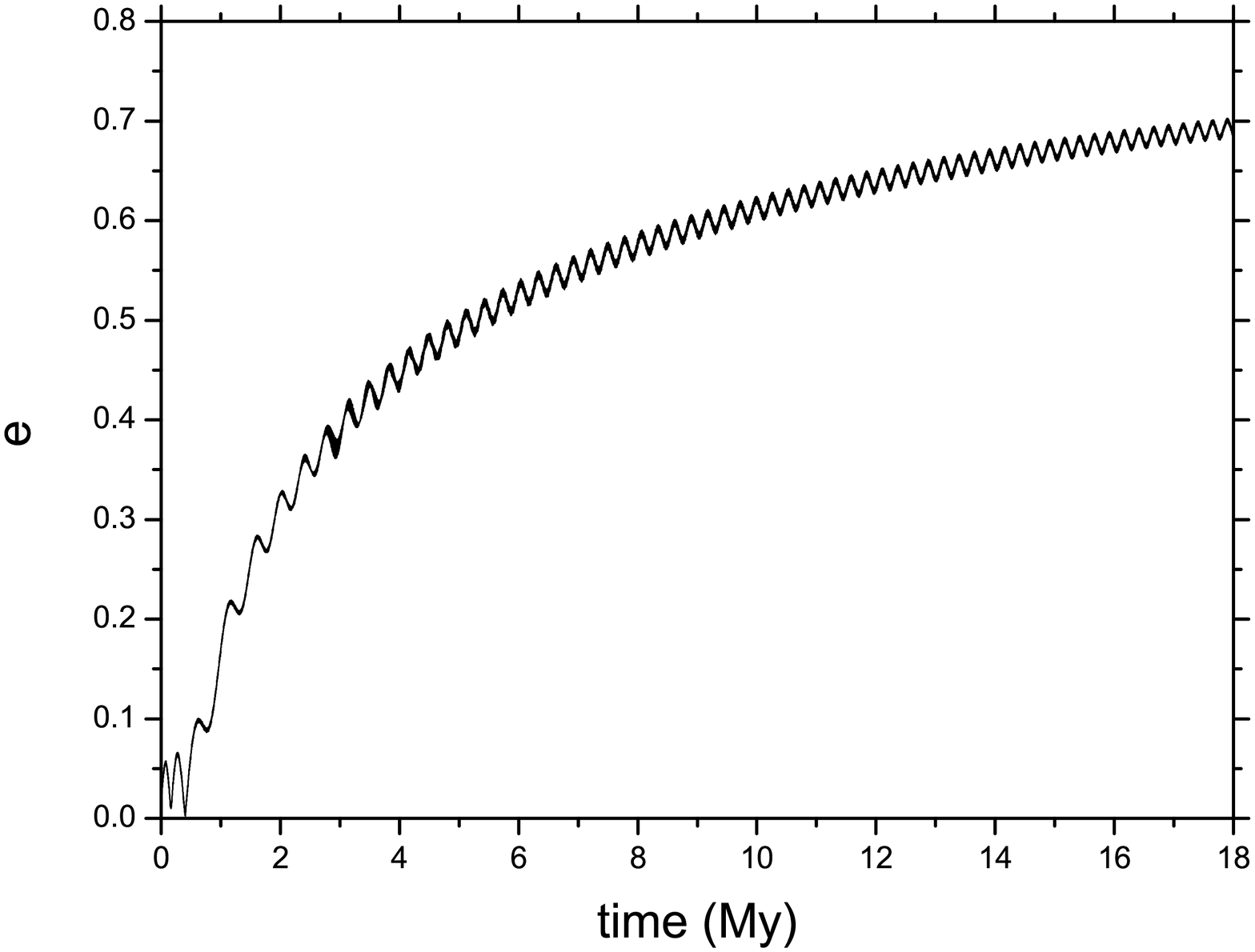}\includegraphics[width=0.5\textwidth]{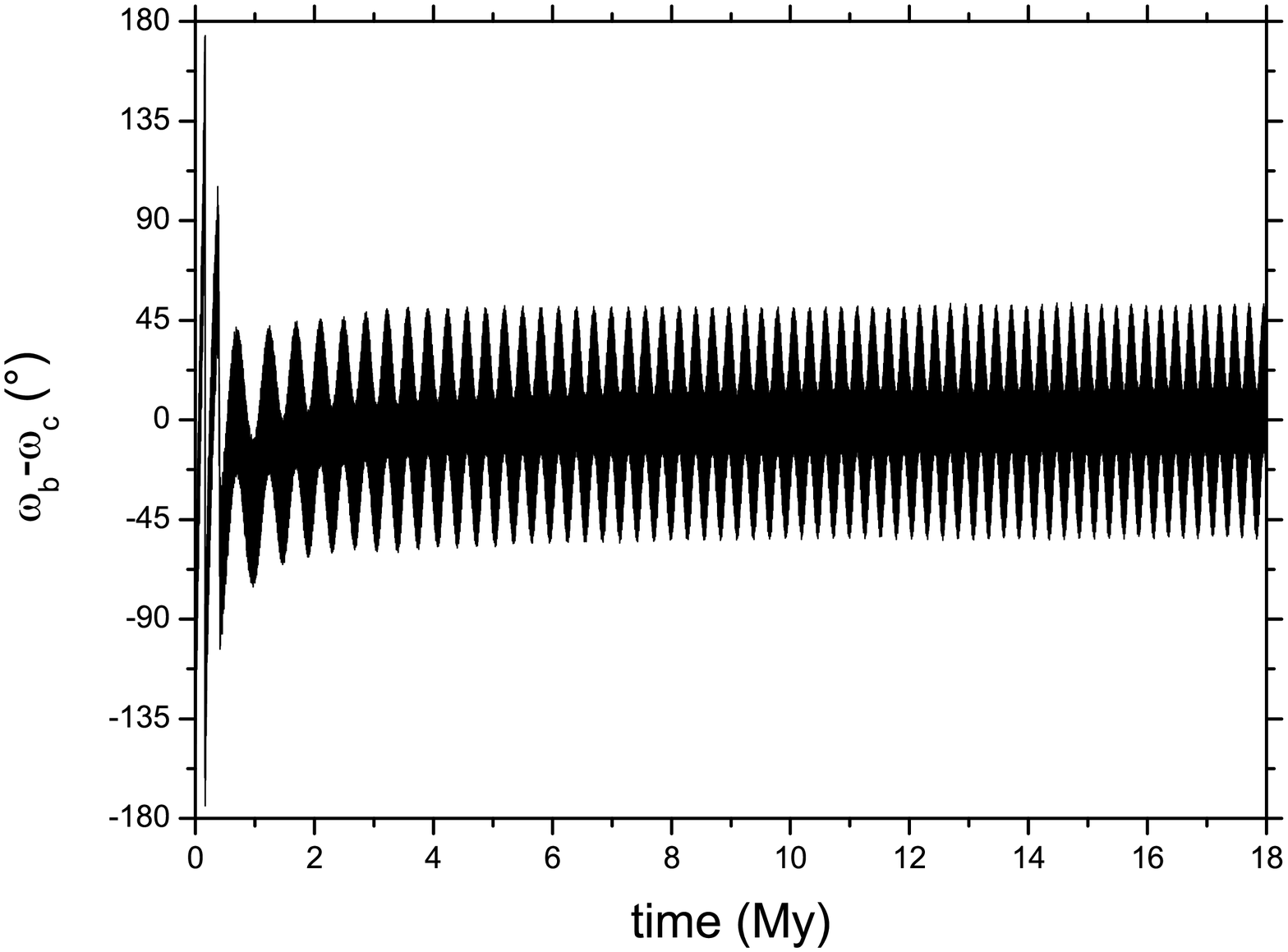}
\caption{The sweeping of secular resonances. Top panels: Evolution of $e_b$ when $\alpha = 1.5 \times 10^{-14}$ 
years in the left panel, and the related evolution of the secular angle $\varpi_b-\varpi_c$ in the right panel.
Note how the angle librates during the passage of the resonance, which corresponds to a sharp increase in the 
orbital eccentricity of planet \textit{b}. Bottom panels: Change in $e_b$ when the spin-down parameter is smaller
by 5.75\% in the left plot, and the evolution of the secular angle $\varpi_b-\varpi_c$ in the right panel.
This time, even if the star keeps on spinning down, planet \textit{b} is locked in the resonance with planet 
\textit{c} as a critical value for $e_b$ is reached which generates the precession rate necessary to maintain 
the resonance. This causes the eccentricity of \textit{b} to grow indefinitely.}
\label{alpha}
\end{figure*}

Results are shown in the top panels of Figure \ref{alpha} for the nominal value of
$\alpha=1.5 \times 10^{-14}$ years. In the right panel it is possible to observe how the 
relative longitudes of pericentre $\varpi_b-\varpi_c$ evolve during the process: 
the initial growth of eccentricity begins when the precession rates of planets 
\textit{b} and \textit{c} match ($P_* \approx 16$ hours). The passage of the resonance is anticipated 
by the orbit of \textit{b} precessing faster initially and then being overtaken by the precession rate 
of \textit{c}. $e_b$ peaks at 0.16 and stabilizes later at a value of $e_b \simeq 0.13$. 
Setting the initial stellar rotation period to 5 hours instead of 14 hours produces the same result,
with the eccentricity peaking when the stellar spin period is 17 hours.

Excitation of orbital eccentricity depends on the mass and eccentricity of the perturber. 
\citet{b11} do not report an uncertainty on the mass of planet $c$, while the quoted errors on
$e_c$ are relatively small with $\sigma_{e_c} = 0.02$. Such a small change in the value of $e_c$
would lead to only small changes in our results. We know the minimum mass of $c$. However, \citet{b3} notes 
that the stable best--fit is found in a dynamically active region of phase space, 
and a value for $\sin i$ noticeably different from 1 would generate instabilities in the system. 
A slightly increased mass for planet $c$ would lead to only a slightly modified secular resonance.

As discussed above, the expected level of eccentricity excitation depends on the rate
of resonant sweeping. A larger value of $e_b$ requires the spin-down rate to be slower,
so we have examined how the evolution changes with smaller values of $\alpha$.
Interestingly, we find that the system behaviour can be divided into two distinct modes
that depend critically on the value of $\alpha$. For spin-down rates that exceed the
critical value ($\alpha_{\rm crit}$ is 5.75 \% smaller than the nominal value) the evolution
is similar to that described above and illustrated in the upper panels of Figure~\ref{alpha}:
temporary capture in the resonance and excitation of $e_b$ to values $e_b \lesssim 0.25$.
Spin-down parameters equal to or smaller than $\alpha_{\rm crit}$ lead to long-term capture in the secular
resonance (apparently indefinite capture) and growth of $e_b$ toward unity. This mode of evolution
is shown in the lower panels of Figure~\ref{alpha} for a run with $\alpha=\alpha_{\rm crit}$, 
where over a run time of 18 Myr $e_b$ reaches a value of 0.7 and $\varpi_b-\varpi_c$ librates around 
zero with a semi-amplitude of $\sim 45$ degrees. Apparently a planet caught within this
mode of evolution is driven to $e_b=1$ and collision with the central star unless tides
are able to intervene for cases where $Q_{\rm p}$ is small enough to drive sufficiently rapid
tidal damping of $e_b$.

\begin{figure*}
\includegraphics[width=0.5\textwidth]{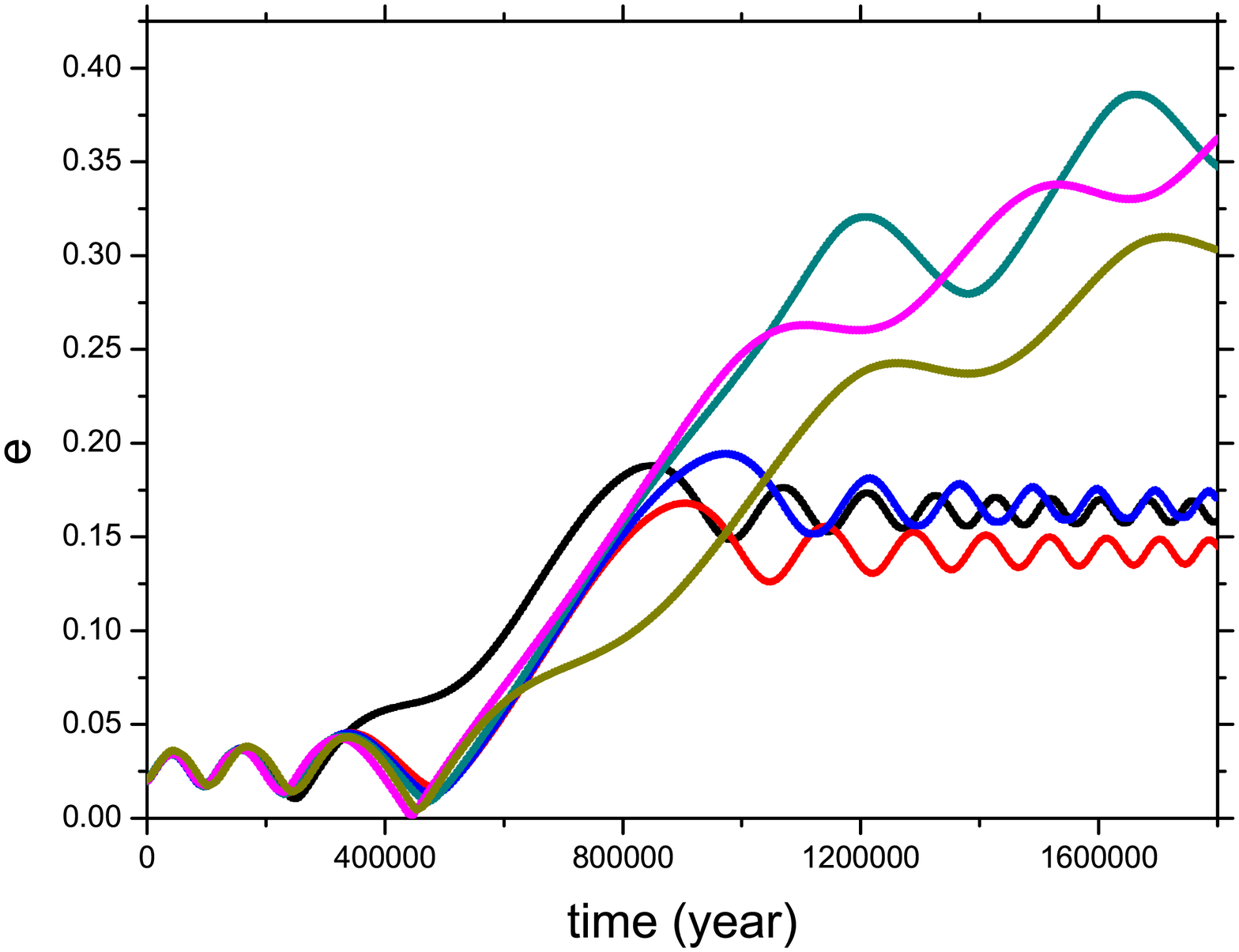}\includegraphics[width=0.5\textwidth]{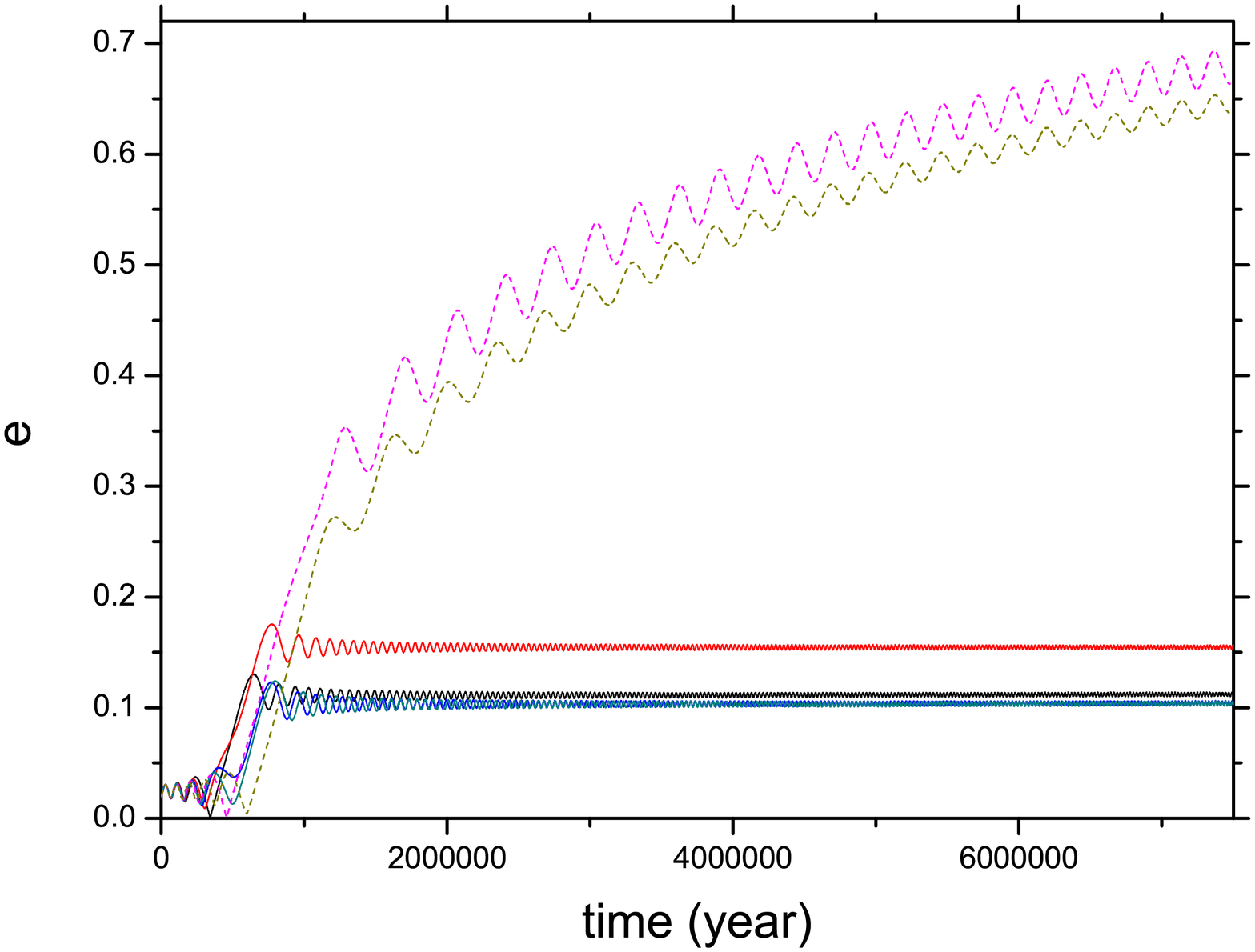}
\caption{The growth of $e_b$. Left: Considering both GR and $J_2$ effects. Values of 10\% larger and 3\% and 
4.5\% smaller than the nominal $\alpha$ stabilize around a value for $e_b$ equal to $0.15$. For values of 
6\%, 12\% and 50\% smaller, $e_b$ grow indefinitely. Right: Considering $J_2$ effects only. For the nominal 
$\alpha$ and values 10\% larger and 9\% and 15\% smaller than it, $e_b$ stabilize around $0.15$. 
For values 22\% and 50\% smaller, $e_b$ grows indefinitely. }
\label{spindownevst}
\end{figure*}

The reason for the existence of these two regimes can be sought in the expressions for the precession 
rate of the longitude of periastron due to GR and $J_2$, $\dot{\varpi}_{GR}$ (see e.g. Misner et al. 1973 
for a derivation) and $\dot{\varpi}_{J_2}$ \citep{b34}, respectively:
\begin{equation}\label{precrate}
{\dot \varpi}_{\rm GR}=\frac{3GM_*}{ac^2(1-e^2)}n\\
{\dot \varpi}_{J_2}=\frac{3}{2}\frac{J_2}{(1-e^2)^2}\left(\frac{R_*}{a}\right)^2n
\end{equation}
where \textit{n} is the mean motion, and both ${ \dot \varpi}_{GR}$ and ${ \dot \varpi}_{J_2}$ depend 
on the eccentricity such that an increase in $e$ leads to an increase in ${\dot \varpi}_{GR}$ and
${\dot \varpi}_{J_2}$. The condition for the resonance to be maintained during spin-down is given by
\begin{equation}
\frac{\partial {\dot \varpi}_{J_2}}{\partial J_2} \frac{d J_2}{dt}
= - \left( \frac{\partial {\dot \varpi}_{J_2}}{\partial e} + 
         \frac{\partial {\dot \varpi}_{\rm GR}}{\partial e} \right) \frac{de}{dt}.
\label{eq:res-condition}
\end{equation}
In other words, the reduction in precession rate due to stellar spin-down needs
to be compensated by the increase in precession rate that occurs as eccentricity grows.
We predict from equation~\ref{eq:res-condition} that removing the effects of GR will
still allow long term secular resonant locking, but for slower values of the spin-down
parameter $\alpha$. We have performed simulations to examine this by omitting the
GR term in the equations of motion, and find that a rotation period of $\simeq 15$ hours is 
required to enter secular resonance and the spin-down parameter needs to be more than 20 \% smaller
than the nominal value to maintain long-term resonant capture, in agreement with our
expectation. The plots in Figure \ref{spindownevst} show the growth of $e_b$ for different 
values of $\alpha$ in cases in which GR effects are included (left panel) and neglected (right panel),
demonstrating long-term resonant capture for $\alpha$ below a threshold value in each case.

\subsubsection{Resonant sweeping with additional exterior planets}
\label{addpl}
The simulations presented in the previous section indicate two modes of behaviour,
but neither of them are able to explain the observed eccentricity of planet \textit{b}.
One results in an eccentricity that is too small, and the other apparently results in
either an eccentricity which is too high or collision with the central star.
One possibility that we explore here is that there may have been additional planets
in the system orbiting relatively close to planet \textit{b} during sweeping
of the secular resonance. If the spin-down is below the critical value required
for long-term capture then interactions with the additional planets when the
eccentricity becomes large may release planet \textit{b} from the secular resonance,
resulting in a final eccentricity of the required magnitude.

We begin by exploring the evolution with one additional Earth-mass planet 
(so-called planet \textit{x}) in the system located outside of the orbit of 
planet \textit{b} on a circular orbit. We ran a suite of 12 simulations where planet 
\textit{x} is located within 2-8 mutual Hill radii from the apocentre of planet \textit{b} 
calculated when $e_b$ is in the range 0.3-0.7. The idea here is to induce planet-planet scattering
when eccentricity growth is already underway; it is equivalent to placing 
planet \textit{x} in the range 0.10-0.16 AU. Including planet \textit{x} modifies the resonance
condition, so for each simulation we have calculated the new spin period required to produce 
the necessary precession rate for \textit{b}. We find that for an
additional Earth-mass planet placed between 0.10-0.16 AU, that a
stellar spin period from 0.75 days up to 5 days will enable resonant
interaction with the exterior giants during stellar spin down. 

\begin{figure*}
\includegraphics[width=0.33\textwidth]{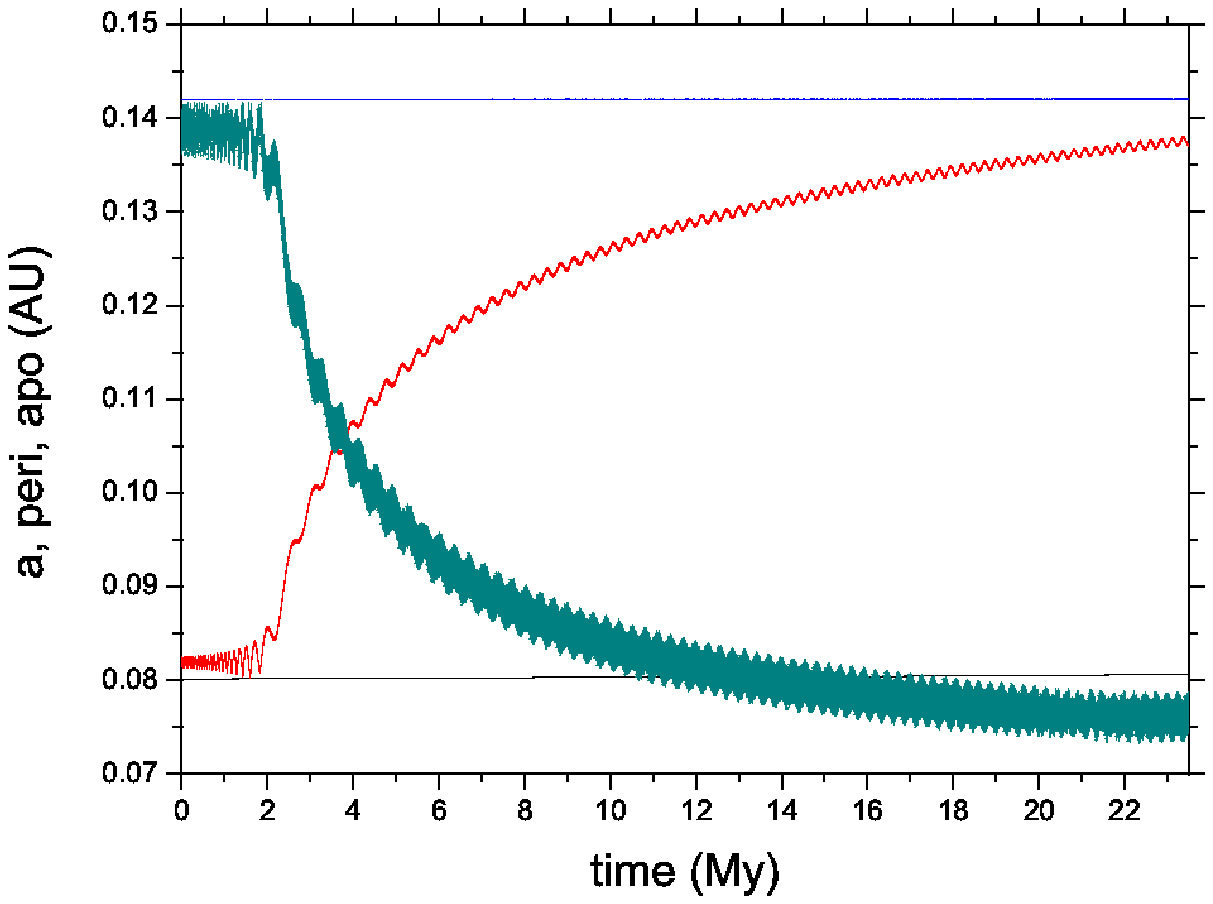}\includegraphics[width=0.33\textwidth]{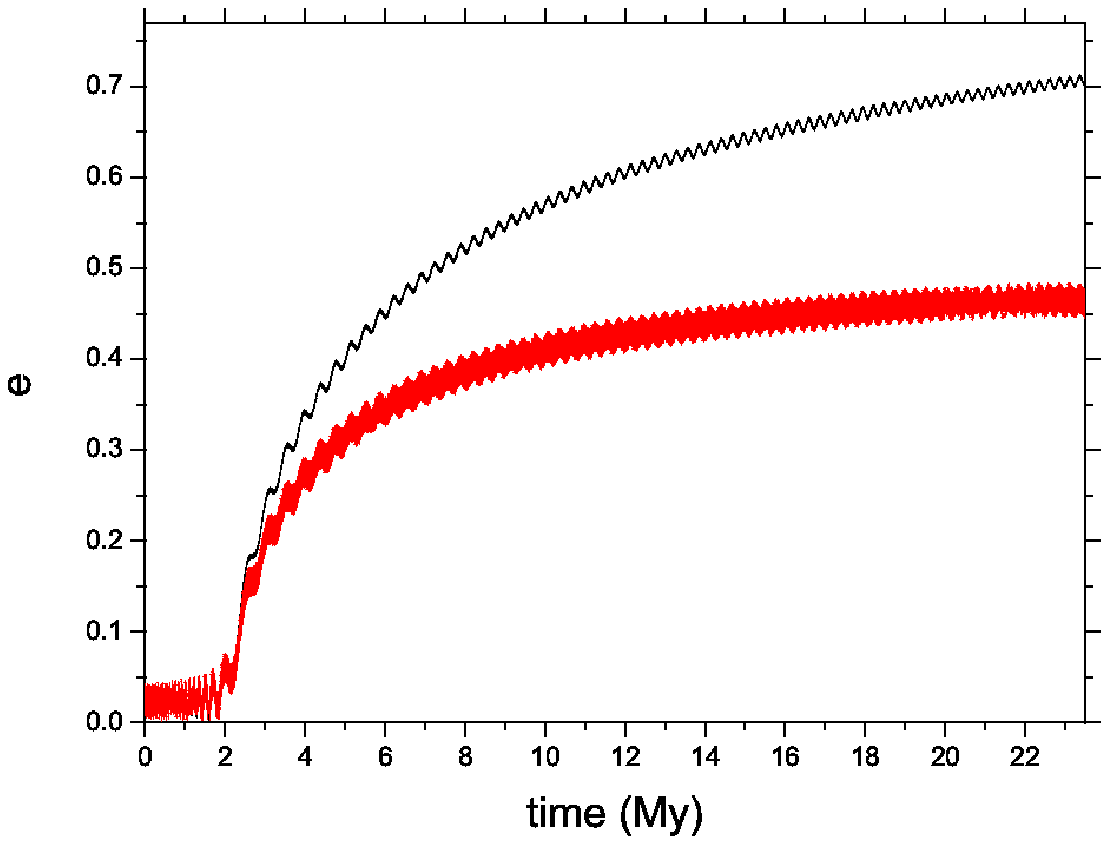}\includegraphics[width=0.33\textwidth]{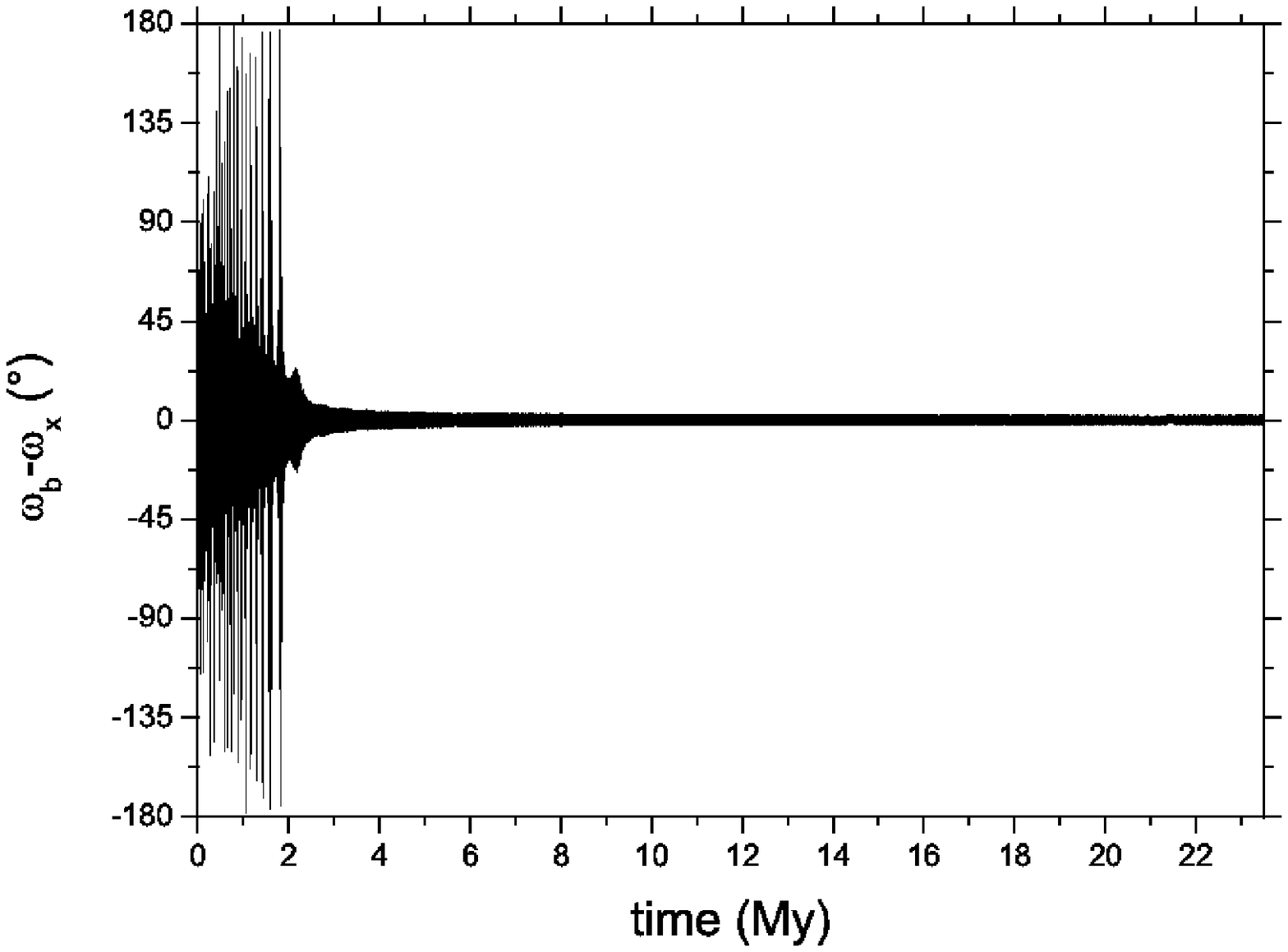}
\caption{The process of sweeping secular resonances for HD 181433 when an Earth-mass planet is added. 
The spin-down parameter is reduced by 6\% with respect to the nominal value. Left: Evolution of semi-major axis and apocentre for planet 
\textit{b}, semi-major axis and pericentre for planet \textit{x}. Centre: Evolution of $e_b$ and $e_x$. 
Right: Evolution of the secular angle $\varpi_b-\varpi_x$, showing the lock into the resonance occurs after $\sim 2$ My.}
\label{1extra}
\end{figure*}

The simulations yield the result that both planets \textit{b} and \textit{x} become 
trapped in the secular resonance with \textit{c}, and each of them experiences eccentricity 
growth without limit. Figure \ref{1extra} reports one example with planet \textit{x} at 
0.14 AU, a stellar spin period $\simeq 20$ hours necessary to generate the resonance, and spin-down 
parameter $\alpha$ that is 6\% smaller than the nominal value.
The left panel illustrates how the orbits of \textit{b} 
and \textit{x} cross as their eccentricities grow continuously (central plot). Collisions are
avoided, however, because planet \textit{x} is trapped in resonance with planet \textit{b},
as demonstrated by the right panel. Planets \textit{b} and \textit{c} are in a
secular apsidal resonance, with $\varpi_b-\varpi_c$ librating around 0$^{\circ}$ with a 
semi-amplitude of $\sim 45^{\circ}$.  This behaviour is a feature of all runs for which
we included one additional Earth-mass planet, and we note that in the absence of
stellar spin-down all of the configurations that we considered were dynamically
stable over $10^6$ years.

For completeness we consider a model with three additional terrestrial planets to see
whether this promotes the sought-after instability when planet \textit{b} has reached
the desired eccentricity $e_b \sim 0.5$. 
We construct systems consisting of three additional 0.5--1 Earth-mass planets. The
inner-most additional body is placed 2--5 mutual Hill radii from the apocentre of the 
planet, calculated when its eccentricity is in the range 0.3--0.7. The
second additional body is placed 2--5 mutual Hill radii from the apocentre of the first additional body, calculated when its eccentricity is in the range 0.3--0.7. The third additional planet is placed 2--5 mutual Hill radii from the apocentre of the second additional body, calculated when its eccentricity is in the range 0.2--0.7.
We note that these three bodies are all placed 
interior 0.26 AU which is a stable zone. In fact, a planet with semi-major axis in the range 0.26 -- 0.35 AU obtains a forced eccentricity $e_{\rm forced} \gtrsim 0.2$ (see Section \ref{stabadd}), which will destabilize any planets in the range 0.1 -- 0.26 AU before the secular resonance is entered.
We consider a spin-down parameter 5.75 \% smaller than the nominal $\alpha$ and we calculate the 
necessary spin period for each case to enter resonance. We set a density of 3 g/cm$^{3}$ for 
the additional planets and a Neptune density (1.638 g/cm$^{3}$) for \textit{b}. 
We ran more than 100 simulations, varying the planetary mutual separations.

\begin{figure*}
\includegraphics[width=0.5\textwidth]{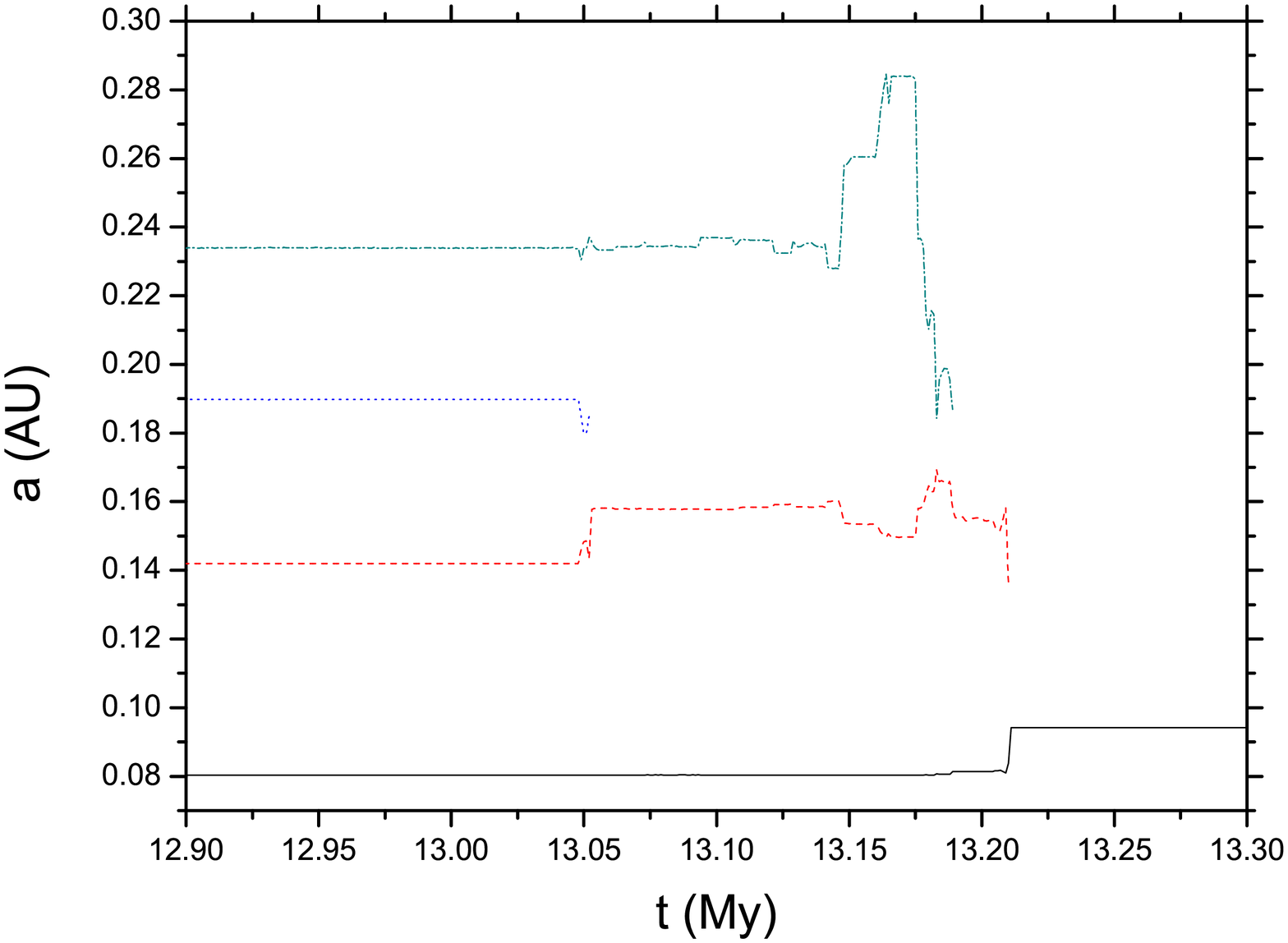}\includegraphics[width=0.5\textwidth]{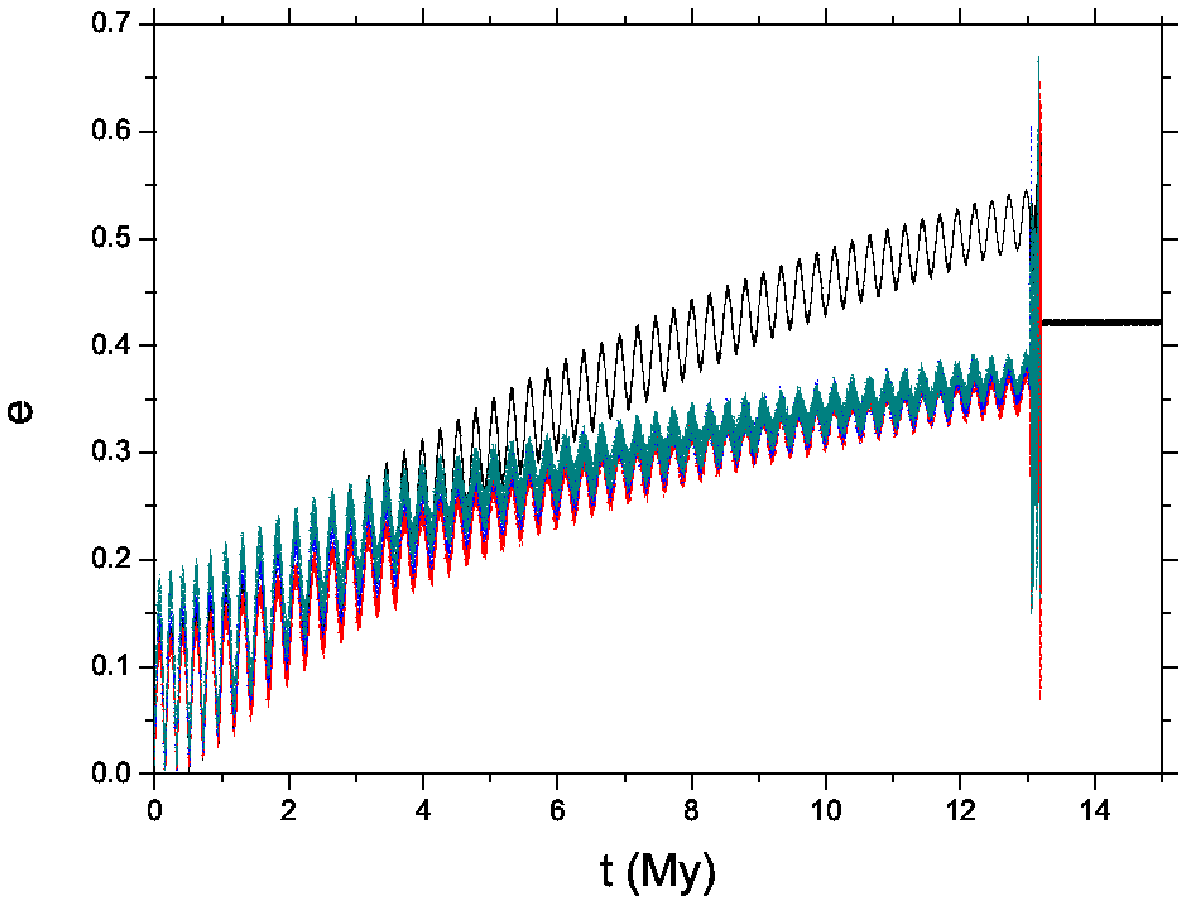}
\caption{The sweeping of secular resonances when three extra Earth-mass planets are included in the inner part of the 
HD 181433 system. Left: Evolution of the semi-major axes of the four inner planets, with planet \textit{b} being the innermost. 
Collisions are observed leading to reduction in the number of planets. Right: Evolution of their eccentricities, 
$e_b$ is the one growing fastest, achieving a final value of 0.43.}
\label{3ext}
\end{figure*}

From the results of the N-body simulations we observe that all four inner planets become involved 
in the resonant trapping, with the eccentricity growing for all of them before instability occurs
and strong mutual interactions take place. The outcomes of these simulations include
mutual close encounters, collisions between the planets, and collisions with the central star.
Occasionally the inner planets can disturb the fragile resonance between the two outer giants 
causing catastrophic ejections from the system. Out of 100 models we find two  models that
replicate the present configuration with a value for $e_b \gtrsim 0.4$. When a terrestrial planet 
survives in the process, the model is still compatible with the detected system because, 
for example using the Systemic Console \citep{b40}, a 1 Earth-mass planet at 0.19 AU would be at 
the noise level of the radial velocity data with an F-test value of $\approx 40\%$.
Figure \ref{3ext} illustrates a successful model for which only planet \textit{b} survives 
the instabilities and is able to achieve the required eccentricity. The necessary stellar rotation period to 
produce the resonance is  $\approx 22.2$ hours. The right panel represents how all the inner 
planets are increasing their eccentricity during the passage of the resonance, leading eventually
to close encounters and collisions which are displayed in the left panel. Orbital inclinations remain 
small in the system. We have tested the evolution of the systems neglecting the effects of stellar
spin-down and found that the eccentricities remain small and the system is stable over runs times of $3.2$ Myr.

\subsubsection{The resonance with an interior planet}
\label{1int}
\begin{figure*}
\includegraphics[width=0.5\textwidth]{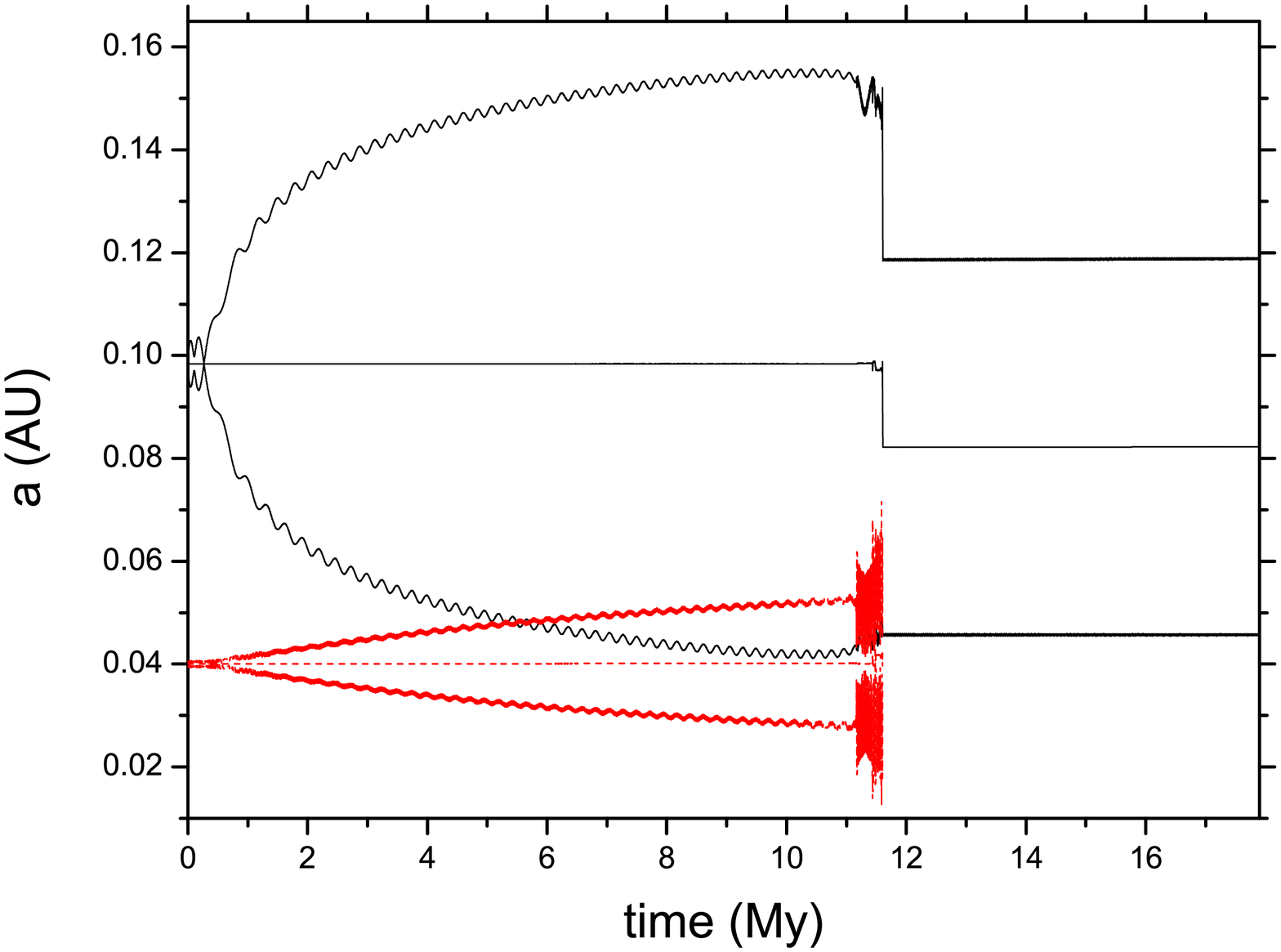}\includegraphics[width=0.5\textwidth]{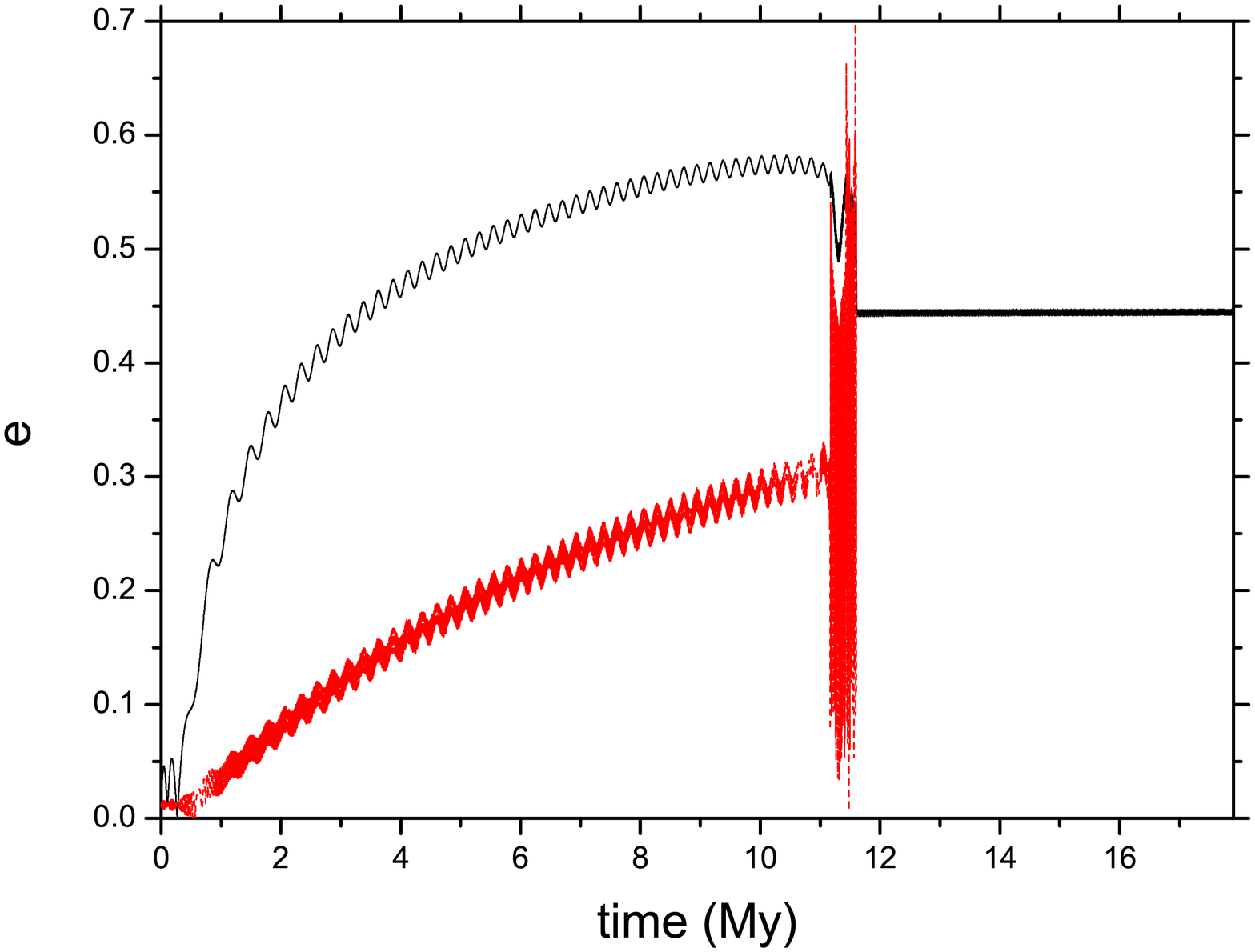}
\caption{The passage of the resonance in the inner part of the HD 181433 system, an Earth-mass planet at 0.04 AU is included. 
Left: Evolution of the semi-major axes, apocentres and pericentres of the two inner planets, with planet \textit{b} initially 
at 0.098 AU. The inner terrestrial planet is destroyed by collision after 11 My. Right: Evolution of the eccentricities.
$e_b$ is the one growing faster, achieving a final value of 0.44.}
\label{int}
\end{figure*}

To test the generality of the results presented in the previous section, 
we evaluate how a hypothetical interior terrestrial planet may have influenced the 
sweeping of the resonance. We consider a single Earth-mass planet in circular orbit with 
semi-major axes in the range 0.03-0.055 AU. We want to assess if the additional planet will 
be trapped long-term into the resonance with planet \textit{b} (as in Sect.~\ref{addpl}),
or if its presence will release \textit{b} from the resonance with the required eccentricity. 
We prepare a set of 100 runs with planet \textit{b} initially located at 0.9-0.11 AU so that 
close encounters may start when a high eccentricity has already been reached. Each sub-set has 
its own stellar spin period for resonance capture. The particular behaviour depends on the individual
run, but in general we observe that planets \textit{b} and \textit{x} are not 
quite trapped into a mutual resonance but instead experience differential precession, with the precession
periods differing by approximately 2000 years. We find that collisions do occur,
but because of the slow differential precession it takes on the order of $10^7$ years for 
them to happen. Figure \ref{int} presents a successful case: the eccentricities of the 
two inner planets grow because of the passage of the resonance, the orbits cross each other, 
after 11 Myr instability arises leading to a collision, leaving planet \textit{b} 
with the required eccentricity. Although planet \textit{b} is initiated with a semi-major
axis of 0.1, the inelastic collision causes the composite planet to effectively migrate to
0.08 AU, which is the observed value. Orbital inclinations stay small in the system. 
To generate the resonance this simulation required a stellar rotation period $\lesssim 13.4$ hours. 
A set of simulations with $a_b=0.105$ AU demands a stellar spin period of 11.3 hours. 
In these configurations the eccentricity stops growing and stabilizes around 0.45 as the secular
resonance is disrupted.

\subsubsection{Resonance passage considering stronger tidal evolution}
In the previous section we showed that sweeping secular resonances due
to stellar spin-down, combined with planet-planet scattering, can cause
eccentricity excitation of planet \textit{b} up to $e_b \simeq 0.4$,
consistent with present day observations. The HD 181433 system, however,
is unusual because of the large eccentricity of the inner planet
in spite of the $\sim 6.7$ Gyr age of the star. The implication is that tidal
dissipation inside the planet is weak compared with that measured for
terrestrial bodies in the Solar System. If the value $e_b=0.4$ is primordial
then it implies a value of $Q_{\rm p} \ge 10^5$.

A value of $Q_{\rm p} \sim 10^4$ is inferred for Neptune and Uranus \citep{b24},
and we showed in Section \ref{tides} that if such a value is adopted
for HD 181433\textit{b} then the planet is approximately half-way through
the process of circularising from an initial eccentricity of $e_b \simeq 0.6$
and semi-major axis $a_b=0.1$. This raises the question of whether or not the 
sweeping secular resonance model with additional terrestrial-mass planets can also 
achieve an eccentricity this large, and we have addressed this issue below using
a further batch of N-body simulations.

Here the planet \textit{b} is placed further out so a shorter stellar
spin period is required to generate the resonance. We infer that the
star had to rotate in $\lesssim 11.0$ hours in order for the resonance to 
sweep the location of \textit{b}. Using the nominal value of $\alpha$ 
(see equation \ref{down}) in a simulation consisting of planets
\textit{b}, \textit{c} and \textit{d} only, $e_b$ peaks at 0.28 and 
stabilizes later around a value of 0.25. If we reduce the stellar spin-down rate 
to be equal to or less than the nominal value by just 5.75 \%,
$e_b$ grows indefinitely as described above.

\begin{figure*}
\includegraphics[width=0.5\textwidth]{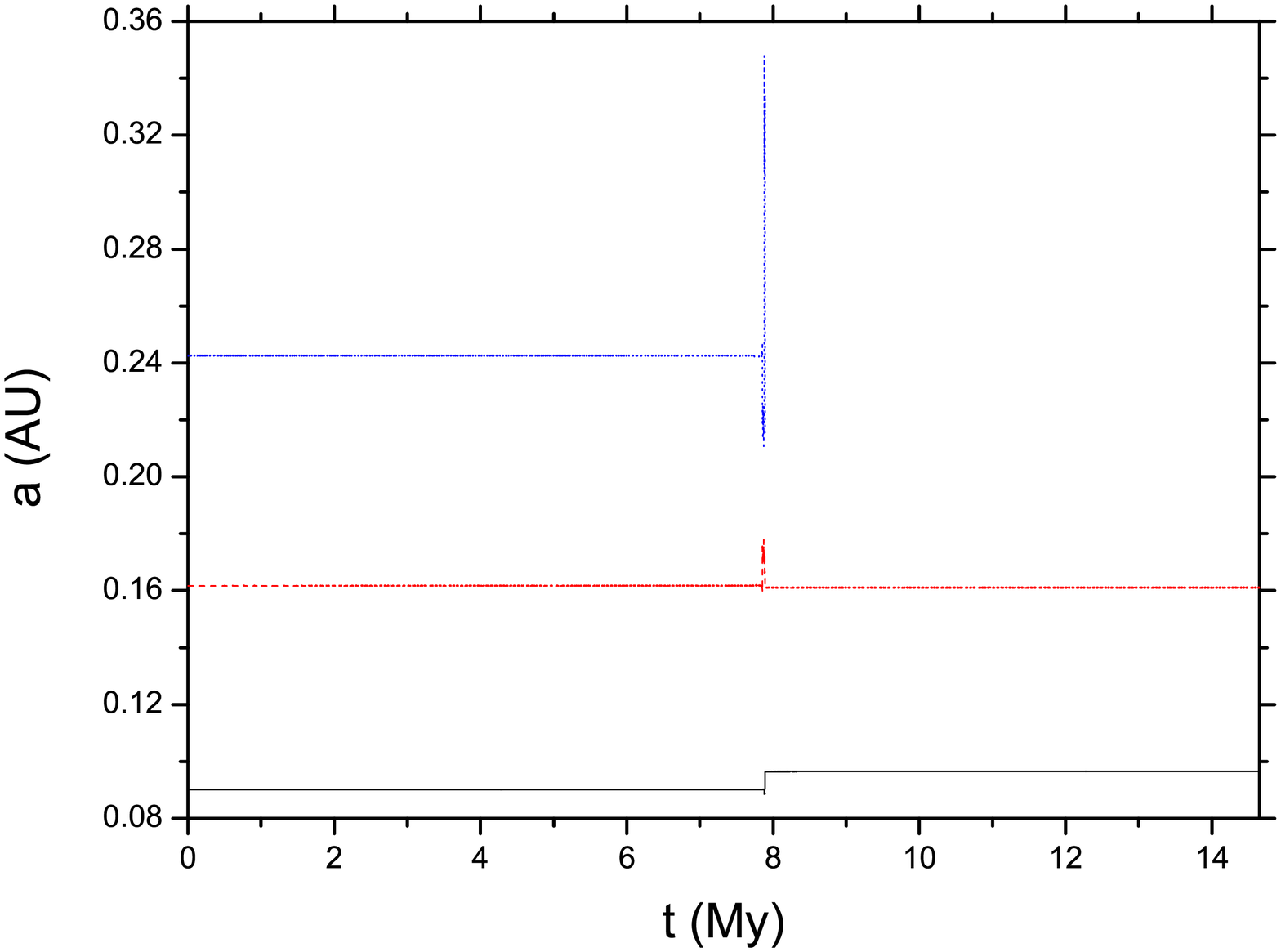}\includegraphics[width=0.5\textwidth]{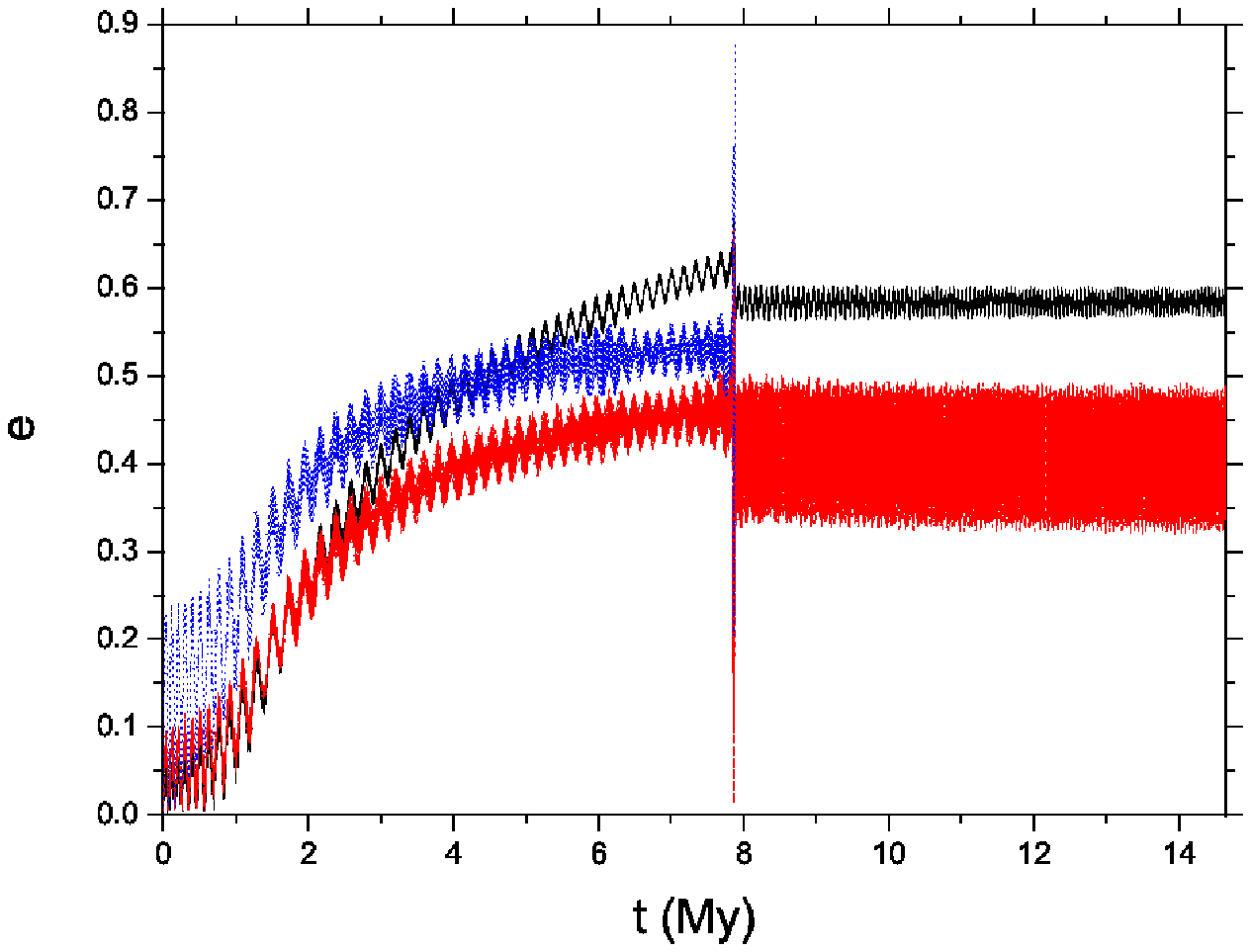}
\caption{The sweeping of secular resonances when two extra Earth-mass planets are introduced in the inner part of the 
HD 181433 system. Left: Evolution of the semi-major axes of the three inner planets, with planet \textit{b} being the 
innermost. The outer terrestrial planet is destroyed through collision after 8 My. Right: Evolution of their eccentricities.
The eccentricity of the outer planet is the one growing fastest at the beginning as it is at the outer limit of the stable zone.
A final value of 0.6 is achieved for $e_b$, while the surviving Earth-mass planet acquires an eccentricity of 0.45.}
\label{2ext}
\end{figure*}

Our original aim was to test the possibility of breaking the secular resonance
for planet \textit{b} by including three additional terrestrial planets orbiting
beyond \textit{b}, but the larger initial semi-major axis and larger final
eccentricity required of $e_b=0.6$ mean that a stable system of three additional
planets cannot be set-up because the outer one orbits outside the stable zone
located within 0.26 AU (see Section \ref{Properties} and \ref{addpl}). For this reason we ran a 
suite of simulations with two Earth-mass planets orbiting outside of planet \textit{b}.
The qualitative evolution is similar to that described in Section \ref{addpl}. 
An example of a successful simulation is displayed in Figure \ref{2ext},
showing the growth of eccentricity through secular resonance, planet-planet
scattering and collision, and two surviving inner planets with planet \textit{b} 
orbiting with $e_b \sim 0.6$ and $a_b \sim 0.1$.
The necessary stellar rotation period to generate the resonance is 11.6 hours. 
The rocky planet which survives at 0.16 AU would be in the noise level of
the measured radial velocities (F-test value of 84\%). Its orbital inclination increases 
up to 12$^{\circ}$ for the smaller survivor while it remains much smaller for planet
\textit{b}. Taking account of tides with $Q_{\rm p} =10^4$, once the initially
chaotic phase of evolution has finished the system will evolve to exhibit characteristics
very similar to those observed today on a time scale of $\sim 6$ Gyr.

Note that we have tested the stability of the system without the resonance and find 
that it remains stable over a time-scale of 1 Myr.

\begin{figure}
\centering
\includegraphics[width=\columnwidth]{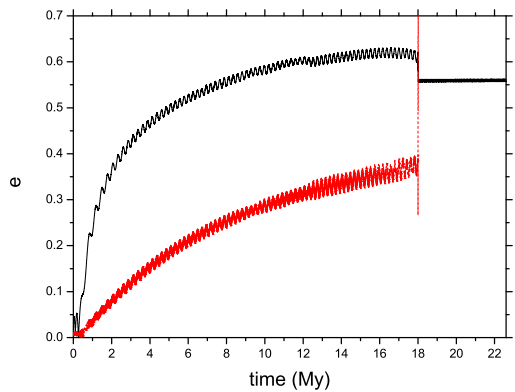}
\caption{The passage of the resonance in the inner part of the HD 181433 system when an Earth-mass planet at 0.04 AU is included. The evolution of the eccentricities is show with $e_b$ growing faster and achieving a final value of about 0.56. The inner terrestrial planet is destroyed after 18 My.}
\label{inttide}
\end{figure}

Finally, we report here also an example of evolution when an Earth-mass body is 
placed on an orbit interior to planet \textit{b}. This scenario has already been 
discussed in Section \ref{1int}. Figure \ref{inttide} shows the increase of the 
eccentricities for the two inner planets (\textit{b} and \textit{x}). Instabilities 
arise after 18 My, only planet \textit{b} survives and it is left with an eccentricity
$e_b=0.56$, close to the required value. To generate the resonance a stellar rotation 
period $\lesssim 13.4$ hours is required.

\subsection{Large forced $e_b$ by an additional planet}
The spin-down of the central star provides a means of
  enhancing the eccentricity of planet \textit{b}. In the absence of a rapidly
rotating star an additional undetected planet in the system can also
enhance the forced eccentricity of planet \textit{b} by strengthening coupling with
the outer giant planets. By carefully choosing of the mass and location
of an additional planet the forced eccentricity of planet \textit{b} may be
enhanced. We focus on the effect of a low mass planet near planet \textit{b} 
below.

\subsubsection{Exterior low mass planet}
\citet{2004ApJ...614..955M}, for example, have investigated the hypothesis that eccentric short-period planets 
have eccentricities that are excited by undetected outer companions. We have already discussed in 
Section \ref{stabadd} that an additional planet can be stable in the region 0.1 -- 0.35 AU. We use the 
secular model to estimate the planetary mass required in this range in order for the precession period of 
$b$ to match that of $c$ or of the hypothetical planet. We then refine the solution by means of numerical 
simulations, setting the eccentricities of the extra planet and \textit{b} to small non-zero values and
examining the evolution of 75 configurations for 10 Myr.

The maximum forced eccentricity obtained for these runs was $e_b=0.28$ for a 12 Earth mass companion
orbiting at 0.35 AU or for a 15 Earth mass companion orbiting at 0.3 AU. 
All other simulations have resulted in values of $e_b < 0.28$. For example, a 11 Earth mass companion orbiting at 0.25 AU resulted in a peak value of $e_b=0.27$ (see left panel of Figure \ref{eccexcit}).
%with the secular angle $\varpi_b-\varpi_c$ librating about 180 degrees with a semi-amplitude of 90 degrees (see left panel of Figure \ref{eccexcit}).
%Lower values of the planet mass are required to tune the secular resonance for smaller values of the semi-major axis down to 0.1 AU, but these simulations always resulted in values of $e_b < 0.28$. For example, a 3 Earth mass companion orbiting at 0.15 AU resulted in a peak value of $e_b=0.12$.

\subsubsection{Interior low mass planet}
\begin{figure*}
\includegraphics[width=0.5\textwidth]{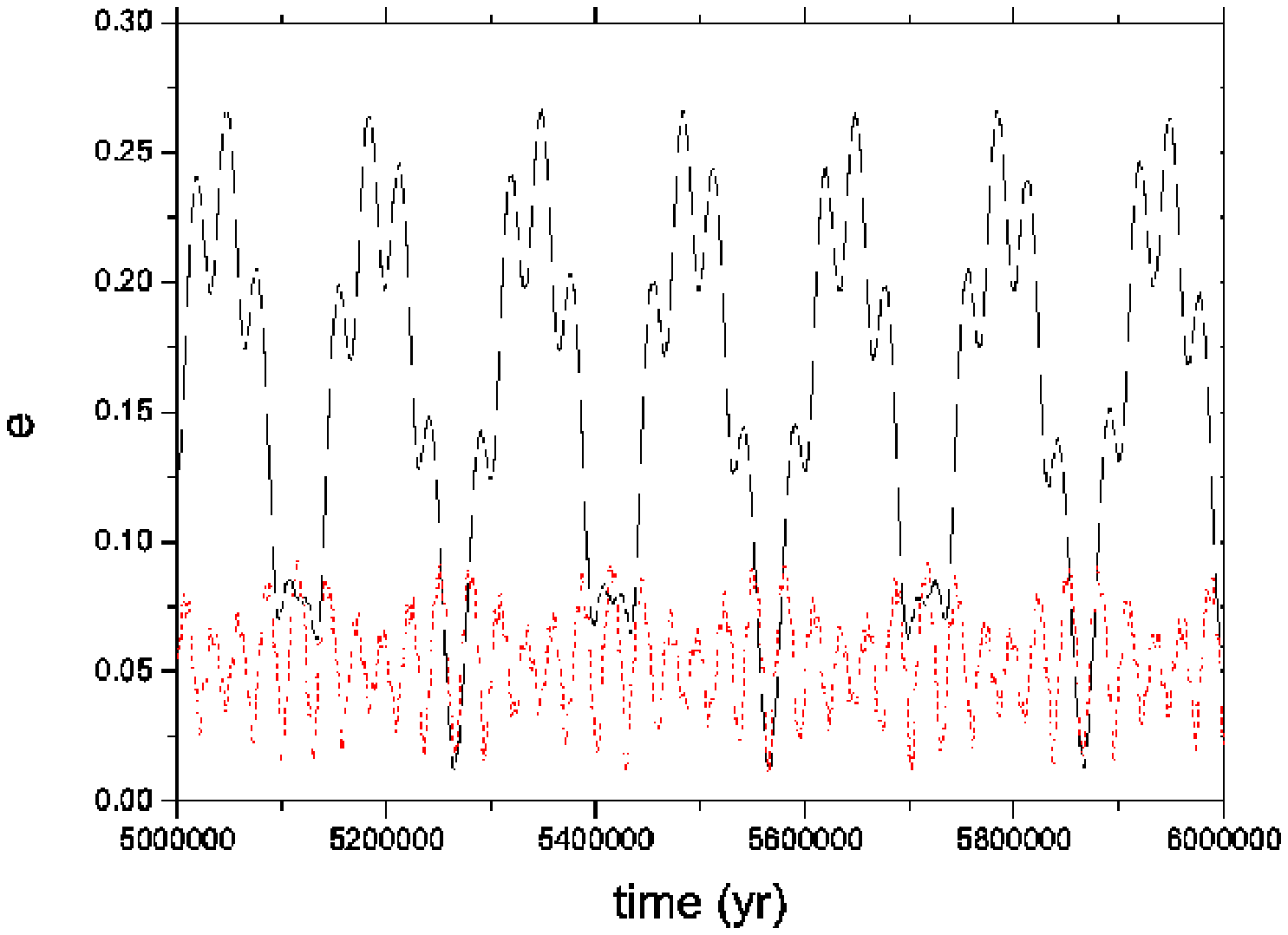}\includegraphics[width=0.5\textwidth]{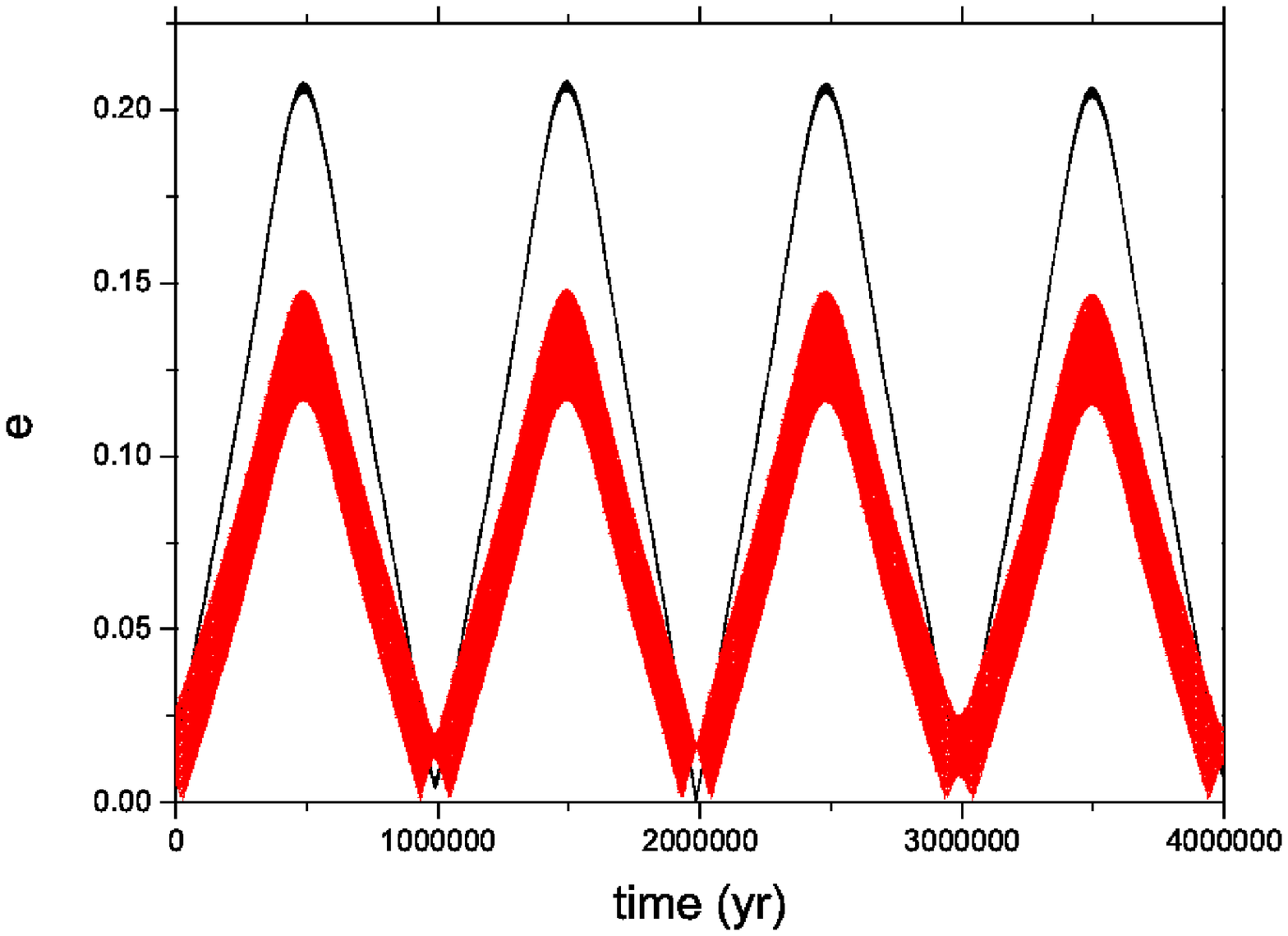}
\caption{Eccentricity excitation. Left panel: Evolution of $e_b$ and $e_x$ for a case with $a_x = 0.25$ AU and 
$m_x = 11 M_\oplus$. Maximum values for $e_b$ and $e_x$ are 0.27 and 0.09, respectively. Right panel: Evolution of $e_b$ and $e_x$ for a case with $a_x = 0.041$ AU and $m_x = 1.5 M_\oplus$. Maximum values for $e_b$ and $e_x$ are 0.21 and 0.15, respectively.}
\label{eccexcit}
\end{figure*}

We now consider the evolution when an extra planet is orbiting interior to $b$. We use the secular theory to
estimate the planetary masses required between 0.008 -- 0.062 AU for planet \textit{b} to experience a significantly enhanced forced eccentricity. Results from this secular analysis indicate that an additional planet located at $a=0.008$ AU must have a mass equal to 45 $m_{\oplus}$. An additional planet located at $a=0.062$ AU must have a mass equal to $0.13 m_{\oplus}$. 
Numerical integration of 40 initial configurations, covering the range of semi-major axes and $m_p$ discussed above, show again that $e_b$ is not forced sufficiently to explain
the currently observed value of $e_b=0.39$. The maximum forced eccentricity obtained was $e_b=0.21$, this happened in two cases: a 1.5 Earth mass companion orbiting at 0.041 AU (right panel of Figure \ref{eccexcit}) and a 0.75 Earth mass companion orbiting at 0.056 AU. 
%In these two cases the secular angle $\varpi_b-\varpi_x$ is seen to librate while $\varpi_b-\varpi_c$ circulates.
% with planet masses able to generate this value being in the range 0.75--1.5 $M_\oplus$ (e.g. bottom-left panel of Figure \ref{eccexcit}).

\subsubsection{Detectability}
We assess the detectability of these hypothetical planets assuming a radial velocity precision of
1 m/s \citep{b11}. In the range 0.1 -- 0.35 AU planets with mass as small as 4--6 $M_\oplus$ would be detectable. 
This excludes the possibility of exciting a sufficiently large $e_b$ with an unseen planet in this region. 
In the inner zone, the detection limit is about 2 Earth-masses. This cannot exclude the generation of a
forced eccentricity of 0.21.

If planet \textit{b} had been formed with a free eccentricity of $\simeq 0.18$ then in principle
the observed value of $e_b=0.39$ could have been obtained through excitation of a forced 
eccentricity equal to 0.21. This value for the free eccentricity, however, is rather
large for either of the two most plausible scenarios for the arrival of planet \textit{b}
at its current location. Gas-disc driven orbital migration of a fully formed planet \textit{b}
to its observed location is likely to have left the planet in an essentially
circular orbit. Formation {\it in situ} through a series of giant impacts may have left the
planet with a remnant free eccentricity, but a value of $e_b=0.18$ requires the scattering
bodies during the giant impacts phase to be approximately 6 Earth masses, which is only
slightly smaller than the measured minimum mass of planet \textit{b}.

\section{Discussion and conclusions}
\label{concl}
We have investigated the dynamical evolution of the three-planet system orbiting the 
main sequence K-dwarf star HD 181433 \citep{b11}. The system consists of a close-in super-Earth
with orbital period 9.37 days, and two sub-Jovian giant planets that orbit at
much larger distance from the host star with orbital periods 975 and 2468 days.
In order to be dynamically stable these outer giants need to be in a 5:2 mean motion
resonance \citep{b3}. The corresponding semi-major axes and eccentricities are 
$a_b=0.08$, $a_c=1.77$ and $a_d=3.29$ AU.
Of particular interest for our study are the large eccentricities displayed by the
system: $e_b=0.39$, $e_c=0.27$, $e_d=0.47$. The value for the short-period super-Earth
in particular raises interesting questions that we have addressed in this paper:
given the large separation between the inner super-Earth and outer giants, what are
the plausible mechanisms that can lead to excitation of $e_b$? Given the age of the system ($\sim 6$ Gyr), what are the implications of the observed value of $e_b$ for tidal evolution in the system and the $Q$-value for the inner body?

We assume that the HD 181433 planetary system attained its currently observed configuration
shortly after dispersal of the protoplanetary disc. Analysis of the secular dynamics
indicates that the eccentricity of the inner body cannot be explained through present day 
interactions between it and the outer giants. This analysis, however, does point to the 
existence of a nearby secular resonance that could have caused eccentricity growth during 
earlier evolution.

The inferred mass of HD 181433\textit{b} is $m_b \sin{i}=7.4$ M$_{\oplus}$. If $\sin{i} \sim 1$
then this planet is a super-Earth. We have analysed the tidal evolution and conclude that
if the tidal dissipation factor $Q_{\rm p} \le 10^3$ then tidal circularisation should
be completed easily within the 6.7 Gyr age of the system. This suggests that 
HD 181433\textit{b} is not a massive terrestrial-like planet. A value of 
$Q_{\rm p} \ge 10^5$ leads to very little tidal evolution, implying that the system 
observed now is similar to its primordial state if this $Q$-value is appropriate. 
A value of  $Q_{\rm p} \simeq 10^4$, characteristic of Neptune and Uranus 
\citep{b24}, would indicate that the system started out with $a_b \simeq 0.1$ AU and $e_b \simeq 0.6$
such that the system is essentially half-way through the tidal circularisation process. This
might also indicate that HD 181433\textit{b} should be considered as a hot-Neptune rather
than a short-period super-Earth.

Given that all bodies in the system have quite large eccentricities, we began our study by analysing
a scenario in which an additional outer giant planet was present 
originally, leading to global dynamical instability and ejection of this additional body. 
This is a natural starting point given the recent work showing that planet-planet
scattering can explain the observed eccentricity distribution of the extrasolar planet
population \citep{b5,2008ApJ...686..603J}. The chaotic dynamics involved
in such a scenario preclude us from obtaining a close-analogue to the HD 181433 system through 
N-body simulations, so we restrict our analysis to addressing the following two issues as a 
means of estimating the likelihood that the scenario may have operated:
can the short-period super-Earth be perturbed onto an eccentric orbit with the required
$e_b \ge 0.4$ during the dynamical instability, given that it orbits close to the central star?; 
what is the likelihood of the two outer planets \textit{c} and \textit{d} landing in the 
5:2 resonance after scattering? From a suite of 300 N-body simulations we find that 
$e_b$ reaches $e_b \simeq 0.4$ in 6\% of the runs, and reaches $e_b=0.6$ in 2\% of them,
indicating that we can obtain values of $e_b$ that are appropriate for a range of tidal
histories. Further analysis of the width of the 5:2 resonance suggests a probability for landing 
in resonance of $\sim 10^{-3}$, in good agreement with the more extensive set of N-body 
simulations of planet-planet scattering reported by \citet{b2}. A naive estimate of the joint 
probability of eccentricity excitation for planet \textit{b} and formation of the 5:2 resonance 
for planets \textit{c} and \textit{d} suggests a value $\sim 6 \times 10^{-5}$, indicating that 
the hypothetical planet-planet scattering scenario is plausible but only occurs as a rather rare 
event in planetary system evolution.

Given this conclusion regarding the planet-planet scattering scenario,
we also considered the possibility that secular resonance 
caused excitation of planet \textit{b}'s eccentricity.
We examined how an additional planet in the system may have
  increase the secular forcing of planet \textit{b}'s eccentricity, but this
  idea is discounted because the eccentricity obtained for masses
  below the detectability threshold was too small.
During pre-main-sequence evolution most host stars are rapid rotators 
\citep[e.g.][]{b35}. Rotational flattening introduces a $J_2$ component in the 
moments of the stellar gravitational potential. 
Precession caused by this effect can be important in determining the evolutionary fate of 
short-period planets \citep[e.g.][]{b30,b10}. The semi-major axis of planet \textit{b} around 
HD 181433 is $\sim$ 0.08 AU, close to a secular resonance.
Using customised N-body simulations that incorporate precession due to GR and stellar oblateness,
and evolution of the stellar spin through magnetic braking, we have tested the hypothesis that 
the resonance swept past the location of \textit{b}, generating the large eccentricity. 
We identify two distinct modes of evolution: for nominal values of the stellar spin-down parameter
planet \textit{b} is trapped in the resonance temporarily, leading to a maximum growth of
eccentricity to $e_b=0.25$; for values of the stellar spin-down parameter that are marginally
smaller than the nominal value (i.e only 5.75\% slower), planet \textit{b} becomes
trapped in the resonance indefinitely with its eccentricity being driven toward unity.
The final fate of the system in this latter case appears to be collision with the
central star. Neither of these two outcomes leads to a system that looks
like HD 181433\textit{b}, but we find that the inclusion of additional short-period
low mass planets in the system, orbiting in the vicinity of planet \textit{b}, can 
perturb it out of long-term resonant capture when its eccentricity has reached large
values. Such a scenario can account successfully for the observed orbital properties 
of planet \textit{b} for a range of tidal histories, but in most cases we require a 
stellar spin period $\lesssim 20$ hours for resonant capture to occur. Given that
we assume the resonant interaction occurs shortly after protoplanetary disc removal,
this scenario requires that HD 181433\textit{a} was a member of the young stellar population
that displays rotation periods of less than 2 days. While these stars are in the minority,
they are by no means rare \citep{b35}, suggesting that this scenario provides a plausible
explanation for the high eccentricity observed for HD 181433\textit{b}.

\section*{Acknowledgements}

G.C. acknowledges the support of a PhD studentship from Queen Mary University of London. We are grateful to the referee Makiko Nagasawa for valuable comments that improved this paper.

\bsp

\label{lastpage}

\end{document}